\documentclass[12pt]{article}  
\usepackage{titling}
\usepackage{booktabs} 
\usepackage{setspace}
\usepackage{footmisc}
\renewcommand{\footnotelayout}{\setstretch{1.5}}
\usepackage{times}
\usepackage{fullpage}
\usepackage[round]{natbib}
\usepackage{comment, commath, float, caption, threeparttable}
\usepackage{amsmath,amssymb, bbm, bm, amsthm, nicefrac, multirow}
\usepackage{graphicx}
\usepackage[a4paper]{geometry}
\usepackage[usenames,dvipsnames]{xcolor}
\definecolor{dukeblue}{rgb}{0.0, 0.0, 0.61}
\usepackage[round]{natbib}
\usepackage{algorithm}
\usepackage[]{algpseudocodex}
\usepackage[hyperfootnotes=false]{hyperref}
\hypersetup{
	colorlinks,
	citecolor=dukeblue,
	linkcolor=dukeblue,
	urlcolor=dukeblue}
\newtheorem {theorem}{Theorem}

\newtheorem {corollary}{Corollary}

\newtheorem*{procedure}{Procedure}

\newtheorem {lemma}{Lemma}

\newtheorem {proposition}{Proposition}

\newtheorem {assumption}{Assumption}
\newcommand\blfootnote[1]{%
	\begingroup
	\renewcommand\thefootnote{}\footnote{#1}%
	\addtocounter{footnote}{-1}%
	\endgroup
}
\graphicspath{{"Images/"}}
\title{Identification of dynamic treatment effects when treatment histories are partially observed}
\author{Akanksha Negi$^\dag$, Didier Nibbering$^\dag$}
\date{June 2025}

\begin{document}
	
\maketitle

\blfootnote{$^\dag$Department of Econometrics and Business Statistics, Monash University. Emails: \href{mailto:akanksha.negi@monash.edu}{akanksha.negi@monash.edu},  \href{mailto:didier.Nibbering@monash.edu}{didier.nibbering@monash.edu}. We would like to thank Christian Cox, Chris Muris, Hidehiko Ichimura, Irene Botosaru, Hashem Pesaran, James Powell, Otavio Bartalotti, Tiemen Woutersen, and Wooyong Lee for comments.} 

	\begin{abstract} 
        This paper presents a general difference-in-differences framework for identifying path-dependent treatment effects when treatment histories are partially observed. We introduce a novel robust estimator that adjusts for missing histories using a combination of outcome, propensity score, and missing treatment models. We show that this approach identifies the target parameter as long as \textit{any two} of the three models are correctly specified. The method delivers improved robustness against competing alternatives under the same set of identifying assumptions. Theoretical results and numerical experiments demonstrate how the proposed method yields more accurate inference compared to conventional and doubly robust estimators, particularly under nontrivial missingness and misspecification scenarios. Two applications demonstrate that the robust method can produce substantively different estimates of path-dependent treatment effects relative to conventional approaches.
\end{abstract}
	
	\textbf{JEL Classification Codes:} C14, C21, C23
	
	\textbf{Keywords:} Parallel trends, Missing treatments, Panel data, Dynamic treatment effects, Robust, Difference-in-differences
	
	\newpage 

\section{Introduction}\label{sec:intro}
Estimating dynamic treatment effects with panel data is often a central goal in applied research. Many empirical settings involve binary time-varying treatments (such as health shocks, medicare churning, or union membership) whose effects may persist well beyond the initial intervention period and depend on the full history of prior exposures. Difference-in-differences (DID) provides a compelling framework for studying such dynamics by leveraging repeated observations to control for unobserved time-invariant heterogeneity. However, identification of such path-dependent effects is challenging if a complete history of treatment decisions is not observed, whether that is due to i) survey non-response  \citep{pepper2001response},  ii) attrition in repeated surveys \citep{ghanem2024correcting}, or iii) not observing some individuals in certain time periods, as in the case of rotating panels \citep{bellego2024chained}. Standard approaches such as complete case analysis or imputation methods are valid only under relatively restrictive assumptions about the missingness mechanism or correct model specification. These conditions can often be violated in practice and lead to biased or inefficient estimates.

To illustrate this challenge, consider a stylized example of a balanced panel constructed from two non-consecutive survey waves. In this setting, a standard pre-post DID analysis that ignores the intermediate history generally fails to identify a causal parameter. We formally show that this estimand (which ignores persistence) identifies a non-convex weighted average of different path-dependent average treatment effects (PDATTs), which may not correspond to a causally meaningful quantity. This failure in identification is related to problems in the literature that discuss incorrect aggregation of heterogeneous causal effects.\footnote{See  \cite{goodman2021difference, ishimaru2021we, callaway2021difference, sun2021estimating, imai2021use, de2020two}, and others.} Moreover, the common alternative of relying on complete cases (CC), which essentially excludes observations with missing histories, only recovers specific PDATTs under strong assumptions that limit the heterogeneity of treatment paths by excluding adoption in period one, excluding dropouts or late-adopters, ruling out persistence, or imposing staggered adoption.\footnote{We show that a specific convex weighted average of PDATTs is partially identified under a monotone treatment response condition \citep{molinari2010missing}.} 

In this paper, we introduce a general framework for identifying path-dependent treatment effects in short panels when treatment histories are partially observed. Our setting allows for binary time-varying treatments which can switch on or off in each period, with no adoption in the initial period, and nests staggered adoption as a special case.\footnote{See \cite{roth2022s} for a recent synthesis of the current DID literature with discussions on staggered adoption, violation of parallel trends, and design-based inference.} We develop a novel identification result for PDATTs under a missing-at-random selection mechanism and straightforward extensions of the DID identifying assumptions. The estimand adjusts for missing treatment histories by re-weighting observations based on the probability of experiencing a particular treatment path (\textit{propensity score}) and the probability of it being observed in the population (\textit{missing data probability}), combined with the conditional mean of outcomes (\textit{outcome regression}). These elements are combined into a new augmented inverse probability weighted (AIPW) type estimand. Importantly, identification holds if \textit{any two} of the three models involved - the outcome model, the propensity score model, or the missing treatment model - are correctly specified. Our identification result also nests missingness-adjusted versions of (i) outcome regression (OR), (ii) inverse probability weighting (IPW), and (iii) doubly robust (DR) estimands. However, each of these alternatives requires a correct missing data model along with at least one additional correctly specified model. In contrast, our proposed estimand identifies the target parameter even if the missing data model is misspecified.

Based on this identification result, we construct a two-step \textit{robust} (R) estimator. The  first step estimates the true nuisance functions (probability weights and outcome regression)\footnote{Our identification results are agnostic about the nature of the first-step estimators, and hence machine learning methods may be employed for estimating the three nuisance functions, especially in situations where large administrative datasets are available. Note that our theoretical results for inference are developed for parametric first-step estimators, and therefore do not cover the choice of machine learning methods or cross-fitting procedures.} and the second step plugs-in the estimated first-stage parameters into the sample analogue of the proposed estimand. We establish formal results on identification, estimation, and inference with the robust procedure. When all three models are correctly specified, the robust estimator attains the semiparametric efficiency bound for the PDATT parameter, making it efficient within the class of missingness-adjusted estimators. This result follows from our derivation of the associated efficiency bound. Moreover, since OR, IPW, and DR are nested within the robust proposal, inference with these methods is also made available.

Although the proposed method identifies the target parameter under misspecification of at-most one model, inference may still be affected. Specifically, the effect of estimating first-stage parameters indexing a misspecified model may propagate into the second step, altering the form of the asymptotic variance of the robust estimator. To reduce the effect of misspecification on inference, we further refine the asymptotic properties of the robust estimator by proposing two alternatives. Our approach builds on the recommendations in \citet{vermeulen2015bias}, who propose minimizing the first-order effect of nuisance parameter estimation on a second-stage parameter of interest, and generalizes \citet{sant2020doubly}'s improved estimation results to settings with persistent treatment effects and/or missing treatment histories.

Numerical experiments help to evaluate the performance of the \textit{robust} estimator against other missingness-adjusted estimators and CC-DID methods. First, we show that the robust estimator remains unbiased if either the missing data model, propensity score model, or outcome regression model is misspecified. In contrast, DR, IPW, and OR are biased whenever the missing data model is misspecified, irrespective of the specification of the other nuisance functions. Second, inference based on the robust estimator has accurate test size across all experiments, whereas the DR, IPW, and OR estimators show considerable size distortions when the missing data model is misspecified. Third, experiments varying the extent of missingness and the degree of misspecification in the missing data model reveal that the bias in CC-DR and DR estimators grows as missingness rates or misspecification severity increases. In contrast, the robust estimator continues to perform well.

We conclude by demonstrating the practical relevance of the robust estimator with two empirical applications. The first investigates the persistent effects of COVID-19 cases on county-level voter turnout in the 2022 U.S. general elections, where case histories are missing for 51\% of the counties. The robust estimator suggests a statistically significant reduction in voter turnout of 0.18\% points for counties that experienced above-average number of cases in 2020 and 2021, while standard CC-DID estimates suggest a negligible and statistically insignificant reduction in voter turnout between 0.01\% and 0.03\% points. The second application uses individual-level data from the Current Population Survey (CPS) to study the effects of worker disability, job certification, and work absence on family income and hours worked. While missingness in treatment histories is modest (under 5\%), a meta-analysis across the three treatments reveals that the robust estimator can yield estimates that differ substantially from those obtained using CC-DR methods. 

\paragraph{Relation to the literature:} We contribute to a growing body of literature which allows outcomes to be affected by the entire treatment path. An early example is \citet{hull2018estimating}, who studies two-way-fixed-effects regressions for mover panels and imposes some version of conditional mean impersistence. \citet{strezhnev2018semiparametric} develops inverse propensity score weighted DID estimators for estimating persistent treatment effects with multiple time periods. \citet{de2022two} and \citet{de2024difference} extend their earlier work on interpretations of two-way-fixed-effects regressions to allow for several treatments and treatment lags, respectively. \citet{viviano2021dynamic} propose a dynamic covariate balancing method for estimating the effects of different treatment trajectories. None of these papers address the challenge of missing treatment histories.

In the panel data literature, our paper is broadly related to the strand studying missing covariates or treatments. \citet{abrevaya2017gmm} present a generalized method of moments (GMM) estimator for dealing with missing regressors. \citet{muris2020efficient} provides a GMM framework for efficient parameter estimation with incomplete data and \citet{botosaru2018difference} propose a proxy-variable solution to address the problem of a missing treatment variable within a standard DID analysis with repeated cross sections. Finally, \citet{coe2019estimation} proposes an inverse probability weighted solution for pooled ordinary least squares and first-differenced moments. 

There is also a rich literature on robust estimation of treatment effects.\footnote{See \cite{robins1994estimation, scharfstein1999adjusting, graham2012inverse, bang2005doubly, sloczynski2018general, lewbel2023over, negi2024doubly} for doubly robust estimators in cross-sectional settings.} In the context of panel data, \citet*{arkhangelsky2021double} develop an augmented doubly robust two-way-fixed effects estimator and \citet{arkhangelsky2022doubly} integrate design-based and model-based identification strategies to construct a doubly robust alternative. Our paper is closely related to the papers by \citet{sant2020doubly} (SZ) and \citet{callaway2021difference} (CS), who propose doubly robust estimators for ATTs in simple and staggered adoption settings, respectively. Our PDATT estimator equals the proposal in SZ and CS in special cases. \citet{yanagi2022doubly} generalizes SZ and CS to allow for general treatment patterns across multiple time periods. Implicitly, these papers assume that treatments are fully observed. Recently, \cite{bellego2024chained} propose a chained DID method that combines short-term treatment effects from many incomplete unbalanced panels to estimate long-run effects in staggered settings.\footnote{In the high-dimensional literature, \cite{farrell2015robust} introduces a doubly robust estimator for constructing confidence intervals for the ATE after model selection and \cite*{chernozhukov2022locally} propose locally robust orthogonal moment conditions that also exhibit doubly robust properties.} 

The statistics literature explores multiply robust estimation of average treatment effects. 
This spans mediation analysis \citep*{xia2023identification, tchetgen2014estimation, jiang2022multiply}, missing outcomes \citep{han2014multiply, han2013estimation}, and missing treatment information \citep*{zhang2016causal}. \citet*{shi2020multiply} propose a multiply robust ATE estimator in the presence of categorical unmeasured confounding and negative controls, while \citet{wang2018bounded} use instrumental variables, and \citet*{wei2023multiply} study nonrandom assignment and missing outcomes. \citet{zhang2016causal} examine robust estimation of ATEs in a cross-sectional setting with missing treatment data with a binary outcome, and propose an estimator which exhibits properties similar to ours under model misspecification. 

The rest of the paper is organized as follows. Section \ref{sec:model} introduces our framework, the parameters of interest, and the identifying assumptions. Section \ref{sec:robust} presents the identification, estimation, and inferential results with the proposed approach. Section \ref{sec:adjusted} discusses other missingness-adjusted estimands that, while less robust, are nested within our theoretical results. Section \ref{sec:sims} presents numerical experiments comparing the different estimators and Section \ref{sec:emp_app} illustrates the estimators in two empirical applications. Section \ref{sec:conclude} concludes. 

\section{General treatment patterns and missing treatments}\label{sec:model} 
Let $Y_t$ be the observed outcome at time period $t$, and $D_t$ be a binary treatment which is equal to one if an individual is treated in period $t$ or zero otherwise. Assume that there is no treatment in the baseline with $D_0=0$. Additionally, we observe a $k$-dimensional vector of pre-treatment characteristics $\mathbf{X}$.\footnote{It is standard in the DID literature to only consider time-invariant covariates or to condition only on the pre-treatment values for any time-varying covariates (see \cite{callaway2019quantile} and references therein).} 
For ease of exposition, consider a setting with three time periods denoted by $t=0,1,2$. The treatment history is then denoted by $\mathbf{D}=(D_1,D_2)$ and we define $\Delta Y= Y_2-Y_0$. Empirically, treatment histories may be partially observed. To formalize this, let $S$ be a binary indicator which is equal to one if $D_1$ is observed and zero otherwise. Extension of this framework to a general short panel setting with an arbitrary number of time periods and general missing treatment history patterns is discussed in Section~\ref{sec:general}. 

\subsection{Causal parameters of interest}
A parameter that is of interest to policy makers is the effect of a particular treatment history on final period ($t=2$) outcomes. We define the average effect of experiencing treatment path $\mathbf{D} = \mathbf{d}$ compared to $\mathbf{d}^\prime$ for individuals who experienced path $\mathbf{d}$ in period $t=2$ as 
\begin{equation}\label{eq:att}
	\tau_{\mathbf{dd^\prime}} = \mathbb{E}[Y_2(\mathbf{d})-Y_2(\mathbf{d^\prime})|\mathbf{D}=\mathbf{d}],
\end{equation} 
where $Y_t(\mathbf{d})$ denotes the potential outcome in period $t$ if the treatment history $\mathbf{D}$ takes the value $\mathbf{d}=(d_1, d_2)\in \{0,1\}^2$. Our parameters of interest always consider $\mathbf{d^\prime} = (0,0)$.\footnote{Supplementary Appendix \ref*{sec:cpt_alt} discusses identification of $\tau_{\mathbf{dd^\prime}}$ which involves comparisons with $\mathbf{d^\prime} = (1,1)$. Such comparisons would necessitate assuming parallel trends in the treated counterfactual distribution - an assumption that is practically never invoked in the literature.} The observed outcome is $Y_t  =  Y_t(\mathbf{D})$. With two treatments, the definition in \eqref{eq:att} allows for three PDATTs. Summaries of these effects may also be of interest. For instance, the average effect of receiving the second treatment ($D_2=1$) compared to not receiving any treatment equals $\tau_{(11)(00)}\mathbb{P}(D_1=1|D_2=1)+\tau_{(01)(00)}\mathbb{P}(D_1=0|D_2=1)$.

Our setup covers a wide range of policy-relevant treatment settings, including both sequential and simultaneous interventions. In an educational context, $D_1$ and $D_2$ could represent college enrollment and graduation, respectively. Here, $\tau_{(11)(00)}$ measures the effect of the whole college program, $\tau_{(10)(00)}$ measures the impact of enrollment for dropouts, and $\tau_{(01)(00)}$ captures the effect of graduation for late-adopters. \citet{nibbering2024instrument} show that in general, treatment programs with dropouts and late enrollment are ubiquitous. Similarly, in a workforce development program \citep{katz2022sectoral}, $D_1$ and $D_2$ might represent an initial job training program followed by an internship, with PDATTs defined analogously.

With staggered treatment adoption, such as an irreversible implementation of state-level minimum wage laws \citep{callaway2021difference}, PDATTs will capture differential effects of early ($\tau_{(11)(00)}$) versus late adoption ($\tau_{(01)(00)}$). One can also use this framework to study simultaneous treatments that may be correlated in time \citep{de2023two}. For instance, with labor market policies, minimum wage regulations and working hours restrictions may be implemented together, making it important to jointly account for them when studying outcomes of interest. The PDATTs remain relevant here for capturing heterogeneous treatment effects corresponding to simultaneous, and potentially, interactive policies.

In order to identify the PDATTs in \eqref{eq:att}, we extend the difference-in-differences assumptions to allow potential outcomes to depend on the full history of treatment decisions.

\begin{assumption}[Difference-in-differences assumptions]\label{ass:did} \, \\ 
For each $\mathbf{d}$, we have
\begin{enumerate}
    \item (No anticipation) $\mathbb{E}\left[Y_0(\mathbf{d})|\mathbf{D}=\mathbf{d},\mathbf{X}\right] = \mathbb{E}\left[Y_0(\mathbf{0})|\mathbf{D}=\mathbf{d},\mathbf{X}\right]$.
    \item (Parallel trends) $\mathbb{E}\left[Y_2(\mathbf{0})-Y_0(\mathbf{0})|\mathbf{D}=\mathbf{d}, \mathbf{X}\right] = \mathbb{E}[Y_2(\mathbf{0})-Y_0(\mathbf{0})|\mathbf{X}]$.
	\item (Overlap) $\mathbb{P}(\mathbf{D}=\mathbf{d}|\mathbf{X})\equiv p_\mathbf{d}(\mathbf{X})$ is bounded away from one.	
\end{enumerate}
\end{assumption}

Assumption~\ref{ass:did}.1 rules out any anticipatory effects of future treatment on outcomes at $t=0$. Violations arise if individuals change their behavior in anticipation of the treatment. Assumption~\ref{ass:did}.2 imposes that the average trend in the untreated potential outcome of the treatment and comparison groups would have evolved in parallel between $t=0$ and $t=2$, conditional on $\mathbf{X}$. This is commonly referred to as conditional parallel trends. Assumption \ref{ass:did}.3 is an overlap or common support condition which bounds the propensity score $p_\mathbf{d}(\mathbf{X})$ away from one.

\subsection{No causal interpretation with standard DID methods}
A natural starting point when only partial information on $D_1$ is available, is to completely ignore $D_1$ and conduct a conditional pre/post DID analysis using the first and final time period, while varying $D_2$. As we show below, this strategy fails to identify a causal parameter. 
\begin{proposition}[A non-convex weighted average of PDATTs]\label{prop:ignoreD1}\ \\
Under Assumption~\ref{ass:did}, 
the DID estimand that conducts a pre/post DID analysis with $D_2$ identifies 
\begin{align}\label{eq:did02_id}
		&\mathbb{E}[D_2]^{-1}\mathbb{E}\left[D_2\left(\mathbb{E}[\Delta Y|D_2=1,\mathbf{X}]-\mathbb{E}[\Delta Y|D_2=0,\mathbf{X}]\right)\right]=\\
        & \tau_{(11)(00)}\cdot \mathbb{P}(D_1=1|D_2=1)
        +\tau_{(01)(00)}\cdot \mathbb{P}(D_1=0|D_2=1) 
		-\tau_{(10)(00)}\cdot\mathbb{P}(D_1=1|D_2=0).\notag
\end{align}
\end{proposition}
Proof is deferred to Appendix~\ref{sec:proof_ignoreD1}. The conditional DID estimand which ignores $D_1$ identifies a non-convex weighted average of three PDATTs, with weights given by different conditional treatment probabilities. This estimand does not have a causal interpretation unless one is willing to impose additional assumptions on treatment adoption. For instance, if dropouts and late-adopters are ruled out, and $D_1=D_2$, the estimand identifies $\tau_{(11)(00)}$. When treatment in period $t=1$ can be ruled out i.e.\ $D_1=0$, we recover $\tau_{(01)(00)}$. Under the assumption of treatment impersistence ($Y_2(0,d_2)=Y_2(1,d_2)$ for each $d_2$) or the assumption of staggered adoption ($D_2\geq D_1$), we identify the weighted average $\tau_{(11)(00)}\cdot \mathbb{P}(D_1=1|D_2=1)+\tau_{(01)(00)}\cdot \mathbb{P}(D_1=0|D_2=1)$. Hence, this DID approach only recovers specific causal parameters under strong assumptions, but cannot identify individual PDATTs that are permissible in the general case.\footnote{Supplementary Appendix~\ref*{sec:partial_DID} shows that a convex weighted average of specific PDATTs is partially identified under a monotone treatment response assumption.}

\subsection{Missing treatment histories}
Participation in treatments may be missing for a variety of reasons. 
First, missingness may arise due to item non-response where survey questions elicit information about sensitive behaviors such as drug use and alcohol consumption \citep{pepper2001response}.  
Second, treatment participation may only be reported partially, thereby obscuring whether individuals adhered to the full treatment program,  dropped-out, or adopted late \citep{silliman2022labor, zimmerman2014returns}. For example, information about when treatment was initiated may be available, but actual adoption timing might be unknown. Additionally, treatment data may also be missing for individuals due to noncompliance with the assigned treatment. 
Finally, when panel data are constructed from repeated surveys, attrition can pose a significant problem \citep{ghanem2024correcting}. If the attrited sample is systematically different from the observed sample, attrition bias can distort treatment effect estimates. We impose the following assumptions on the missing treatment mechanism.
\begin{assumption}[Missingness assumptions]\label{ass:mar} \ 
	\begin{enumerate}
		\item (Missing at random) $ S \perp (D_1, \Delta Y)| D_2, \mathbf{X}$.
		\item (Partial observability) $0< \mathbb{P}(S=1|D_2=d_2,\mathbf{X}) \equiv q_{d_2}(\mathbf{X}) \leq1$. 
	\end{enumerate}
\end{assumption}
Assumption~\ref{ass:mar}.1 is a novel missing-at-random (MAR) assumption tailored to our DID setting with missing treatments. It permits missingness in $D_1$ to be correlated with the fully-observed treatment and covariates, and subsumes both the stronger version of MAR, $S \perp (\Delta Y, \mathbf{D})|\mathbf{X}$, and the missing completely at random (MCAR) version, $S \perp (\Delta Y, \mathbf{D}, \mathbf{X})$.\footnote{Supplementary Appendix~\ref*{sec:weakmar} discusses an alternative MAR assumption that allows missingness to be correlated with $\Delta Y$, explains why existing estimands and those proposed in this paper are biased in this case, and proposes a novel estimand that remains unbiased under certain conditions.} 
Assumption 2.2 ensures that for each group defined by $(D_2,\mathbf{X})$, there is a positive probability of observing $D_1$. 

From Assumption~\ref{ass:mar}.1, it follows that $\mathbbm{E}[\Delta Y|\mathbf{D}=\mathbf{d}, \mathbf{X}, S=s] = \mathbbm{E}[\Delta Y|\mathbf{D}=\mathbf{d}, \mathbf{X}]$, which has implications for the potential outcomes. For the comparison group with $\mathbf{D}=\mathbf{0}$, this imposes conditional parallel trends between the observed and unobserved comparison groups: $\mathbbm{E}[Y_2(\mathbf{0})-Y_0(\mathbf{0})|\mathbf{D}=\mathbf{0}, \mathbf{X}, S=s] = \mathbbm{E}[Y_2(\mathbf{0})-Y_0(\mathbf{0})|\mathbf{D}=\mathbf{0}, \mathbf{X}]$. 
This assumption is made on the outcome trends instead of levels, and therefore the latter can still depend on the missingness mechanism. 
For the treatment groups, we have $\mathbbm{E}[Y_2(\mathbf{d})-Y_2(\mathbf{0})+Y_2(\mathbf{0})-Y_0(\mathbf{0})|\mathbf{D}=\mathbf{d}, \mathbf{X}, S=s] = \mathbbm{E}[Y_2(\mathbf{d})-Y_2(\mathbf{0})+Y_2(\mathbf{0})-Y_0(\mathbf{0})|\mathbf{D}=\mathbf{d}, \mathbf{X}]$ which requires (i) conditional parallel trends between the observed and unobserved treated groups and (ii) conditional independence between the treatment effects and the missingness mechanism. 
In particular, Assumption~\ref{ass:mar} can be violated if treatment effects vary with the missingness mechanism even after conditioning on observables, despite conditional trends being the same across observed and unobserved groups. \citet{shin2024difference} discusses similar assumptions in a DID setting with missing outcomes. 

\subsection{No causal interpretation with complete case DID methods}
A common empirical strategy to deal with missing data is to restrict the analysis to the subset of observations for whom treatment histories are fully observed, also known as complete-case (CC) analysis. In the current setting, this entails conducting the DID analysis on the set of observations for whom $S=1$. While this approach is simple and avoids the need for imputation or weighting adjustments, it will produce inconsistent estimates of PDATT unless the missingness mechanism satisfies the stricter MCAR assumption. The following proposition provides an explicit expression of selection bias introduced by this method within our setup.
\begin{proposition}[Bias with CC-DID]\label{prop:ccbias}\ \\
Under Assumptions \ref{ass:did} and \ref{ass:mar}, the DID estimand that uses the observed sample (also known as complete cases), identifies
\begin{align}\label{eq:ss}
      &\mathbb{E}\left[S\mathbbm{1}[\mathbf{D}=\mathbf{d}]\right]^{-1}\mathbb{E}\left[S\mathbbm{1}[\mathbf{D}=\mathbf{d}]\left(\mathbb{E}[\Delta Y|\mathbf{D}=\mathbf{d},S=1, \mathbf{X}]-\mathbb{E}[\Delta Y|\mathbf{D}=\mathbf{d}^\prime, S=1, \mathbf{X}]\right)\right]=\notag\\
      &\qquad\tau_{\mathbf{dd^\prime}} + \mathbb{P}(S=0|\mathbf{D}=\mathbf{d})\times\notag\\&\int_\mathbf{X} \mathbb{E}[Y_2(\mathbf{d})-Y_2(\mathbf{d}')|\mathbf{D}=\mathbf{d},\mathbf{X}]\left(\mathbb{P}(\mathbf{X}|\mathbf{D}=\mathbf{d},S=1)-\mathbb{P}(\mathbf{X}|\mathbf{D}=\mathbf{d},S=0) \right)d\mathbf{X}.
\end{align}
\end{proposition}

Proof is in Appendix \ref{sec:proof_prop1}. Proposition~\ref{prop:ccbias} shows that in the presence of missing treatments ($\mathbb{P}(S=0|\mathbf{D}=\mathbf{d})\neq 0$), the CC estimand is biased if the covariate distributions for the groups experiencing treatment path $\mathbf{d}$ are different between the observed and unobserved subpopulations: $\mathbb{P}(\mathbf{X}|\mathbf{D}=\mathbf{d},S=1)\neq\mathbb{P}(\mathbf{X}|\mathbf{D}=\mathbf{d},S=0)$. Such selection bias arises due to the fact that the PDATTs require the integration over $\mathbf{X}$ under Assumption~\ref{ass:did}.2, despite the fact that $\Delta Y$ does not depend on $S$ given $D_2$ and $\mathbf{X}$ under Assumption~\ref{ass:mar}.1. This bias disappears if $D_1$ is MCAR, but the efficiency loss from discarding incomplete observations may be substantial.

\subsection{Multiple time periods and general missingness patterns}\label{sec:general}
For ease of exposition, we consider a setting with three time periods and a partially missing $D_1$ throughout this paper. Our main results hold in a setting with a general number of time periods in which we allow for general missing treatment history patterns. We briefly discuss this setup here, and defer the details to Appendix~\ref{A:mainproofs}.

First, we extend our framework to $1<T<<n$ time periods with $n$ the number of individuals or units. The treatment history is denoted by $\mathbf{D} = (D_1, \ldots, D_T)$, and we maintain that no unit receives treatment in $t=0$. The PDATT in \eqref{eq:att} now generalises to 
$\tau_{\mathbf{dd^\prime}} = \mathbbm{E}[Y_T(\mathbf{d}) - Y_T(\mathbf{d}^\prime)|\mathbf{D} = \mathbf{d}]$,
with $Y_T=Y_T(\mathbf{d})$ and $\mathbf{d}^\prime = (0,\ldots,0) \equiv \mathbf{0}_T$. 

Second, we generalize our missing treatment mechanism as follows. Partition $\mathbf{D}$ into two vectors $\mathbf{D}_{-h}$ and $\mathbf{D}_{h}$, such that the elements in $\mathbf{D}_{-h}$ are observed with probability one and the elements in $\mathbf{D}_{h}$ may be missing. With $T=2$ and $D_1$ partially missing, this boils down to $\mathbf{D}_{-h}=D_2$ and $\mathbf{D}_{h}=D_1$. However, we can also capture $D_2$ missing by setting $\mathbf{D}_{-h}=D_1$ and $\mathbf{D}_{h}=D_2$. With $T>2$, multiple time periods of the treatment history may be missing, from which follows that $\mathbf{D}_{h}$ may include multiple time periods. 

The binary indicator $S$ now indicates whether all elements in $\mathbf{D}_h$ are observed, and the missing at random assumption imposes that this indicator is independent of $\mathbf{D}_{h}$ and $\Delta Y=Y_T-Y_0$, given $\mathbf{D}_{-h}$ and $\mathbf{X}$. When $T$ is large, a stronger assumption may be invoked to make estimation of a missing data model feasible. For instance, one that only requires conditioning on time periods adjacent to the ones in $\mathbf{D}_{h}$ instead of all time periods in $\mathbf{D}_{-h}$. Alternatively, more flexible missingness patterns can be allowed by defining separate missingness indicators for each time period in $\mathbf{D}_{h}$. 

\section{Robust estimation of treatment effects}\label{sec:robust} 

\subsection{A robust causal estimand}
We consider three models corresponding to the true unknown outcome means, propensity scores, and missing treatment probabilities, respectively. More precisely, $\mu_{\mathbf{d}}(\mathbf{X})$ represents a model for the outcome mean $m_{\mathbf{d}}(\mathbf{X}) \equiv \mathbb{E}[\Delta Y|\mathbf{D}=\mathbf{d},\mathbf{X}]$. The models
$\pi_{d_1|d_2}(\mathbf{X})$ and $\pi_{d_2}(\mathbf{X})$ represent the propensity scores $p_{d_1|d_2}(\mathbf{X})\equiv \mathbb{P}({D_1}={d_1}|D_2=d_2,\mathbf{X})$ and $p_{d_2}(\mathbf{X})\equiv \mathbb{P}({D_2}={d_2}|\mathbf{X})$, respectively. Finally, $\phi_{d_2}(\mathbf{X})$ is a model for the missing treatment probability, $q_{d_2}(\mathbf{X})= \mathbb{P}(S=1|D_2=d_2,\mathbf{X})$. Our first result shows how correctly specified $\mu_{\mathbf{d}}(\mathbf{X})$ and $\pi_{d_1|d_2}(\mathbf{X})$ can be identified under the MAR assumption, even when $D_1$ is missing.
\begin{lemma}[Identification of outcome and propensity score models with missing treatments]\label{lem:mu_pi_identification}\ \\
	Under Assumptions \ref{ass:did} and \ref{ass:mar}, it holds that
	\begin{enumerate}
		\item (Identification of outcome model) \\ 
		If $\mu_{\mathbf{d}}(\mathbf{X})=m_{\mathbf{d}}(\mathbf{X})$, then $\mu_{\mathbf{d}}(\mathbf{X}) = \mathbb{E}[\Delta Y|\mathbf{D}=\mathbf{d},\mathbf{X},S=1]$.
		\item (Identification of propensity score) \\
		If $\pi_{d_1|d_2}(\mathbf{X})=p_{d_1|d_2}(\mathbf{X})$, then $\pi_{d_1|d_2}(\mathbf{X})=\mathbb{P}(D_1=d_1|D_2=d_2,\mathbf{X},S=1)$.
	\end{enumerate}
\end{lemma}
The proof is deferred to Appendix~\ref{sec:proof_mu_pi_identification}. The expressions on the right-hand sides of the equations in Lemma~\ref{lem:mu_pi_identification} only depend on observables which implies that outcome models and propensity score models are identified. We combine these models with the missing treatment probability models into a single estimand. This estimand identifies the PDATT in \eqref{eq:att} even if one of the three models is misspecified. This robust estimand is given as
\begin{align}\label{eq:robust_tau}
		\tau_{\mathbf{dd^\prime}}^{\textup{R}} = &\mathbb{E}\left[  
		\left(w_1(S,\mathbf{D},\mathbf{X})-w_2(S,\mathbf{D},\mathbf{X})\right)\left( \Delta Y -\mu_{\mathbf{d^\prime}}(\mathbf{X}) \right)\right]+\notag\\
		&\mathbb{E}\left[
		\left(w_3(D_2,\mathbf{X})-w_4(S,D_2,\mathbf{X})\right)\left(\mu_{\mathbf{d}}(\mathbf{X}) -\mu_{\mathbf{d^\prime}}(\mathbf{X}) \right)
		\right], 
\end{align}
where the \citet{hajek1971discussion}-type weights are defined as 
{\normalsize \begin{align}\label{eq:weights}
		w_1(S,\mathbf{D},\mathbf{X}) =& \frac{\frac{S}{\phi_{d_2}(\mathbf{X})}\mathbbm{1}[\mathbf{D}=\mathbf{d}]}{\mathbb{E}\left[\frac{S}{\phi_{d_2}(\mathbf{X})}\mathbbm{1}[\mathbf{D}=\mathbf{d}]\right]}, &  
		w_2(S,\mathbf{D},\mathbf{X}) =& \frac{\frac{S}{\phi_{d^\prime_2}(\mathbf{X})}\frac{\pi_{\mathbf{d}}(\mathbf{X})}{\pi_{\mathbf{d^\prime}}(\mathbf{X})}\mathbbm{1}[\mathbf{D}=\mathbf{d}']}{\mathbb{E}\left[\frac{S}{\phi_{d^\prime_2}(\mathbf{X})}\frac{\pi_{\mathbf{d}}(\mathbf{X})}{\pi_{\mathbf{d}'}(\mathbf{X})}\mathbbm{1}[\mathbf{D}=\mathbf{d}']\right]}, \nonumber \\
		w_3(D_2,\mathbf{X}) =& \frac{\pi_{d_1|d_2}(\mathbf{X})\mathbbm{1}[D_2=d_2]}{\mathbb{E}\left[\pi_{d_1|d_2}(\mathbf{X})\mathbbm{1}[D_2=d_2]\right]}, &  
		w_4(S,D_2,\mathbf{X}) =& \frac{\frac{S}{\phi_{d_2}(\mathbf{X})}\pi_{d_1|d_2}(\mathbf{X})\mathbbm{1}[D_2=d_2]}{\mathbb{E}\left[\frac{S}{\phi_{d_2}(\mathbf{X})}\pi_{d_1|d_2}(\mathbf{X})\mathbbm{1}[D_2=d_2]\right]}.
\end{align}}
The following result states the robustness property of the estimand:
\begin{theorem}[Robust identification of PDATT with missing treatments]\label{thm:robust}\, \\
	Under Assumptions~\ref{ass:did} and \ref{ass:mar}, $\tau_{\mathbf{dd^\prime}}^{\textup{R}}=\tau_{\mathbf{dd^\prime}}$ for each $\mathbf{d} \in \{(1,1), (0,1), (1,0)\}$ if either
	\begin{enumerate}
		\item (Propensity score and outcome models are correct) $\pi_{\mathbf{d}}(\mathbf{X})=p_\mathbf{d}(\mathbf{X})$ and $\mu_{\mathbf{d}}(\mathbf{X})=m_{\mathbf{d}}(\mathbf{X})$; 
		\item (Missing data and propensity score models correct)
		$\phi_{d_2}(\mathbf{X})=q_{d_2}(\mathbf{X})$ and $\pi_{\mathbf{d}}(\mathbf{X})=p_\mathbf{d}(\mathbf{X})$;
		\item (Missing data and outcome models are correct) $\phi_{d_2}(\mathbf{X})=q_{d_2}(\mathbf{X})$ and $\mu_{\mathbf{d}}(\mathbf{X})=m_{\mathbf{d}}(\mathbf{X})$; 
	\end{enumerate}
	where $\pi_{\mathbf{d}}(\mathbf{X})=\pi_{d_1|d_2}(\mathbf{X})\cdot\pi_{d_2}(\mathbf{X})$ and $p_\mathbf{d}(\mathbf{X})=p_{d_1|d_2}(\mathbf{X})\cdot p_{d_2}(\mathbf{X})$.
\end{theorem}
Proof is deferred to Appendix~\ref{sec:proof_robust}. The intuition behind this result follows from the fact that when any two of the three models are replaced by their true counterparts, certain components of the estimand —best described as adjustment terms— vanish in expectation, and the remaining term equals the true parameter. Consider the following decomposition of the estimand:
 \begin{align}\label{eq:decomp}
		\tau_{\mathbf{dd^\prime}}^{\text{R}} = 
		&\underbrace{\mathbb{E}\left[w_1(S,\mathbf{D},\mathbf{X})\Delta Y\right]}_{(\textup{I})}-
		\underbrace{\mathbb{E}\left[w_1(S,\mathbf{D},\mathbf{X})\mu_{\mathbf{d^\prime}}(\mathbf{X})\right]}_{(\textup{II})} \notag\\
		&-
		\underbrace{\mathbb{E}\left[w_2(S,\mathbf{D},\mathbf{X})\Delta Y\right]}_{(\textup{III})}+
		\underbrace{\mathbb{E}\left[w_2(S,\mathbf{D},\mathbf{X})\mu_{\mathbf{d^\prime}}(\mathbf{X})\right]}_{(\textup{IV})}\notag\\
		&+\underbrace{\mathbb{E}\left[w_3(D_2,\mathbf{X})\left(\mu_{\mathbf{d}}(\mathbf{X}) -\mu_{\mathbf{d^\prime}}(\mathbf{X}) \right)\right]}_{(\textup{V})}-
		\underbrace{\mathbb{E}\left[w_4(S,D_2,\mathbf{X})\left(\mu_{\mathbf{d}}(\mathbf{X}) -\mu_{\mathbf{d^\prime}}(\mathbf{X}) \right)\right]}_{(\textup{VI})}.
\end{align}
The proof of Theorem~\ref{thm:robust} shows that with correct propensity score and outcome models, (I)-(II)=(VI) and (III)-(IV)=0. The term (V) is only a function of propensity score and outcome models, and provided that these models are correctly specified, identifies the target parameter:

\begin{corollary}[Identification of PDATT with propensity score and outcome models]\label{cor:ATT_identification}\, \\
	Under Assumption~\ref{ass:did}, ${\mathbb{E}\left[p_{\mathbf{d}}(\mathbf{X})\right]}^{-1}\mathbb{E}\left[(m_{\mathbf{d}}(\mathbf{X})-m_{\mathbf{d^\prime}}(\mathbf{X}))p_\mathbf{d}(\mathbf{X})\right]= \tau_{\mathbf{dd^\prime}}$ for each $\mathbf{d}$ and $\mathbf{d^\prime} = (0, 0)$, and with $p_\mathbf{d}(\mathbf{X})=p_{d_1|d_2}(\mathbf{X})\cdot p_{d_2}(\mathbf{X})$.
\end{corollary}

In case one of the correctly specified models is the missing treatment model $\phi_{d_2}(\mathbf{X})$, the proof of Theorem~\ref{thm:robust} shows that (V)=(VI). If, in addition, the outcome model is correct we have (III)-(IV)=0 and (I)-(II) identifies the PDATT, or the propensity score model is correct and (II)-(IV)=0 and (I)-(III) identifies the PDATT:
\begin{corollary}[Identification of PDATT with correct missing treatment model]\label{cor:ATT_identification_md}\, \\
	Under Assumptions~\ref{ass:did} and \ref{ass:mar}, it holds for each $\mathbf{d}$ and $\mathbf{d^\prime}=(0,0)$ that
	\begin{enumerate}
		\item ${\mathbb{E}\left[\frac{S}{q_{d_2}(\mathbf{X})}\mathbbm{1}[\mathbf{D}=\mathbf{d}]\right]}^{-1}\mathbb{E}\left[ {\frac{S}{q_{d_2}(\mathbf{X})}\mathbbm{1}[\mathbf{D}=\mathbf{d}]}\left( \Delta Y -m_{\mathbf{d^\prime}}(\mathbf{X}) \right)\right]= \tau_{\mathbf{dd^\prime}}.$
		\item ${\mathbb{E}\left[p_{\mathbf{d}}(\mathbf{X})\right]}^{-1}\mathbb{E}\left[\frac{S}{q_{d_2}(\mathbf{X})}\left({\mathbbm{1}[\mathbf{D}=\mathbf{d}]}-{\frac{p_{\mathbf{d}}(\mathbf{X})}{p_{\mathbf{d^\prime}}(\mathbf{X})}\mathbbm{1}[\mathbf{D}=\mathbf{d}']}\right) \Delta Y \right]= \tau_{\mathbf{dd^\prime}}.$
	\end{enumerate}
\end{corollary}
Corollaries \ref{cor:ATT_identification} and \ref{cor:ATT_identification_md} directly follow from the proof of Theorem~\ref{thm:robust}. Note that all results in this section are derived in Appendix~\ref{A:mainproofs} for the general case discussed in Section~\ref{sec:general}.

The main novelty of the estimand in Theorem~\ref{thm:robust} is that it enables identification of PDATTs with partially observed treatment histories, even with misspecification in the missing treatment model. We highlight this contribution with three examples. 
First, consider a setting in which the missing treatment model depends on the observed covariates $\mathbf{X}$ in an unknown way. If Assumption~\ref{ass:did} holds, and the propensity score and outcome models are correctly specified in $\mathbf{X}$, the robust estimand would identify the PDATTs.  
Second, suppose that the missingness mechanism depends on a different set of covariates than those required for the conditional parallel trends. Let $\mathbf{X}=(\mathbf{X}_1,\mathbf{X}_2)$, where Assumption~\ref{ass:did}, the propensity score, and the outcome models depend on $\mathbf{X}_1$, but the missing treatment model depends on $\mathbf{X}_2$. In such case, the researcher only requires knowledge on how the propensity scores and the outcome models vary with $\mathbf{X}_1$ to identify the PDATTs.
Third, consider a situation where missingness is driven by unobserved factors. If Assumption~\ref{ass:did}, the propensity score, and outcome models depend on $\mathbf{X}$, while Assumption~\ref{ass:mar} depends on unobservables, we still achieve identification.

\subsection{Semiparametric efficiency bound}\label{sec:seb}
To investigate the conditions under which the robust estimand is efficient, we first derive the semiparametric efficiency bound for the PDATT parameter in \eqref{eq:att} in the presence of missing treatments. The semiparametric efficiency bound serves as a benchmark for the asymptotic variance of any $\sqrt{n}$-consistent estimator of $\tau_{\mathbf{dd^\prime}}$. In spirit, one can think of this as the semiparametric analogue of the Cramer-Rao lower bound for parametric models. 

\begin{theorem}[Semiparametric efficiency bound for $\tau_{\mathbf{dd^\prime}}$ with missing treatments]\label{thm:seb}\, \\
	Under Assumptions \ref{ass:did} and \ref{ass:mar}, the semiparametric efficiency bound for all regular estimators of $\tau_{\mathbf{dd^\prime}}$ is given by $\Omega^\ast = \mathbb{E}[F_{\tau_{\mathbf{dd^\prime}}}( \mathbf{W})^2]$, with efficient influence function for $\tau_{\mathbf{dd^\prime}}$ defined as
	\begin{align*}
		F_{\tau_{\mathbf{dd^\prime}}}(\mathbf{W}) =& w_1(S, \mathbf{D}, \mathbf{X})\left(\Delta Y-m_{\mathbf{d^\prime}}(\mathbf{X})-\tau_{\mathbf{dd^\prime}}\right)- w_2(S, \mathbf{D}, \mathbf{X})  \left(\Delta Y - m_{\mathbf{d^\prime}}(\mathbf{X})\right) \notag\\&+ \left(w_3(D_2, \mathbf{X})-w_4(S, D_2, \mathbf{X})\right)\big(m_{\mathbf{d}}(\mathbf{X})-m_{\mathbf{d^\prime}}(\mathbf{X})-\tau_{\mathbf{dd^\prime}}\big),
	\end{align*}
	where the weights depend on the true unknown functions $m_{\mathbf{d}}(\mathbf{X})$, $q_{d_2}(\mathbf{X})$, and $p_{\mathbf{d}}(\mathbf{X})$ instead of $\mu_{\mathbf{d}}(\mathbf{X})$, $\phi_{d_2}(\mathbf{X})$, and $\pi_{\mathbf{d}}(\mathbf{X})$, respectively.
\end{theorem}
Proof is deferred to Appendix \ref{sec:proof_seb}. The derivation of the bound for the data $(Y_2, Y_0, \mathbf{D}, \mathbf{X})$ is self-contained and can be seen to follow previous results in the literature (see for example, \citet{hahn1998role} and \citet{sant2020doubly}). From there on, we employ the result in Theorem 7.2 in  \citet{tsiatis2006semiparametric} to derive the bound under our MAR assumption. 

\subsection{Inference}
The expression in \eqref{eq:robust_tau} suggests that the robust estimand can be estimated with a two-step procedure, provided that a random sample is available. 
\begin{assumption}[Random sampling]\label{ass:rs} \ \\
	$\left\{\mathbf{W}_i=(Y_{i0}, Y_{i2}, S_i, S_iD_{1i}, D_{2i}, \mathbf{X}_i);i=1,\ldots, n\right\}$ are $i.i.d$ draws from an infinite population.
\end{assumption}
Assumption \ref{ass:rs} covers a setting in which panel data are available.\footnote{The $i.i.d$ assumption can be relaxed to allow for intra-cluster correlations in cases where data have a clustering dimension. Our identification and estimation results will continue to hold in such case, while inference will have to be adjusted to account for such correlation structure.}  
Estimation can then proceed as follows. First, the models for the true unknown outcome means, propensity scores, and missing data probabilities are estimated. Second, the predicted values for these estimated models are plugged into the sample analogue of $\tau_{\mathbf{dd^\prime}}^{\textup{R}}$. 

The first step requires a choice of models and estimators for the outcome means, propensity scores, and missing data probabilities. So far, we have simply postulated the existence of models for each of these functions but have not committed to it either being parametric or non-parametric in nature. We derive the asymptotic behavior of the estimator for $\tau_{\mathbf{dd^\prime}}^{\textup{R}}$ assuming parametric first-stage estimators, which allows us to derive asymptotic theory for general parametric estimators. These estimators are often preferred in applied work due to their simplicity, and due to the fact that nonparametric estimators may suffer from challenges such as the curse of dimensionality or tuning parameter selection.   

Let $\mu(\bm{\beta}_\mathbf{d})$, $\pi(\bm{\gamma}_{\mathbf{d}})$, and $\phi(\bm{\delta}_{d_2})$ be parametric models for $m_{\mathbf{d}}(\mathbf{X})$, $p_{\mathbf{d}}(\mathbf{X})$, and $q_{d_2}(\mathbf{X})$, respectively, where we suppress the dependence of these models on data for notational convenience. Define the pseudo-true parameter values as $\bm{\beta^\ast}_{\mathbf{d}}$, $\bm{\gamma^\ast}_{\mathbf{d}} =  (\bm{\gamma^\ast}_{d_1|d_2}, \bm{\gamma^\ast}_{d_2})$, and $\bm{\delta^\ast}_{d_2}$. Let $\bm{\widehat{\beta}}_{\mathbf{d}}$, $\bm{\widehat{\gamma}}_{\mathbf{d}}$,  $\bm{\widehat{\delta}}_{d_2}$ denote $\sqrt{n}$-consistent estimators of these pseudo-true values. The estimator of the robust estimand $\widehat{\tau}^\textup{R}_{\mathbf{dd^\prime}}$ is given by
\begin{align}\label{eq:tau_hat}
	\widehat{\tau}_{\mathbf{dd^\prime}}^{\text{R}}  &= \mathbb{E}_n\left[  
	\left(\widehat{w}_1(\bm{\widehat{\delta}}_{d_2})-\widehat{w}_2(\bm{\widehat{\gamma}}, \bm{\widehat{\delta}}_{d_2^\prime})\right)\left(\Delta Y -\mu(\bm{\widehat{\beta}}_{\mathbf{d^\prime}})\right)\right] \nonumber \\
	&+\mathbb{E}_n\left[\left(\widehat{w}_3(\bm{\widehat{\gamma}}_{d_1|d_2})-\widehat{w}_4(\bm{\widehat{\gamma}}_{d_1|d_2}, \bm{\widehat{\delta}}_{d_2})\right)\left( \mu(\bm{\widehat{\beta}}_{\mathbf{d}})-\mu(\bm{\widehat{\beta}}_{\mathbf{d^\prime}})\right)\right],
\end{align}
where $\mathbb{E}_n(\cdot)$ denotes the empirical mean and the weights are estimated as
{\normalsize \begin{align}\label{eq:weightSat}
	\widehat{w}_1(\bm{\widehat{\delta}}_{d_2}) = & \frac{\frac{S}{\phi(\bm{\widehat{\delta}}_{d_2})}\mathbbm{1}[\mathbf{D}=\mathbf{d}]}{\mathbb{E}_n\left[\frac{S}{\phi(\bm{\widehat{\delta}}_{d_2})}\mathbbm{1}[\mathbf{D}=\mathbf{d}]\right]}, \ \ 
	 \widehat{w}_2(\bm{\widehat{\gamma}},\bm{\widehat{\delta}}_{d_2'}) = \frac{\frac{S}{\phi(\bm{\widehat{\delta}}_{d_2^\prime})}\frac{\pi(\bm{\widehat{\gamma}}_{\mathbf{d}})}{\pi(\bm{\widehat{\gamma}}_{\mathbf{d^\prime}})}\mathbbm{1}[\mathbf{D}=\mathbf{d}']}{\mathbb{E}_n\left[\frac{S}{\phi(\bm{\widehat{\delta}}_{d_2^\prime})}\frac{\pi(\bm{\widehat{\gamma}}_{\mathbf{d}})}{\pi(\bm{\widehat{\gamma}}_{\mathbf{d^\prime}})}\mathbbm{1}[\mathbf{D}=\mathbf{d}']\right]},  \\
	\widehat{w}_3(\bm{\widehat{\gamma}}_{d_1|d_2}) =& \frac{\pi(\bm{\widehat{\gamma}}_{d_1|d_2})\mathbbm{1}[D_2=d_2]}{\mathbb{E}_n\left[\pi(\bm{\widehat{\gamma}}_{d_1|d_2})\mathbbm{1}[D_2=d_2]\right]}, \ \ 
	\widehat{w}_4(\bm{\widehat{\gamma}}_{d_1|d_2},\bm{\widehat{\delta}}_{d_2}) = \frac{\frac{S}{\phi(\bm{\widehat{\delta}}_{d_2})}\pi(\bm{\widehat{\gamma}}_{d_1|d_2})\mathbbm{1}[D_2=d_2]}{\mathbb{E}_n\left[\frac{S}{\phi(\bm{\widehat{\delta}}_{d_2})}\pi(\bm{\widehat{\gamma}}_{d_1|d_2})\mathbbm{1}[D_2=d_2]\right]}, \notag
\end{align}}
with $\bm{\widehat{\gamma}}=(\bm{\widehat{\gamma}}_{\mathbf{d}},\bm{\widehat{\gamma}}_{\mathbf{d^\prime}})$, and the dependence of these weights on the data is suppressed. Define  $\bm{\beta^\ast}=(\bm{\beta^\ast}_{\mathbf{d}},\bm{\beta^\ast}_{\mathbf{d^\prime}})$, $\bm{\gamma^\ast}=(\bm{\gamma^\ast}_{\mathbf{d}},\bm{\gamma^\ast}_{\mathbf{d^\prime}})$, and $\bm{\delta^\ast}=(\bm{\delta^\ast}_{d_2}$, $\bm{\delta^\ast}_{d_2^\prime})$.
Theorem~\ref{thm:asyvar} derives the asymptotic properties of $\widehat{\tau}^\textup{R}_{\mathbf{dd^\prime}}$ using some weak high-level conditions on the estimators for the generic parametric models, which are outlined in Appendix~\ref{sec:conditions}:
	
\begin{theorem}[Asymptotic behavior of $\widehat{\tau}^\textup{R}_{\mathbf{dd^\prime}}$]\label{thm:asyvar}\, \\
	Under Assumptions \ref{ass:did}-\ref{ass:rs}, Conditions 1-5 in Appendix \ref{sec:conditions}, and provided that either 
	$\mu(\bm{\beta^\ast}_\mathbf{d})=m_{\mathbf{d}}(\mathbf{X})$ and $\pi(\bm{\gamma^\ast}_{\mathbf{d}})=p_{\mathbf{d}}(\mathbf{X})$;
	$\phi(\bm{\delta^\ast}_{d_2})=q_{d_2}(\mathbf{X})$ and 
	$\pi(\bm{\gamma^\ast}_{\mathbf{d}})=p_{\mathbf{d}}(\mathbf{X})$; or
	$\phi(\bm{\delta^\ast}_{d_2})=q_{d_2}(\mathbf{X})$ and 
	$\mu(\bm{\beta^\ast}_\mathbf{d})=m_{\mathbf{d}}(\mathbf{X})$
	, 
	as $n\rightarrow \infty$,
	\begin{align*}
		\sqrt{n}(\widehat{\tau}^\textup{R}_{\mathbf{dd^\prime}}-\tau^{\textup{R}}_{\mathbf{dd^\prime}}) =  \frac{1}{\sqrt{n}}\sum_{i=1}^{n}\xi(\mathbf{W}_i,\bm{\beta^\ast}, \bm{\gamma^\ast}, \bm{\delta^\ast})+o_p(1)  \rightsquigarrow N(0, \Omega),
	\end{align*} 
	where $\Omega = \mathbb{E}[\xi(\mathbf{W},\bm{\beta^\ast}, \bm{\gamma^\ast}, \bm{\delta^\ast})^2]$ and $\xi(\mathbf{W},\bm{\beta^\ast}, \bm{\gamma^\ast}, \bm{\delta^\ast})$ is provided in Appendix~\ref{sec:proof_asyvar}.
\end{theorem}
Proof is deferred to Appendix~\ref{sec:proof_asyvar}. Theorem~\ref{thm:asyvar} shows that $\widehat{\tau}^\textup{R}_{\mathbf{dd^\prime}}$ is $\sqrt{n}$-consistent and asymptotically normal provided that at least two of the three models are correct. This result suggests that we can use the sample analogue to $\Omega$ to conduct asymptotically valid inference. 
Estimation of the nuisance parameters affects the asymptotic variance of the robust estimator. This effect is proportionate to the average change in the influence function of the robust estimator from locally perturbing the first-stage parameters around their probability limits. When these probability limits index a correctly specified population model, small changes in $(\bm{\beta}, \bm{\gamma}, \bm{\delta})$ have no effect on the influence function of the robust estimator, causing the estimation effect from the first stage to disappear. 
In the special case when all three models are correctly specified, we show that $\widehat{\tau}^\textup{R}_{\mathbf{dd^\prime}}$ achieves the semiparametric efficiency bound.

\begin{corollary}[Semi-parametric efficiency of $\widehat{\tau}^\textup{R}_{\mathbf{dd^\prime}}$]\label{cor:efficient}\, \\
	Under Assumptions \ref{ass:did}-\ref{ass:rs}, Conditions 1-5 in Supplementary Appendix \ref*{sec:conditions}, and provided that 
	$\mu(\bm{\beta^\ast}_\mathbf{d})=m_{\mathbf{d}}(\mathbf{X})$, $\pi(\bm{\gamma^\ast}_{\mathbf{d}})=p_{\mathbf{d}}(\mathbf{X})$, and
	$\phi(\bm{\delta^\ast}_{d_2})=q_{d_2}(\mathbf{X})$, then  
    $\Omega = \Omega^\ast$. 
\end{corollary}

Proof is deferred to Appendix \ref{sec:proof_cor_efficient}. A practical implication of Corollary~\ref{cor:efficient} is that when all three models are correct, the choice of first-step estimators does not influence the asymptotic variance of the robust estimator. However, this property is lost as soon as one of the working models is misspecified. In this case, the expression for $\Omega$ in Theorem~\ref{thm:asyvar} includes terms that depend on the first-stage estimators, making inference sensitive to such choice. Supplementary Appendix \ref*{sec:inferencerobust} explores two inference-robust alternatives whose asymptotic variance remains unaffected under misspecification of any one model. The search for such an alternative is inspired from \cite{sant2020doubly} who propose improved DID estimators in the standard DID setup with similar robustness properties. Building on the insights from \cite{vermeulen2015bias}, our inference-robust proposals use first-stage estimators that are specifically designed to minimize the effect of the nuisance parameters on the robust estimator.  

\section{Alternative missingness-adjusted estimation approaches}\label{sec:adjusted}
Corollary~\ref{cor:ATT_identification_md} presents estimands based on a correct missing data model along with either a correct propensity score or mean outcome model thereby giving us missingness-adjusted outcome regression (OR) and inverse probability weighting (IPW) estimands. These are given by 
 \begin{equation}\label{or_adj}
	\tau^\textup{OR}_{\mathbf{dd^\prime}} 
	\equiv \mathbb{E}[w_1(S,\mathbf{D},\mathbf{X})(\Delta Y - \mu_{\mathbf{d^\prime}}(\mathbf{X}))], 
\end{equation} and 
\begin{equation}\label{ipw_adj}
	\tau^\textup{IPW}_{\mathbf{dd^\prime}} \equiv \mathbb{E}[\left(w_1(S,\mathbf{D},\mathbf{X})-w_2(S,\mathbf{D},\mathbf{X})\right)\Delta Y].
\end{equation}
It is important to note that unlike the standard OR method, which only depends on a correct outcome model, the missingness-adjusted OR estimand given in \eqref{or_adj} depends on both a correct missing treatment and outcome model. In a similar spirit, the adjusted IPW estimand in \eqref{ipw_adj} depends not only on a correct propensity score but also a correct missing treatment model, thereby requiring both probability weights to be correct to identify the target parameter. 

For our setting, we can also combine the two results in Corollary~\ref{cor:ATT_identification_md}  to give us the missingness-adjusted DR estimand, which is given by
 \begin{equation}\label{dr_adj}
	\tau^\textup{DR}_{\mathbf{dd^\prime}} 
	\equiv \mathbb{E}\left[\left(w_1(S,\mathbf{D},\mathbf{X})-w_2(S,\mathbf{D},\mathbf{X})\right)\left( \Delta Y -\mu_{\mathbf{d^\prime}}(\mathbf{X}) \right)\right].
\end{equation}
It follows from Theorem~\ref{thm:robust}, and the discussion around the decomposition in \eqref{eq:decomp}, that this estimand identifies $\tau_{\mathbf{dd^\prime}}$ when either the missing data model and the outcome model are correct, or the missing data model and the propensity score model are correct. While our proposed approach (R) is robust to misspecification in the missing data model, the missingness-adjusted OR, IPW, and DR strategies presented above will not identify the target parameter if the missing data model is incorrect, making it a less preferred alternative compared to R. 

\begin{proposition}[Identification]\ \label{prop:tau_est_missing} \\
	Under Assumptions \ref{ass:did} and \ref{ass:mar}, for each $\mathbf{d}$ and $\mathbf{d^\prime}= (0,0)$, 
	$\tau^{\textup{OR}}_{\mathbf{dd^\prime}}$, $\tau^{\textup{IPW}}_{\mathbf{dd^\prime}}$, and $\tau^{\textup{DR}}_{\mathbf{dd^\prime}}$ identify $\tau_{\mathbf{dd^\prime}}$ if either $\phi_{d_2}(\mathbf{X}) = q_{d_2}(\mathbf{X})$ and $\mu_{\mathbf{d}}(\mathbf{X}) = m_{\mathbf{d}}(\mathbf{X})$; $\phi_{d_2}(\mathbf{X}) = q_{d_2}(\mathbf{X})$ and $\pi_{\mathbf{d}}(\mathbf{X}) = p_{\mathbf{d}}(\mathbf{X})$; $\phi_{d_2}(\mathbf{X}) = q_{d_2}(\mathbf{X})$ and either $\mu_{\mathbf{d}}(\mathbf{X}) = m_{\mathbf{d}}(\mathbf{X})$ or $\pi_{\mathbf{d}}(\mathbf{X}) = p_{\mathbf{d}}(\mathbf{X})$, respectively.
\end{proposition}
The proof follows directly from Corollary \ref{cor:ATT_identification_md} and Theorem \ref{thm:robust} for OR, IPW, and DR estimands, respectively. Replacing the population models in \eqref{or_adj}-\eqref{dr_adj} with their estimated counterparts allows us to propose  estimators which are given by
\begin{align}
		\widehat{\tau}^\textup{OR}_{\mathbf{dd^\prime}} & = \mathbb{E}_n[\widehat{w}_1(\bm{\widehat{\delta}}_{d_2})(\Delta Y - \mu(\bm{\widehat{\beta}}_{\mathbf{d^\prime}}))], \label{eq:orhat} \\
		\widehat{\tau}^\textup{IPW}_{\mathbf{dd^\prime}} & = \mathbb{E}_n[(\widehat{w}_1(\bm{\widehat{\delta}}_{d_2})-w_2(\bm{\widehat{\gamma}}, \bm{\widehat{\delta}}_{d_2^\prime}))\Delta Y], \label{eq:ipwhat}\\
		\widehat{\tau}^\textup{DR}_{\mathbf{dd^\prime}} & = \mathbb{E}_n[(\widehat{w}_1(\bm{\widehat{\delta}}_{d_2})-w_2(\bm{\widehat{\gamma}}, \bm{\widehat{\delta}}_{d_2^\prime}))(\Delta Y - \mu(\bm{\widehat{\beta}}_{\mathbf{d^\prime}}))]. \label{eq:drhat}
\end{align}
Since these are incomplete versions of the robust estimator, their asymptotic variances can be easily and directly obtained from the variance of $\widehat{\tau}^\textup{R}_{\mathbf{dd^\prime}}$. Supplementary Appendix~\ref*{sec:simsadd} presents the asymptotic influence function representations for all three alternatives. 

\section{Numerical experiments}\label{sec:sims}
In this section, we first conduct a Monte Carlo study which analyzes the finite sample performance of different estimators for PDATTs. We then study how varying amounts of missingness and degrees of misspecification in the missing data model affect their performance.

\subsection{Set-up}
The data generating process is defined as
\begin{align}\label{eq:dgp}
D_2 =& \mathbbm{1}[\Lambda(\mathbf{X}_p\bm{\gamma}_{1})\geq U_1],\\
D_1 =& D_2\cdot \mathbbm{1}[\Lambda(\mathbf{X}_p\bm{\gamma}_{1|1})\geq U_2]+(1-D_2)\cdot\mathbbm{1}[\Lambda(\mathbf{X}_p\bm{\gamma}_{1|0})\geq U_2],\\ 
 S =& D_2\cdot\mathbbm{1}[\Lambda(\mathbf{X}_m\bm{\delta}_{1})\geq U_3]+(1-D_2)\cdot\mathbbm{1}[\Lambda(\mathbf{X}_m\bm{\delta}_{0})\geq U_3],  \\
 \Delta Y =& \mathbf{X}_o(\bm{\beta}_{11}D_1D_2+\bm{\beta}_{10}D_1(1-D_2)+\bm{\beta}_{01}(1-D_1)D_2+\bm{\beta}_{00})+\varepsilon,
\end{align}
where $U_1$, $U_2$, and $U_3$ are three independently distributed random variables with a standard uniform distribution, and $\varepsilon$ has a standard normal distribution. 

The covariates in the propensity scores, missing treatment probabilities, and outcome means are denoted by $\mathbf{X}_p$, $\mathbf{X}_m$, and $\mathbf{X}_o$, respectively. We set $\mathbf{X}_g=\eta_g\mathbf{X}+(1-\eta_g)\mathbf{{Z}}$ with $\eta_g=0,1$ and $g=p,m,o$. The vector $\mathbf{X}$ includes an intercept and four independently distributed standard normal random variables $X_1, \ldots, X_4$. We then use the transformations defined in \citet{kang2007demystifying}:  $\tilde{Z}_1= \text{exp}(0.5 X_1)$, $\tilde{Z}_2= 10+X_2/(1+\text{exp}(X_1))$, $\tilde{Z}_3 = \left(0.6+X_1 X_3/25\right)^3$ and $\tilde{Z}_4 = \left(20+X_2+X_4\right)^2$. This gives us the vector $\mathbf{Z}$ which includes an intercept and $\tilde{Z}_1, \ldots, \tilde{Z}_4$ that are standardized to have mean 0 and variance 1. Since $\mathbf{X}$ is treated as the vector of observed covariates, setting $\eta_g=0$ results in a misspecified working model. 
The values for the parameters in \eqref{eq:dgp} are provided in Table~\ref{tab:params}. The percentage of missing values for $D_1$ is governed by $c$, where $c=0$ corresponds to approximately 50\% missingness. 

\begin{table}[tb!]
	\centering
	\begin{threeparttable}
		\caption{Parameter values}
		\label{tab:params}%
		\begin{tabular}{rrrrrrrrr}
		\toprule
		\multicolumn{1}{c}{$\bm{\gamma}_{1}$} & \multicolumn{1}{c}{$\bm{\gamma}_{1|1}$}& \multicolumn{1}{c}{$\bm{\gamma}_{1|0}$} & \multicolumn{1}{c}{$\bm{\beta}_{11}$} & $\bm{\beta}_{10}$  & $\bm{\beta}_{01}$  & $\bm{\beta}_{00}$  & \multicolumn{1}{c}{$\bm{\delta}_{1}$} & $\bm{\delta}_{0}$ \\
		\midrule
     0.00  & 0.00  & 0.00  & 1.50  & 1.00  & 1.00  & 0.00  & c  & c \\
    -0.50 & -0.50 & 0.50  & -0.25 & -0.25 & 0.25  & 0.25  & -0.50 & 0.50 \\
    -0.50 & -0.50 & 0.50  & 0.25  & -0.25 & 0.25  & 0.25  & -0.50 & 0.50 \\
    -0.50 & 0.50  & -0.50 & 0.25  & 0.25  & -0.25 & 0.25  & 0.50  & 0.50 \\
    -0.50 & 0.50  & -0.50 & 0.25  & 0.25  & -0.25 & 0.25  & 0.50  & -0.50 \\
		  \bottomrule
		\end{tabular}%
		\begin{tablenotes}[flushleft]
			\footnotesize
			\item \textit{Notes:} This table shows the values for the parameters in \eqref{eq:dgp}, where $c=0$ corresponds to approximately 50\% missingness for $D_1$.
		\end{tablenotes}
	\end{threeparttable}%
\end{table}%

We estimate the PDATTs $\tau_{(11)(00)}$, $\tau_{(10)(00)}$, and $\tau_{(01)(00)}$ using the estimators $\widehat{\tau}_{\mathbf{dd^\prime}}^{\textup{R}} $, $\widehat{\tau}^\textup{OR}_{\mathbf{dd^\prime}}$, $\widehat{\tau}^\textup{IPW}_{\mathbf{dd^\prime}}$, and $\widehat{\tau}^\textup{DR}_{\mathbf{dd^\prime}}$ defined in \eqref{eq:tau_hat}, \eqref{eq:orhat}, \eqref{eq:ipwhat}, and \eqref{eq:drhat}, respectively. The asymptotic distributions of the OR, IPW, and DR estimators are derived in Supplementary Appendix~\ref*{sec:simsdet}. Additionally, we estimate the PDATTs using the CC-OR, CC-IPW, and CC-DR estimators which rely on the observed sample, without adjusting for missing histories through a missing data model. 
All estimators use working models as specified in Appendix~\ref{sec:routine}.

\subsection{Finite sample performance of PDATT estimators}
We study the finite sample performance of the estimators with four Monte Carlo experiments. Each experiment consists of 10,000 replications with each a sample of $(\Delta Y_i, S_i, S_iD_{1i},D_{2i},\mathbf{X}_i)$ with 10,000 observations generated from \eqref{eq:dgp} with $c=0$. In these experiments, either only the missing data model (M), only the propensity score (P), only the outcome regression (OR), or none of the models are misspecified (None). The four scenarios in which more than one working model is misspecified are discussed in Supplementary Appendix~\ref*{sec:simsresults}. Since each scenario conditions on different covariates, the values for the PDATTs vary across the experiments; ${\tau}_{(11)(00)}$ varies between 1.68 and 1.71, ${\tau}_{(10)(00)}$ between 0.58 and 0.63, and ${\tau}_{(01)(00)}$ between 1.15 and 1.42. 

\begin{figure}[tb!]
	\caption{Monte Carlo experiments: Bias}
	\begin{minipage}{\columnwidth}
		\centering	
		\includegraphics[width=\textwidth,trim=1.8cm 0 2cm 0,clip]{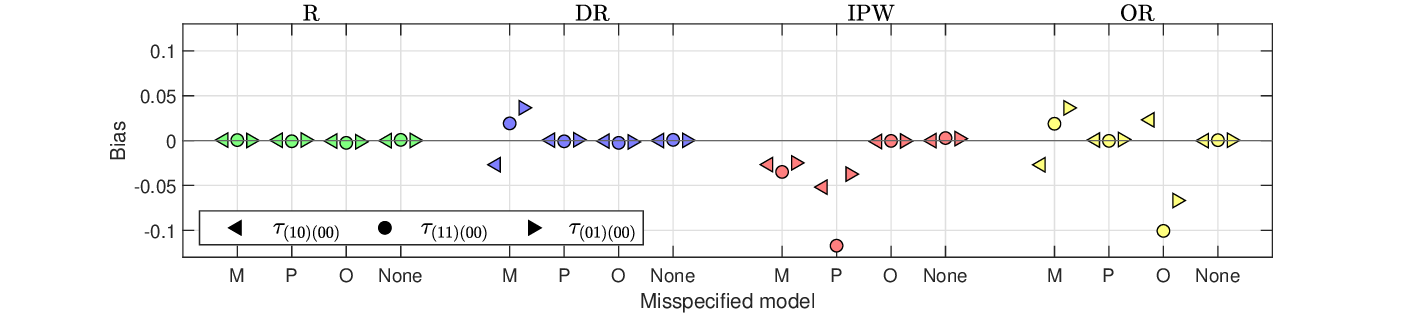} \\ 
         \vspace{1em}   
		\begin{minipage}{0.9\textwidth}
			\footnotesize \textit{Notes:} This figure shows the  bias of different missingness-adjusted estimators (R, DR, IPW, OR) for the PDATTs ${\tau}_{(11)(00)}$ (circle), ${\tau}_{(10)(00)}$ (left-pointing triangle), and ${\tau}_{(01)(00)}$ (right-pointing triangle). The x-axis shows the four different experiments in which either only the mssing data model (M), only the propensity score (P), only the outcome regression (O), or none of the models are misspecified (None).
		\end{minipage}
	\end{minipage}
	\label{fig:bias}
\end{figure}

Figure~\ref{fig:bias} shows the bias of the missingness-adjusted estimators for the PDATTs across the four experiments. We find that the proposed robust estimator (R) is unbiased across all settings under consideration. For the other estimators, there is a bias when the missing data model is misspecified. As expected, IPW also shows bias when the propensity score model is misspecified, OR when the outcome regression is misspecified, and all missingness-adjusted estimators are unbiased when all models are correctly specified. 

For the CC estimators, the bias becomes substantial compared to the bias caused by misspecified working models in the case of missingness-adjusted estimators. This indicates that selection bias can have a relatively large impact on the accuracy of the PDATT estimates. Detailed results for these estimators are reported in Supplementary Appendix~\ref*{sec:simsresults}. 

Figure~\ref{fig:size} shows the test size of testing the null-hypothesis that a PDATT equals its true value at a nominal level of 5\% across the four experiments. We find that the robust estimator obtains nominal test size for all PDATTs across all experiments. The tests corresponding to the other estimators are oversized when the missing data model is misspecified. In addition, we find major distortions for the OR and IPW estimators if the outcome regression or propensity score models are incorrect, respectively. When all models are correctly specified, all four estimators appropriately control size. In this case, the asymptotic variance of R is close to the semiparametric efficiency bound. Supplementary Appendix~\ref*{sec:simsresults} provides the estimates for the asymptotic variances of the estimators together with the semiparametric efficiency bounds.

\begin{figure}[tb!]
	\caption{Monte Carlo experiments: Test size}
	\begin{minipage}{\columnwidth}
		\centering	
		\includegraphics[width=\textwidth,trim=1.8cm 0 2cm 0,clip]{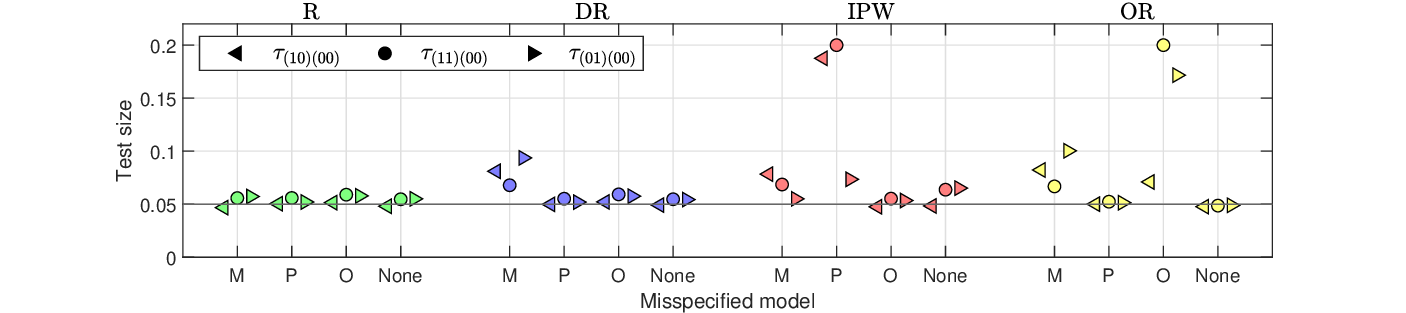} \\
        \vspace{1em}
		\begin{minipage}{0.9\textwidth}
			\footnotesize \textit{Notes:} 
This figure shows the test size of testing the null-hypothesis that a PDATT equals its true value at a nominal level of 5\%. The panels correspond to different missingness-adjusted estimators (R, DR, IPW, OR) with test size truncated at 0.2 for the PDATTs ${\tau}_{(11)(00)}$ (circle), ${\tau}_{(10)(00)}$ (left-pointing triangle), and ${\tau}_{(01)(00)}$ (right-pointing triangle). The x-axis shows the four different experiments in which either only the missing data model (M), only the propensity score (P), only the outcome regression (O), or none of the models are misspecified (None). 
		\end{minipage}
	\end{minipage}
	\label{fig:size}
\end{figure}

\begin{figure}[tb!]
	\caption{Monte Carlo experiments: Statistical power}
	\begin{minipage}{\columnwidth}
		\centering	
		\includegraphics[width=\textwidth,trim=1.8cm 0 2cm 0,clip]{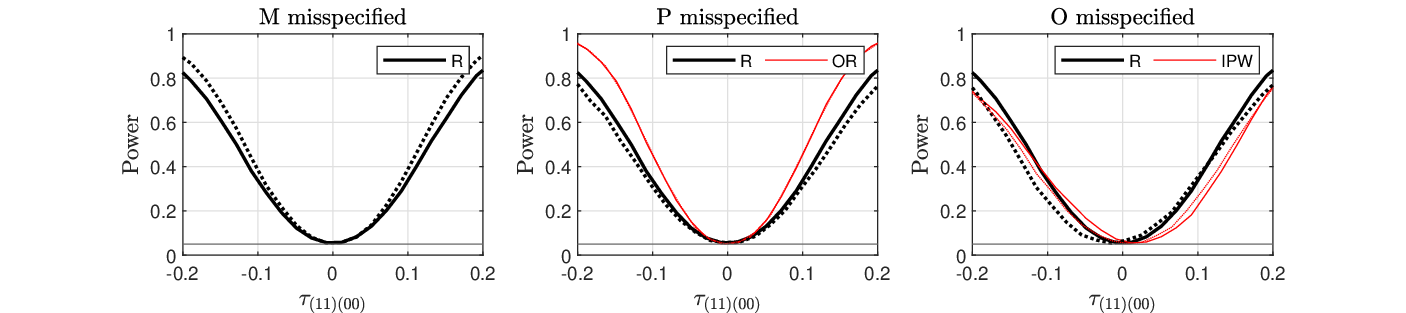} \\
        \vspace{1em}
		\begin{minipage}{0.9\textwidth}
			\footnotesize \textit{Notes:} 
            This figure shows the statistical power of the test of $H_0:\tau_{(11)(00)}=0$ with nominal test size of 5\%. The panels correspond to experiments with different misspecified models. Each panel shows the power curves of the estimators that theoretically should control size under the misspecification at hand by dotted lines, and the power curves in the experiments with none of the models misspecified by solid lines. The power curves of R and DR are identical in the second and third panel, and hence the latter are not displayed. The x-axis shows the value of $\tau_{(11)(00)}$. 
		\end{minipage}
	\end{minipage}
	\label{fig:power}
\end{figure}

Figure~\ref{fig:power} shows the statistical power of the test of $H_0:\tau_{(11)(00)}=0$ with nominal test size of 5\%. The panels correspond to experiments with different misspecified models. Each panel shows the power curves of the estimators that theoretically should control size under the misspecification at hand by dotted lines, and the power curves in the experiments with none of the models misspecified by solid lines. Note that the power curves of R and DR are identical in the second and third panel, and hence the latter are not displayed. We find that the power curves of the R estimator are close to the curves of the other estimators across all experiments. This indicates that the potential power loss of using the most robust estimator relative to the OR, IPW, and DR estimators is small. Moreover, since the power curves in the experiments with only correctly specified models are close to the power curves with misspecified models, the power loss due to misspecification also seems small. 

\subsection{Varying degree of missingness}
Second, we illustrate the importance of appropriately accounting for missing treatment histories at varying rates of missingness. The percentage of missingness in the Monte Carlo experiments equals approximately 50\%. By varying $c$, the intercept in the missing data models, we explore how the proportion of missingness influences the bias of different estimators. We assume that only the missing data model is misspecified with $\eta_p=\eta_o=1-\eta_m=1$ and generate one million observations from \eqref{eq:dgp} with $c \in \{-5,-4.5,\dots,4.5,5\}$. 

\begin{figure}[tb!]
	\caption{Bias with increasing percentage of missingness}
	\begin{minipage}{\columnwidth}
		\centering	
		\includegraphics[width=\textwidth,trim=1.8cm 0 2cm 0,clip]{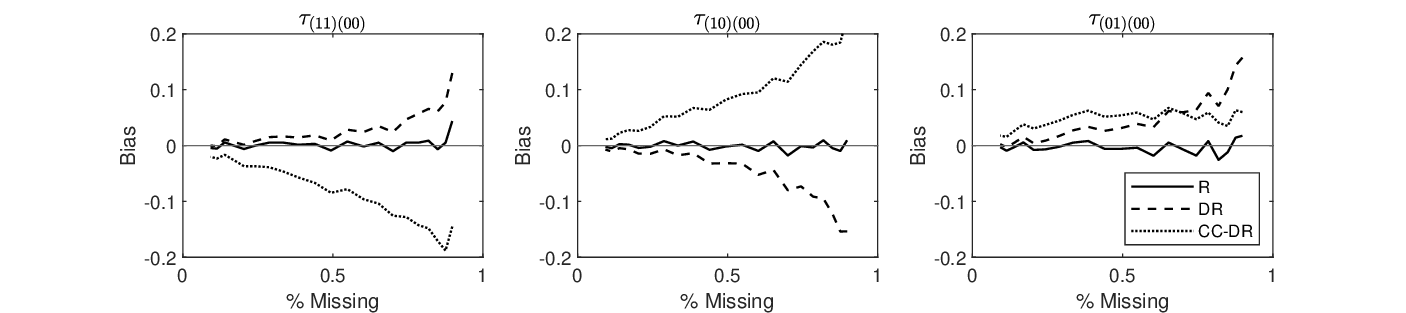} \\
        \vspace{1em}
		\begin{minipage}{0.9\textwidth}
			\footnotesize \textit{Notes:} 
            This figure shows the bias for an increasing percentage of missingness for $D_1$ for the estimators R (solid line), DR (dashed line), and CC-DR (dotted line). The panels correspond to the PDATTs ${\tau}_{(11)(00)}$, ${\tau}_{(10)(00)}$, and ${\tau}_{(01)(00)}$, respectively.
		\end{minipage}
	\end{minipage}
	\label{fig:missing}
\end{figure}

Figure~\ref{fig:missing} shows the bias in R, DR, and CC-DR for an increasing percentage of missingness. We find that the estimates from R have negligible bias, even when the amount of missingness is large. However, both DR and CC-DR show a bias that increases in the percentage of missingness. These biases follow from a misspecified missing data model in DR or from sample selection in CC-DR, and may already affect estimates when only a small percentage of treatment histories is missing.

\subsection{Varying degree of misspecification in the missing data model}
Third, we examine the effect of different degrees of misspecification in the missing data model on the bias of different estimators. Since $\eta_m=1$ corresponds to a correctly specified model, we can explore the effect of an increasing amount of misspecification by decreasing $\eta_m \in \{1,0.9,\dots,0.1,0\}$. We assume that only the missing data model is misspecified with $\eta_p=\eta_o=1$ and $c=0$, and generate one million observations from \eqref{eq:dgp} for each value of $\eta_m$.

\begin{figure}[tb!]
	\caption{Bias with increasing degree of misspecification}
	\begin{minipage}{\columnwidth}
		\centering	
		\includegraphics[width=\textwidth,trim=1.8cm 0 2cm 0,clip]{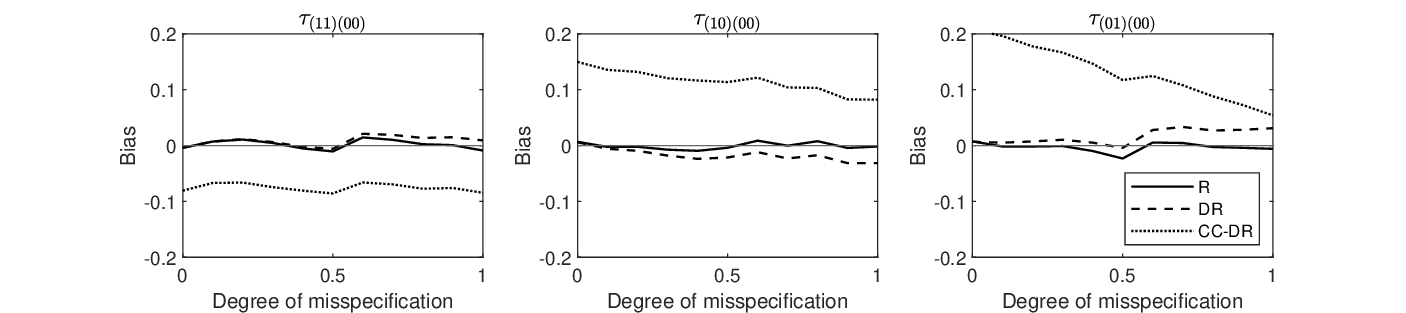} \\
        \vspace{1em}
		\begin{minipage}{0.9\textwidth}
			\footnotesize \textit{Notes:} 
            This figure shows the bias for an increasing degree of misspecification in the missing data model for the estimators R (solid line), DR (dashed line), and CC-DR (dotted line). The x-axis shows $1-\eta_m$. The panels correspond to the PDATTs ${\tau}_{(11)(00)}$, ${\tau}_{(10)(00)}$, and ${\tau}_{(01)(00)}$, respectively.
		\end{minipage}
	\end{minipage}
	\label{fig:degree}
\end{figure}

Figure~\ref{fig:degree} shows the bias in R, DR, and CC-DR for an increasing degree of misspecification. Again, we find negligible bias in the estimates from R across all degrees. The bias in DR increases in the amount of misspecification, but is small compared to the bias in CC-DR, which does not necessarily depend on the degree of misspecification. Both findings align with our theoretical results, which show that the DR estimator requires the missing data model to be correct and CC-DR requires the sample selection bias to be negligible. The first directly hinges on the degree of misspecification, while the latter also depends on other features of the data generating process. The illustrations in Figures~\ref{fig:missing} and \ref{fig:degree} show that minor violations of these assumptions can already cause substantial biases in these PDATT estimators. 

\section{Empirical applications}\label{sec:emp_app}
In this section, we demonstrate the proposed method by applying it to two distinct empirical settings. The first application investigates the effects of covid case surges on voter turnout in the 2022 U.S. general elections.\footnote{Numerous studies have investigated the effects of COVID cases and COVID-led policies on a range of outcomes \citep{callaway2023evaluating, callaway2023policy, kim2021impact, morgenstern2022interaction, badinlou2024trajectories, reuschke2024impacts,herrnson2023impact}.} This setting involves aggregated county-level data and has treatment histories missing for around 50\% of the sample. The second application conducts a comprehensive meta-analysis utilizing individual-level household data from CPS to examine the effects of worker disability, job certification, and work absenteeism on family income and hours worked. Treatment histories are generally missing here for less than 5\% of the sample.

\subsection{Political engagement in the U.S. during COVID-19}\label{sec:app}
We obtain daily county-level COVID-19 transmission data from the CDC from 2020 to 2022.\footnote{Following \citet{callaway2023evaluating}, we obtain data on weekly number of covid cases from \url{https://data.cdc.gov/Public-Health-Surveillance/United-States-COVID-19-County-Level-of-Community-T/8396-v7yb/about_data}.} This is combined with county-level voter turnout data between 2004-2022 from the National Neighborhood Data Archive at Inter-university Consortium for Political and Social Research. We supplement these data with county-level covariates from the US Census Bureau and United State's Department of Agriculture Economic Research Service database. 

Our binary treatment measures whether, during a given month, a county's weekly number of confirmed cases per 100,000 people ever exceeds the national weekly average in a given year. We refer to our treatment variable as ``above-average'' cases in a particular year. To illustrate our method, we consider August across three years; 2018 is the pre-treatment period, 2020 is the middle period, and 2021 is the final period.\footnote{While we could have used 2022 to be the last treatment period, COVID cases had declined significantly that year with mass vaccination already underway.} In the data, 51\% of the counties were missing confirmed cases data in at least one week of August 2020 (middle period). Even though the last treatment period is 2021, turnout rates are only available in 2022 since there were no general elections in the previous year. Our outcome is defined as changes in county-level voter turnout between the first and final period. The final sample has 3,096 counties.

We condition on county-level covariates\footnote{See Supplementary Appendix \ref*{sec:emp_sc} for additional details.} that can account for differential trends in voter turnout. Turnout patterns are typically stable over time at the county level, and the timing of case surges is plausibly exogenous to underlying electoral dynamics. Since covid cases are likely to be missing due to factors like public health infrastructure and reporting practices, which could be correlated with the county characteristics in the observed covariates, our MAR assumption is also plausible in this setting.

Table \ref{app:tab} reports the estimated effects of having above-average number of cases on turnout rates in the 2022 general elections along with standard errors and estimated confidence intervals. 
Based on the robust method, we find that having above-average cases in 2020 and 2021 reduces turnout rates for counties that experienced it by 0.18\% points, on average. Unlike the estimates obtained using adjusted-DR or CC methods, this estimate is statistically significant.
In general, the CC-OR, CC-IPW, and CC-DR estimates are smaller than their adjusted counterparts with narrower confidence intervals. 

\begin{table}[!ht]
	\centering
	\scalebox{0.90}{
		\begin{threeparttable}
        \caption{Results for the effects of above-average COVID cases on turnout rates}
        \label{app:tab}
        \begin{tabular}{ccccccccc}
        \toprule
        PDATT & \multicolumn{2}{c}{11-00} &       & \multicolumn{2}{c}{10-00} &       & \multicolumn{2}{c}{01-00} \\
        \midrule
        \multirow{3}[2]{*}{R} & \multicolumn{2}{c}{-0.176} &       & \multicolumn{2}{c}{-0.117} &       & \multicolumn{2}{c}{-0.080} \\
              & \multicolumn{2}{c}{(0.066)} &       & \multicolumn{2}{c}{(0.106)} &       & \multicolumn{2}{c}{(0.063)} \\
              & [-0.305 & -0.047] &       & [-0.325 & 0.091] &       & [-0.204 & 0.043] \\
    \cmidrule{1-3}\cmidrule{5-6}\cmidrule{8-9}    \multirow{3}[2]{*}{DR} & \multicolumn{2}{c}{-0.138} &       & \multicolumn{2}{c}{-0.143} &       & \multicolumn{2}{c}{-0.039} \\
              & \multicolumn{2}{c}{(0.079)} &       & \multicolumn{2}{c}{(0.080)} &       & \multicolumn{2}{c}{(0.070)} \\
              & [-0.293 & 0.018] &       & [-0.300 & 0.013] &       & [-0.177 & 0.099] \\
    \cmidrule{1-3}\cmidrule{5-6}\cmidrule{8-9}    \multirow{3}[2]{*}{IPW} & \multicolumn{2}{c}{-0.114} &       & \multicolumn{2}{c}{-0.116} &       & \multicolumn{2}{c}{-0.031} \\
              & \multicolumn{2}{c}{(0.054)} &       & \multicolumn{2}{c}{(0.051)} &       & \multicolumn{2}{c}{(0.043)} \\
              & [-0.219 & -0.009] &       & [-0.216 & -0.017] &       & [-0.116 & 0.054] \\
    \cmidrule{1-3}\cmidrule{5-6}\cmidrule{8-9}    \multirow{3}[2]{*}{OR} & \multicolumn{2}{c}{-0.013} &       & \multicolumn{2}{c}{-0.010} &       & \multicolumn{2}{c}{0.008} \\
              & \multicolumn{2}{c}{(0.018)} &       & \multicolumn{2}{c}{(0.015)} &       & \multicolumn{2}{c}{(0.019)} \\
              & [-0.047 & 0.022] &       & [-0.040 & 0.020] &       & [-0.029 & 0.046] \\
        \midrule
        \midrule
        \multirow{3}[2]{*}{CC-DR} & \multicolumn{2}{c}{-0.034} &       & \multicolumn{2}{c}{-0.033} &       & \multicolumn{2}{c}{-0.029} \\
              & \multicolumn{2}{c}{(0.027)} &       & \multicolumn{2}{c}{(0.026)} &       & \multicolumn{2}{c}{(0.010)} \\
              & [-0.086 & 0.019] &       & [-0.084 & 0.019] &       & [-0.048 & -0.010] \\
    \cmidrule{1-3}\cmidrule{5-6}\cmidrule{8-9}    \multirow{3}[2]{*}{CC-IPW} & \multicolumn{2}{c}{-0.029} &       & \multicolumn{2}{c}{-0.030} &       & \multicolumn{2}{c}{-0.031} \\
              & \multicolumn{2}{c}{(0.023)} &       & \multicolumn{2}{c}{(0.023)} &       & \multicolumn{2}{c}{(0.008)} \\
              & [-0.075 & 0.017]&       & [-0.075 & 0.014] &       & [-0.046 & -0.016] \\
    \cmidrule{1-3}\cmidrule{5-6}\cmidrule{8-9}    \multirow{3}[2]{*}{CC-OR} & \multicolumn{2}{c}{-0.013} &       & \multicolumn{2}{c}{-0.013} &       & \multicolumn{2}{c}{-0.025} \\
              & \multicolumn{2}{c}{(0.017)} &       & \multicolumn{2}{c}{(0.013)} &       & \multicolumn{2}{c}{(0.009)} \\
              & [-0.046 & 0.020] &       & [-0.037 & 0.012] &       & [-0.042 & -0.008] \\
        \bottomrule
        \bottomrule
        \end{tabular}%
        \begin{tablenotes}[flushleft]
            \footnotesize
            \item [a] Standard errors are reported in parentheses and the 95\% confidence intervals are reported in brackets. 
        \end{tablenotes}
    \end{threeparttable}}%
\end{table}%

\subsection{Effect of labor market conditions on income and hours worked}
We use publicly available household survey data from the CPS which is accessed through the Integrated Public Use Microdata Series (IPUMS). The CPS is a nationally representative monthly survey conducted jointly by the U.S. Census Bureau and the Bureau of Labor Statistics, and serves as the official source of labor force statistics for the U.S. population. 

We construct a three-period panel by using the monthly observations within a given year with the household head (HH) as the unit of analysis. We define the initial and final periods as the first and final months a household is observed. Treatment in the middle period is defined as whether the HH receives the treatment during the intermediate month(s). Disability, job certification, and work absence treatment status may be missing in the middle period for various reasons: the household may not have been surveyed during those months due to CPS rotation design, responses may be missing due to item or unit non-response, or responses may be unknown (which CPS codes as NIU). Overall, missingness in the middle period remains low, affecting fewer than 5\% of the samples.\footnote{See Supplementary Appendix \ref*{sec:emp_sc} for additional details and sample construction.} The CPS also collects extensive demographic information on the household members which includes region, race, sex, marital status, level of education, nativity, which are standardized before before being used in estimation. When missingness is uncorrelated with treatment status in the middle time period and income or hours worked, conditional on the covariates, our MAR assumption holds in this setting. 

Based on the robust estimates, we find PDATTs align with economic intuition and vary across time. For example, in 2009, having a disability in both periods reduced hours worked by 3.559 hours (with a standard error of 1.149), while being disabled in only one period in 2009 does not have a statistically significant effect. 
In contrast, having job certification in both periods or the second period in 2017 has a statistically significant positive effect on family income: the effect of job certification in both periods is \$668.652, only in the first period is \$148.938, and only in the second period is \$391.694, with standard errors equal to 297.751, 225.406, and 73.537, respectively. 
For work absence, we find estimates indicating that only the effect in the final period is significant: being absent from work in both periods increases family income in 2009 by \$20.421 (13.576), while the effects of absence in the first and second period equal a reduction in income of \$10.949 (7.441) and \$78.985 (17.976).

The differences between the robust estimates and the estimates from the DR and CC-DR estimators can be substantial. Table~\ref{tab:percentdiff} reports the mean, median, and maximum values of the absolute percent differences in the PDATT estimates of these methods across all outcome-by-year combinations for each treatment variable. Consider $\tau_{(01)(00)}$ for the disability treatment. On average, the R estimate is 16\% larger than the DR estimate with a maximum percent difference of 57\%, and 36\% larger than the CC-DR estimate with a maximum percent difference of 97\%. For job certification, the mean differences between R and DR and CC-DR for $\tau_{(01)(00)}$ are 1.5\% and 18\%. 
Similarly, the mean differences for absence are smaller compared to disability. 
The maximum differences show that the estimates from R can be very different from CC-DR, with percentage differences reaching up to 157\%.

\begin{table}[!ht]
  \centering
  \scalebox{0.90}{\begin{threeparttable}
  \caption{Absolute percent differences in R, DR, and CC-DR estimates}
    \begin{tabular}{cccccc}
    \toprule
    Treatment & PDATT & Summary & R vs. DR & R vs. CC-DR & DR vs. CC-DR \\
    \midrule
    \multirow{9}[6]{*}{Disability} & \multirow{3}[2]{*}{11-00} & Mean  & 1.582 & 6.942 & 6.186 \\
          &       & Median & 1.047 & 6.452 & 5.785 \\
          &       & Max   & 4.222 & 13.628 & 11.955 \\
    \cmidrule{2-6}          & \multirow{3}[2]{*}{10-00} & Mean  & 0.440 & 0.721 & 0.353 \\
          &       & Median & 0.340 & 0.572 & 0.305 \\
          &       & Max   & 0.942 & 1.596 & 0.660 \\
    \cmidrule{2-6}          & \multirow{3}[2]{*}{01-00} & Mean  & 16.133 & 36.375 & 19.276 \\
          &       & Median & 3.846 & 23.815 & 14.313 \\
          &       & Max   & 56.755 & 96.933 & 48.195 \\
    \midrule
    \multirow{9}[6]{*}{Job certification} & \multirow{3}[2]{*}{11-00} & Mean  & 0.527 & 5.429 & 5.623 \\
          &       & Median & 0.406 & 4.066 & 4.454 \\
          &       & Max   & 1.189 & 14.551 & 13.900 \\
    \cmidrule{2-6}          & \multirow{3}[2]{*}{10-00} & Mean  & 10.598 & 11.119 & 0.890 \\
          &       & Median & 1.586 & 2.106 & 1.029 \\
          &       & Max   & 44.591 & 43.979 & 1.732 \\
    \cmidrule{2-6}          & \multirow{3}[2]{*}{01-00} & Mean  & 1.525 & 17.861 & 17.981 \\
          &       & Median & 1.600 & 9.258 & 11.631 \\
          &       & Max   & 2.685 & 60.766 & 59.683 \\
    \midrule
    \multirow{9}[6]{*}{Absence} & \multirow{3}[2]{*}{11-00} & Mean  & 0.355 & 4.534 & 4.342 \\
          &       & Median & 0.161 & 0.770 & 0.794 \\
          &       & Max   & 1.942 & 40.097 & 39.213 \\
    \cmidrule{2-6}          & \multirow{3}[2]{*}{10-00} & Mean  & 0.315 & 2.043 & 1.958 \\
          &       & Median & 0.157 & 0.750 & 0.732 \\
          &       & Max   & 2.821 & 12.915 & 12.976 \\
    \cmidrule{2-6}          & \multirow{3}[2]{*}{01-00} & Mean  & 0.303 & 9.687 & 9.624 \\
          &       & Median & 0.080 & 1.460 & 1.203 \\
          &       & Max   & 3.591 & 156.969 & 154.994 \\
    \bottomrule
    \end{tabular}%
  	\label{tab:percentdiff}%
  	\begin{tablenotes}[flushleft]
  	\footnotesize
  	\item \textit{Notes:} This table presents the absolute percent differences between two sets of estimates, calculated as \textbar(estimate 2 - estimate 1) / estimate 1\textbar $\times$ 100. For each treatment variable, we report the mean, median, and maximum percent differences across 4, 5, and 25 outcome-by-year samples for the disability, job certification, and absence treatments, respectively.
  \end{tablenotes}
\end{threeparttable}}
\end{table}%

\section{Conclusion}\label{sec:conclude} 
In this paper, we consider a difference-in-differences framework with a binary time-varying treatment, no treated units in the pre-treatment period, but otherwise no restrictions on treatment-path heterogeneity. We identify and estimate the effect of each treatment history on final period outcomes with treatment histories partially observed. We propose a novel AIPW estimand which identifies the target parameter as long as \textit{any two} of the outcome, propensity score, or missing treatment models are correctly specified. This method generalizes and improves upon other missingness-adjusted alternatives (such as IPW, OR, and DR) which require the missing treatment model to be correctly specified alongside another correctly specified component.

We present numerical experiments which compare the performance of the missingness-adjusted estimators and their complete-case counterparts. We find that the robust estimand remains unbiased and controls size across the three cases of model misspecification whereas the other adjusted estimators exhibit bias and size distortions once the missing treatment model is misspecified. By varying the degree of missingness and misspecification, we show that the bias in R remains negligible compared to the bias in DR and CC-DR estimators. We further demonstrate the applicability of the missingness-adjusted methods compared to the practice of dropping data through two empirical applications. First, we find an economically and statistically significant treatment effect of covid cases on voter turnout across U.S.\ counties in the presence of 51\% missingness using the proposed estimator. Second, a meta-analysis on CPS household data shows that the proposed method can produce estimates very different from existing methods in a wide range of settings even with missingness below 5\%.

\bibliographystyle{ecta.bst}
\bibliography{GDID_dynamic}

\singlespacing
\appendix
\numberwithin{equation}{section}
\numberwithin{table}{section}
\numberwithin{figure}{section}
\numberwithin{lemma}{section}
\numberwithin{proposition}{section}
\numberwithin{assumption}{section}
\section*{\centering Appendix}
\section{Standard DID approaches}

\subsection{Proof Proposition \ref{prop:ignoreD1}}\label{sec:proof_ignoreD1}
\begin{proof}
    First consider $\mathbbm{E}[\Delta Y|D_2=1, \mathbf{X}]-\mathbbm{E}[\Delta Y|D_2=0, \mathbf{X}] = \mathbbm{E}[Y_2-Y_0|D_2=1, \mathbf{X}] - \mathbbm{E}[Y_2-Y_0|D_2=0, \mathbf{X}]$. 
    Now, 
    {\normalsize \begin{align*}
        \mathbb{E}[Y_2|D_2=1, \mathbf{X}] 
        & = \mathbb{E}[Y_2(1,1)|D_1=1,D_2=1,\mathbf{X}]\cdot \mathbbm{P}(D_1=1|D_2=1,\mathbf{X})\\
        &+\mathbb{E}[Y_2(0,1)|D_1=0, D_2=1,\mathbf{X}]\cdot \mathbbm{P}(D_1=0|D_2=1,\mathbf{X}) \\
        \mathbb{E}[Y_2|D_2=0,\mathbf{X}] 
        & = \mathbb{E}[Y_2(1,0)|D_1=1, D_2=0,\mathbf{X}]\cdot \mathbbm{P}(D_1=1|D_2=0,\mathbf{X}) \\
        &+ \mathbb{E}[Y_2(0,0)|D_1=0, D_2=0,\mathbf{X}]\cdot \mathbbm{P}(D_1=0|D_2=0,\mathbf{X})  \\
        \mathbb{E}[Y_0|D_2=1,\mathbf{X}] 
        & = \mathbb{E}[Y_0(0,0)|D_1=1,D_2=1,\mathbf{X}]\cdot \mathbbm{P}(D_1=1|D_2=1,\mathbf{X})\\
        &+\mathbb{E}[Y_0(0,0)|D_1=0, D_2=1,\mathbf{X}]\cdot \mathbbm{P}(D_1=0|D_2=1,\mathbf{X}) \tag{Assumption \ref{ass:did}.1}\\
        \mathbb{E}[Y_0|D_2=0,\mathbf{X}] 
        & = \mathbb{E}[Y_0(0,0)|D_1=1, D_2=0,\mathbf{X}]\cdot \mathbbm{P}(D_1=1|D_2=0,\mathbf{X})\\
        &+\mathbb{E}[Y_0(0,0)|D_1=0, D_2=0,\mathbf{X}]\cdot \mathbbm{P}(D_1=0|D_2=0,\mathbf{X}). \tag{Assumption \ref{ass:did}.1}
    \end{align*}}
	Combining the above results, we get $\mathbbm{E}[\Delta Y|D_2=1,\mathbf{X}]-\mathbbm{E}[\Delta Y|D_2=0,\mathbf{X}]$
	{\normalsize \begin{align*}
        & = \mathbb{E}[Y_2(1,1)-Y_2(0,0)|D_1=1, D_2=1,\mathbf{X}]\cdot \mathbbm{P}(D_1=1|D_2=1,\mathbf{X})\\
        &+\mathbb{E}[Y_2(0,1)-Y_2(0,0)|D_1=0, D_2=1,\mathbf{X}]\cdot \mathbbm{P}(D_1=0|D_2=1,\mathbf{X}) \\
		& - \mathbb{E}[Y_2(1,0)-Y_2(0,0)|D_1=1, D_2=0,\mathbf{X}]\cdot \mathbbm{P}(D_1=1|D_2=0,\mathbf{X}),
	\end{align*}}
where we use Assumption~\ref{ass:did}.2. Now $\mathbb{E}[D_2]^{-1}\mathbb{E}\left[D_2\left(\mathbb{E}[\Delta Y|D_2=1,\mathbf{X}]-\mathbb{E}[\Delta Y|D_2=0,\mathbf{X}]\right)\right]$ equals $\tau_{(11)(00)}\cdot \mathbb{P}(D_1=1|D_2=1)
        +\tau_{(01)(00)}\cdot \mathbb{P}(D_1=0|D_2=1) 
		-\tau_{(10)(00)}\cdot\mathbb{P}(D_1=1|D_2=0)$, 
where we use that $\mathbbm{P}(D_1|D_2,\mathbf{X})\mathbbm{P}(D_2|\mathbf{X})\mathbbm{P}(\mathbf{X})/\mathbbm{P}(D_2)=\mathbbm{P}(\mathbf{X}|D_1,D_2)\mathbbm{P}(D_1|D_2)$.
\end{proof}

\subsection{Proof Proposition \ref{prop:ccbias}}\label{sec:proof_prop1}
 \begin{proof}
Under Assumption \ref{ass:mar}, it follows from Appendix~\ref{sec:proof_mu_pi_identification} that $\mathbb{E}[\Delta Y|\mathbf{D}=\mathbf{d},S=1, \mathbf{X}]-\mathbb{E}[\Delta Y|\mathbf{D}=\mathbf{d}^\prime, S=1, \mathbf{X}] = \mathbb{E}[\Delta Y|\mathbf{D}=\mathbf{d}, \mathbf{X}]-\mathbb{E}[\Delta Y|\mathbf{D}=\mathbf{d}^\prime, \mathbf{X}]= \mathbb{E}[Y_2(\mathbf{d})-Y_2(\mathbf{d}')|\mathbf{D}=\mathbf{d},\mathbf{X}]$, where the second equality uses \eqref{eq:m2taux} with Assumption \ref{ass:did}. It follows that 
{\normalsize\begin{align*}
    &\mathbb{E}\left[S\mathbbm{1}[\mathbf{D}=\mathbf{d}]\right]^{-1}\mathbb{E}\left[S\mathbbm{1}[\mathbf{D}=\mathbf{d}]\left(\mathbb{E}[\Delta Y|\mathbf{D}=\mathbf{d},S=1, \mathbf{X}]-\mathbb{E}[\Delta Y|\mathbf{D}=\mathbf{d}^\prime, S=1, \mathbf{X}]\right)\right]=\\
    &\mathbb{E}\left[\mathbb{E}[Y_2(\mathbf{d})-Y_2(\mathbf{d}')|\mathbf{D}=\mathbf{d},\mathbf{X}]\frac{\mathbb{P}(\mathbf{D}=\mathbf{d},S=1|\mathbf{X})}{\mathbb{P}(\mathbf{D}=\mathbf{d},S=1)}\right]=\\
    &\int_\mathbf{X} \mathbb{E}[Y_2(\mathbf{d})-Y_2(\mathbf{d}')|\mathbf{D}=\mathbf{d},\mathbf{X}] d \mathbb{P}(\mathbf{X}|\mathbf{D}=\mathbf{d},S=1).
\end{align*}}
From \eqref{eq:mp2tau} follows that $\tau_{\mathbf{dd^\prime}} = \int_\mathbf{X} \mathbb{E}[Y_2(\mathbf{d})-Y_2(\mathbf{d}')|\mathbf{D}=\mathbf{d},\mathbf{X}] d \mathbb{P}(\mathbf{X}|\mathbf{D}=\mathbf{d})$. Hence,
{\normalsize \begin{align*}
    &\mathbb{E}\left[S\mathbbm{1}[\mathbf{D}=\mathbf{d}]\right]^{-1}\mathbb{E}\left[S\mathbbm{1}[\mathbf{D}=\mathbf{d}]\left(\mathbb{E}[\Delta Y|\mathbf{D}=\mathbf{d},S=1, \mathbf{X}]-\mathbb{E}[\Delta Y|\mathbf{D}=\mathbf{d}^\prime, S=1, \mathbf{X}]\right)\right]=\\
    &\tau_{\mathbf{dd^\prime}}+\int_\mathbf{X} \mathbb{E}[Y_2(\mathbf{d})-Y_2(\mathbf{d}')|\mathbf{D}=\mathbf{d},\mathbf{X}] \left(\mathbb{P}(\mathbf{X}|\mathbf{D}=\mathbf{d},S=1)-\mathbb{P}(\mathbf{X}|\mathbf{D}=\mathbf{d})\right)d\mathbf{X}=\\
    &\tau_{\mathbf{dd^\prime}}+\mathbb{P}(S=0|\mathbf{D}=\mathbf{d})\int_\mathbf{X} \mathbb{E}[Y_2(\mathbf{d})-Y_2(\mathbf{d}')|\mathbf{D}=\mathbf{d},\mathbf{X}] \big(\mathbb{P}(\mathbf{X}|\mathbf{D}=\mathbf{d},S=1)\\
    &-\mathbb{P}(\mathbf{X}|\mathbf{D}=\mathbf{d},S=0)\big)d\mathbf{X},
\end{align*}}
 where we use that $\mathbb{P}(\mathbf{X}|\mathbf{D}=\mathbf{d})=\mathbb{P}(\mathbf{X}|\mathbf{D}=\mathbf{d},S=1)\mathbb{P}(S=1|\mathbf{D}=\mathbf{d})+\mathbb{P}(\mathbf{X}|\mathbf{D}=\mathbf{d},S=0)\mathbb{P}(S=0|\mathbf{D}=\mathbf{d})$.
\end{proof}

\section{Robust identification of PDATTs}\label{A:mainproofs}
First, we generalize Assumptions~\ref{ass:did} and \ref{ass:mar} to the setting discussed in Section~\ref{sec:general}. This setting boils down to the exposition in the paper with three periods and $D_1$ partially missing by setting $T=2$ and $\mathbf{D}_h=D_1$. Second, this appendix presents the proofs for our main results in Section~\ref{sec:robust} in the general setting, which also apply to the setting with $T=2$ and $\mathbf{D}_h=D_1$.

For Assumption~\ref{ass:did}, replace $Y_2(\mathbf{D})$ by $Y_T(\mathbf{D})$ and note that $\mathbf{D}$ is now a $T$-dimensional vector:
\begin{assumption}[Difference-in-differences assumptions]\label{ass:did_extn} For each $\mathbf{d}$, we have
\begin{enumerate}
    \item (No anticipation) $\mathbb{E}\left[Y_0(\mathbf{d})|\mathbf{D}=\mathbf{d},\mathbf{X}\right] = \mathbb{E}\left[Y_0(\mathbf{0})|\mathbf{D}=\mathbf{d},\mathbf{X}\right]$.
    \item (Parallel trends) $\mathbb{E}\left[Y_T(\mathbf{0})-Y_0(\mathbf{0})|\mathbf{D}=\mathbf{d}, \mathbf{X}\right] = \mathbb{E}[Y_T(\mathbf{0})-Y_0(\mathbf{0})|\mathbf{X}]$.
	\item (Overlap) $\mathbb{P}(\mathbf{D}=\mathbf{d}|\mathbf{X})\equiv p_\mathbf{d}(\mathbf{X})$ is bounded away from one.	
\end{enumerate}
\end{assumption}
Let $\Delta Y \equiv Y_T-Y_0$, and replace $D_1$ and $D_2$ in Assumption~\ref{ass:mar} by $\mathbf{D}_h$ and $\mathbf{D}_{-h}$, respectively:
\begin{assumption}[Missingness assumptions]\label{ass:mar_extn} \, 
	\begin{enumerate}
		\item (Missing at random) $ S \perp (\mathbf{D}_h, \Delta Y_t)| \mathbf{D}_{-h}, \mathbf{X}$.
		\item (Partial observability) $0<q_{\mathbf{d}_{-h}}(\mathbf{X}) \equiv \mathbb{P}(S=1|\mathbf{D}_{-h}=\mathbf{d}_{-h},\mathbf{X})\leq 1$.
	\end{enumerate}
\end{assumption}
We generalize the notation for the models as follows. Let, $\pi_{\mathbf{d}_{h}|\mathbf{d}_{-h}}(\mathbf{X})$ and $\pi_{\mathbf{d}_{-h}}(\mathbf{X})$ represent the propensity scores $p_{\mathbf{d}_{h}|\mathbf{d}_{-h}}(\mathbf{X})\equiv \mathbb{P}({\mathbf{D}_h}={\mathbf{d}_h}|\mathbf{D}_{-h}=\mathbf{d}_{-h},\mathbf{X})$ and $p_{\mathbf{d}_{-h}}(\mathbf{X})\equiv \mathbb{P}(\mathbf{D}_{-h}=\mathbf{d}_{-h}|\mathbf{X})$, respectively, and $\phi_{\mathbf{d}_{-h}}(\mathbf{X})$ denote the missing treatment probability $q_{\mathbf{d}_{-h}}(\mathbf{X})$.

\subsection{Proof Lemma~\ref{lem:mu_pi_identification}}\label{sec:proof_mu_pi_identification}
If $\mu_{\mathbf{d}}(\mathbf{X})=m_{\mathbf{d}}(\mathbf{X})=\mathbb{E}[\Delta Y|\mathbf{D}=\mathbf{d},\mathbf{X}]$, it holds that
\begin{align}
		\mu_{\mathbf{d}}(\mathbf{X}) = &\frac{\mathbb{E}[\Delta Y|\mathbf{D}=\mathbf{d},\mathbf{X}]\mathbb{P}(\mathbf{D}=\mathbf{d}|\mathbf{X})}{\mathbb{P}(\mathbf{D}=\mathbf{d}|\mathbf{X})}=\frac{\mathbb{E}[\mathbbm{1}[\mathbf{D}_h=\mathbf{d}_h]\Delta Y|\mathbf{D}_{-h}=\mathbf{d}_{-h},\mathbf{X}]\mathbb{P}(\mathbf{D}_{-h}=\mathbf{d}_{-h}|\mathbf{X})}{\mathbb{P}(\mathbf{D}_h=\mathbf{d}_h|\mathbf{D}_{-h}=\mathbf{d}_{-h},\mathbf{X})\mathbb{P}(\mathbf{D}_{-h}=\mathbf{d}_{-h}|\mathbf{X})} \nonumber \\
		=&\frac{\mathbb{E}[\mathbbm{1}[\mathbf{D}_h=\mathbf{d}_h]\Delta Y|\mathbf{D}_{-h}=\mathbf{d}_{-h},\mathbf{X},S=1]}{\mathbb{P}(\mathbf{D}_h=\mathbf{d}_h|\mathbf{D}_{-h}=\mathbf{d}_{-h},\mathbf{X},S=1)} \times \frac{\mathbb{P}(S=1|\mathbf{D}_{-h}=\mathbf{d}_{-h},\mathbf{X})}{\mathbb{P}(S=1|\mathbf{D}_{-h}=\mathbf{d}_{-h},\mathbf{X})} \nonumber\\ {=} &
		\frac{\mathbb{E}[S\mathbbm{1}[\mathbf{D}_h=\mathbf{d}_h]\Delta Y|\mathbf{D}_{-h}=\mathbf{d}_{-h},\mathbf{X}]}{\mathbb{P}(S\mathbbm{1}[\mathbf{D}_h=\mathbf{d}_h]=1|\mathbf{D}_{-h}=\mathbf{d}_{-h},\mathbf{X})} =
		\mathbb{E}[\Delta Y|\mathbf{D}=\mathbf{d},\mathbf{X},S=1],
\end{align}
where the third equality uses Assumption~\ref{ass:mar_extn} to write $\mathbb{E}[\mathbbm{1}[\mathbf{D}_h=\mathbf{d}_h]\Delta Y|\mathbf{D}_{-h}=\mathbf{d}_{-h},\mathbf{X}]=\mathbb{E}[\mathbbm{1}[\mathbf{D}_h=\mathbf{d}_h]\Delta Y|\mathbf{D}_{-h}=\mathbf{d}_{-h},\mathbf{X},S=1]$ and $\mathbb{P}(\mathbf{D}_h=\mathbf{d}_h|\mathbf{D}_{-h}=\mathbf{d}_{-h},\mathbf{X})=\mathbb{P}(\mathbf{D}_h=\mathbf{d}_h|\mathbf{D}_{-h}=\mathbf{d}_{-h},\mathbf{X},S=1)$. 

If $\pi_{\mathbf{d}_{h}|\mathbf{d}_{-h}}(\mathbf{X})=p_{\mathbf{d}_{h}|\mathbf{d}_{-h}}(\mathbf{X})=\mathbb{P}(\mathbf{D}_h=\mathbf{d}_h|\mathbf{D}_{-h}=\mathbf{d}_{-h},\mathbf{X})$, it holds that
\begin{align}
	\pi_{\mathbf{d}_{h}|\mathbf{d}_{-h}}(\mathbf{X}) =
	 \mathbb{P}(\mathbf{D}_h=\mathbf{d}_h|\mathbf{D}_{-h}=\mathbf{d}_{-h},\mathbf{X},S=1),
\end{align}
where we use Assumption~\ref{ass:mar_extn} to write $\mathbb{P}(D_h=d_h|\mathbf{D}_{-h}=\mathbf{d}_{-h},\mathbf{X})=\mathbb{P}(D_h=d_h|\mathbf{D}_{-h}=\mathbf{d}_{-h},\mathbf{X},S=1)$. 
It follows then that correctly specified outcome and propensity score models can be identified using the observed sample. \qedsymbol

\subsection{Proof Theorem~\ref{thm:robust}}\label{sec:proof_robust}
First, we derive expressions for the four terms in the first part of the estimand: 
\begin{align}
\mathbb{E}\left[  
\left(w_{1}(S,\mathbf{D},\mathbf{X})-w_{2}(S,\mathbf{D},\mathbf{X})\right)\left( \Delta Y -\mu_{\mathbf{d}^\prime}(\mathbf{X}) \right)\right].
\end{align}
We invoke LIE and Assumption~\ref{ass:mar_extn} to write
\begin{align}
		\mathbb{E}[w_{1}(S,\mathbf{D},\mathbf{X}) \Delta Y] =&  {\mathbb{E}\left[\frac{S}{\phi_{\mathbf{d}_{-h}}(\mathbf{X})}\mathbbm{1}[\mathbf{D}=\mathbf{d}]\right]}^{-1}\times {\mathbb{E}\left[\frac{S}{\phi_{d_{-h}}(\mathbf{X})}\mathbbm{1}[\mathbf{D}=\mathbf{d}]\Delta Y\right]} \nonumber \\
		=& {\mathbb{E}\left[\frac{q_{\mathbf{d}_{-h}}( \mathbf{X})}{\phi_{\mathbf{d}_{-h}}(\mathbf{X})}p_{\mathbf{d}}(\mathbf{X})\right]}^{-1}\times {\mathbb{E}\left[\frac{q_{\mathbf{d}_{-h}}( \mathbf{X})}{\phi_{\mathbf{d}_{-h}}(\mathbf{X})}p_{\mathbf{d}}(\mathbf{X})m_{\mathbf{d}}(\mathbf{X})\right]}.
		\label{eq:w1dy}
\end{align}
Similarly, we have
{\normalsize \begin{align}
    \mathbb{E}[w_{1}(S,\mathbf{D},\mathbf{X}) \mu_{\mathbf{d}^\prime}(\mathbf{X})] =&
    {\mathbb{E}\left[\frac{q_{\mathbf{d}_{-h}}( \mathbf{X})}{\phi_{\mathbf{d}_{-h}}(\mathbf{X})}p_{\mathbf{d}}(\mathbf{X})\right]}^{-1}\times {\mathbb{E}\left[\frac{q_{\mathbf{d}_{-h}}( \mathbf{X})}{\phi_{\mathbf{d}_{-h}}(\mathbf{X})}p_{\mathbf{d}}(\mathbf{X})\mu_{\mathbf{d}^\prime}(\mathbf{X})\right]},\label{eq:w1mu}\\
    \mathbb{E}[w_{2}(S,\mathbf{D},\mathbf{X}) \Delta Y] =&  {\mathbb{E}\left[\frac{q_{\mathbf{d}_{-h}^\prime}( \mathbf{X})}{\phi_{\mathbf{d}_{-h}^\prime}(\mathbf{X})}\frac{\pi_{\mathbf{d}}(\mathbf{X})}{\pi_{\mathbf{d^\prime}}(\mathbf{X})}p_{\mathbf{\mathbf{d}^\prime}}(\mathbf{X})\right]}^{-1}\times {\mathbb{E}\left[\frac{q_{\mathbf{d}_{-h}^\prime}( \mathbf{X})}{\phi_{\mathbf{d}_{-h}^\prime}(\mathbf{X})}\frac{\pi_{\mathbf{d}}(\mathbf{X})}{\pi_{\mathbf{d^\prime}}(\mathbf{X})}p_{\mathbf{d^\prime}}(\mathbf{X})m_{\mathbf{d}^\prime}(\mathbf{X})\right]}, \label{eq:w2dy}\\
    \mathbb{E}[w_{2}(S,\mathbf{D},\mathbf{X}) \mu_{\mathbf{d}^\prime}(\mathbf{X})] =&  {\mathbb{E}\left[\frac{q_{\mathbf{d}_{-h}^\prime}( \mathbf{X})}{\phi_{\mathbf{d}_{-h}^\prime}(\mathbf{X})}\frac{\pi_{\mathbf{d}}(\mathbf{X})}{\pi_{\mathbf{d^\prime}}(\mathbf{X})}p_{\mathbf{d^\prime}}(\mathbf{X})\right]}^{-1}\times {\mathbb{E}\left[\frac{q_{\mathbf{d}_{-h}^\prime}( \mathbf{X})}{\phi_{\mathbf{d}_{-h}^\prime}(\mathbf{X})}\frac{\pi_{\mathbf{d}}(\mathbf{X})}{\pi_{\mathbf{d^\prime}}(\mathbf{X})}p_{\mathbf{d^\prime}}(\mathbf{X})\mu_{\mathbf{d}^\prime}(\mathbf{X})\right]}. \label{eq:w2mu}
\end{align}}
Second, we show that for each of the three cases, $\tau_{\mathbf{dd^\prime}}^{\text{R}}={\mathbb{E}\left[p_{\mathbf{d}}(\mathbf{X})\right]}^{-1}\mathbb{E}\left[(m_{\mathbf{d}}(\mathbf{X})-m_{\mathbf{d^\prime}}(\mathbf{X}))p_\mathbf{d}(\mathbf{X})\right].$

\subsubsection*{Missing data model correct}
If $\phi_{\mathbf{d}_{-h}}(\mathbf{X})=q_{\mathbf{d}_{-h}}(\mathbf{X})$, it holds that $\mathbb{E}\left[\frac{S}{\phi_{\mathbf{d}_{-h}}(\mathbf{X})}\pi_{\mathbf{d}_h|\mathbf{d}_{-h}}(\mathbf{X})\mathbbm{1}[\mathbf{D}_{-h}=\mathbf{d}_{-h}]\big|\mathbf{D}_{-h}=\mathbf{d}_{-h}, \mathbf{X}\right]$ equals
\begin{equation}
		 \frac{q_{\mathbf{d}_{-h}}(\mathbf{X})}{\phi_{\mathbf{d}_{-h}}(\mathbf{X})}\pi_{\mathbf{d}_h|\mathbf{d}_{-h}}(\mathbf{X})\mathbbm{1}[\mathbf{D}_{-h}=\mathbf{d}_{-h}] 
		=  \pi_{\mathbf{d}_h|\mathbf{d}_{-h}}(\mathbf{X})\mathbbm{1}[\mathbf{D}_{-h}=\mathbf{d}_{-h}].
\end{equation}
It follows that $\mathbb{E}\left[w_{3}(\mathbf{D}_{-h},\mathbf{X})|\mathbf{D}_{-h},\mathbf{X}\right]=\mathbb{E}\left[w_{4}(S,\mathbf{D}_{-h},\mathbf{X})|\mathbf{D}_{-h},\mathbf{X}\right]$, and hence
\begin{align}
	&\mathbb{E}[\left(w_{3}(\mathbf{D}_{-h},\mathbf{X})-w_{4}(S,\mathbf{D}_{-h},\mathbf{X})\right)\left( \mu_{\mathbf{d}}(\mathbf{X}) -\mu_{\mathbf{d}^\prime}(\mathbf{X}) \right)] =\nonumber \\
	&\qquad \qquad \mathbb{E}\left\{\mathbb{E}\left[w_{3}(\mathbf{D}_{-h},\mathbf{X})-w_{4}(S,\mathbf{D}_{-h},\mathbf{X})|\mathbf{D}_{-h},\mathbf{X}\right]\left( \mu_{\mathbf{d}}(\mathbf{X}) -\mu_{\mathbf{d}^\prime}(\mathbf{X}) \right)\right\} =0. \label{eq:w3minw4}
\end{align}

\subsubsection*{1. Missing data model and propensity score correct}
If $\phi_{\mathbf{d}_{-h}}(\mathbf{X})=q_{\mathbf{d}_{-h}}(\mathbf{X})$ and $\pi_{\mathbf{d}}(\mathbf{X}) = p_{\mathbf{d}}(\mathbf{X})$, it follows from \eqref{eq:w1mu} and \eqref{eq:w2mu} that
\begin{equation}
	\mathbb{E}\left[w_{1}(S,\mathbf{D},\mathbf{X})\mu_{\mathbf{d}^\prime}(\mathbf{X})\right]=\mathbb{E}\left[w_{2}(S,\mathbf{D},\mathbf{X})\mu_{\mathbf{d}^\prime}(\mathbf{X})\right],
\end{equation}
which combined with \eqref{eq:w3minw4} implies that $\tau_{\mathbf{dd^\prime}}^{\text{R}} = \mathbb{E}\left[  
\left(w_{1}(S,\mathbf{D},\mathbf{X})-w_{2}(S,\mathbf{D},\mathbf{X})\right) \Delta Y \right] = {\mathbb{E}\left[p_{\mathbf{d}}(\mathbf{X})\right]}^{-1}\mathbb{E}\left[(m_{\mathbf{d}}(\mathbf{X})-m_{\mathbf{d}^\prime}(\mathbf{X}))p_{\mathbf{d}}(\mathbf{X})\right]$, 
where the second equality uses \eqref{eq:w1dy} and \eqref{eq:w2dy} with $\phi_{\mathbf{d}_{-h}}(\mathbf{X})=q_{\mathbf{d}_{-h}}(\mathbf{X})$ and $\pi_{\mathbf{d}}(\mathbf{X}) = p_{\mathbf{d}}(\mathbf{X})$. 

\subsubsection*{2. Missing data model and outcome model correct}
If $\phi_{\mathbf{d}_{-h}}(\mathbf{X})=q_{\mathbf{d}_{-h}}(\mathbf{X})$ and $\mu_{\mathbf{d}}(\mathbf{X})=m_{\mathbf{d}}(\mathbf{X})$, it follows from \eqref{eq:w2dy} and \eqref{eq:w2mu} that
\begin{align}
	\mathbb{E}\left[w_{2}(S,\mathbf{D},\mathbf{X})\Delta Y\right]=\mathbb{E}\left[w_{2}(S,\mathbf{D},\mathbf{X})\mu_{\mathbf{d}^\prime}(\mathbf{X})\right],
\end{align}
which combined with \eqref{eq:w3minw4} implies that
{\normalsize \begin{equation}
		\tau_{\mathbf{dd^\prime}}^{\text{R}} = \mathbb{E}\left[  
		w_{1}(S,\mathbf{D},\mathbf{X})\left( \Delta Y -\mu_{\mathbf{d}^\prime}(\mathbf{X}) \right)\right] = {\mathbb{E}\left[p_{\mathbf{d}}(\mathbf{X})\right]}^{-1}\mathbb{E}\left[(m_{\mathbf{d}}(\mathbf{X})-m_{\mathbf{d}^\prime}(\mathbf{X}))p_{\mathbf{d}}(\mathbf{X})\right],
\end{equation}}
where the second equality uses \eqref{eq:w1dy} and \eqref{eq:w1mu} with $\phi_{\mathbf{d}_{-h}}(\mathbf{X})=q_{\mathbf{d}_{-h}}(\mathbf{X})$ and $\mu_{\mathbf{d}}(\mathbf{X})=m_{\mathbf{d}}(\mathbf{X})$.

\subsubsection*{3. Propensity score and outcome model correct}
Consider the following term in the estimand:
\begin{align}
	&\mathbb{E}[w_{4}(S,\mathbf{D}_{-h},\mathbf{X})\left( \mu_{\mathbf{d}}(\mathbf{X}) -\mu_{\mathbf{d}^\prime}(\mathbf{X}) \right)] =
	{\mathbb{E}\left[\frac{S}{\phi_{\mathbf{d}_{-h}}(\mathbf{X})}\pi_{\mathbf{d}_h|\mathbf{d}_{-h}}(\mathbf{X})\mathbbm{1}[\mathbf{D}_{-h}=\mathbf{d}_{-h}]\right]}^{-1}\times \nonumber \\&
	\mathbb{E}\left[{\frac{S}{\phi_{\mathbf{d}_{-h}}(\mathbf{X})}\pi_{\mathbf{d}_h|\mathbf{d}_{-h}}(\mathbf{X})\mathbbm{1}[\mathbf{D}_{-h}=\mathbf{d}_{-h}]}\left( \mu_{\mathbf{d}}(\mathbf{X}) -\mu_{\mathbf{d}^\prime}(\mathbf{X}) \right)\right] \nonumber\\
	=&{\mathbb{E}\left[\frac{q_{\mathbf{d}_{-h}}(\mathbf{X})}{\phi_{\mathbf{d}_{-h}}(\mathbf{X})}p_{\mathbf{d}}(\mathbf{X})\right]}^{-1}\times 
	\mathbb{E}\left[{\frac{q_{\mathbf{d}_{-h}}(\mathbf{X})}{\phi_{\mathbf{d}_{-h}}(\mathbf{X})}p_{\mathbf{d}}(\mathbf{X})}\left(\mu_{\mathbf{d}}(\mathbf{X}) -\mu_{\mathbf{d}^\prime}(\mathbf{X}) \right)\right] ,\label{eq:Aw1minw4_extn}
\end{align} where the second equality uses LIE, Assumption~\ref{ass:mar_extn} and $\pi_{\mathbf{d}}(\mathbf{X}) = p_{\mathbf{d}}(\mathbf{X})$. Similarly, 
\begin{align}
		&\mathbb{E}[w_{1}(S,\mathbf{D}_{-h},\mathbf{X})\left( \Delta Y -\mu_{\mathbf{d}^\prime}(\mathbf{X}) \right)] =\mathbb{E}\left[\frac{q_{\mathbf{d}_{-ht}}(\mathbf{X})}{\phi_{\mathbf{d}_{-h}}(\mathbf{X})}p_{\mathbf{d}}(\mathbf{X})\right]^{-1}\times \nonumber \\
        &\mathbb{E}\left[\frac{q_{\mathbf{d}_{-h}}(\mathbf{X})}{\phi_{\mathbf{d}_{-h}}( \mathbf{X})}\left(m_{\mathbf{d}}(\mathbf{X})-\mu_{\mathbf{d}^\prime}(\mathbf{X})\right)p_{\mathbf{d}}(\mathbf{X})\right] =\mathbb{E}[w_{4}(S,\mathbf{D}_{-h},\mathbf{X})\left( \mu_{\mathbf{d}}(\mathbf{X}) -\mu_{\mathbf{d}^\prime}(\mathbf{X}) \right)], \label{eq:w1minw4}
\end{align}
where the first equality uses LIE and Assumption~\ref{ass:mar_extn} and the second equality uses $m_{\mathbf{d}}(\mathbf{X})=\mu_{\mathbf{d}}(\mathbf{X})$.
With $\pi_{\mathbf{d}}(\mathbf{X}) = p_{\mathbf{d}}(\mathbf{X})$ and $\mu_{\mathbf{d}}(\mathbf{X})=m_{\mathbf{d}}(\mathbf{X})$, it follows from \eqref{eq:w2dy} and \eqref{eq:w2mu} that
\begin{align}
	\mathbb{E}\left[w_{2}(S,\mathbf{D},\mathbf{X})\Delta Y\right]=\mathbb{E}\left[w_{2}(S,\mathbf{D},\mathbf{X})m_{\mathbf{d}^\prime}(\mathbf{X})\right],
\end{align}
which combined with \eqref{eq:w1minw4} implies that $\tau_{\mathbf{dd^\prime}}^{\text{R}} =\mathbb{E}[w_{3}(\mathbf{D}_{-h},\mathbf{X})\left( \mu_{\mathbf{d}}(\mathbf{X}) -\mu_{\mathbf{d}^\prime}(\mathbf{X}) \right)] = $
\begin{align}
	 =&{\mathbb{E}\left[\pi_{\mathbf{d}_h|\mathbf{d}_{-h}}(\mathbf{X})\mathbbm{1}[\mathbf{D}_{-h}=\mathbf{d}_{-h}]\right]}^{-1}\times \mathbb{E}\left[{\pi_{\mathbf{d}_h|\mathbf{d}_{-h}}(\mathbf{X})\mathbbm{1}[\mathbf{D}_{-h}=\mathbf{d}_{-h}]}\left( \mu_{\mathbf{d}}(\mathbf{X}) -\mu_{\mathbf{d}^\prime}(\mathbf{X}) \right)\right] \nonumber\\=
	&{\mathbb{E}\left[p_{\mathbf{d}}(\mathbf{X})\right]}^{-1}\mathbb{E}\left[\left( m_{\mathbf{d}}(\mathbf{X}) -m_{\mathbf{d}^\prime}(\mathbf{X}) \right)p_{\mathbf{d}}(\mathbf{X})\right],
\end{align}
where the final equality uses the fact that $\pi_{\mathbf{d}}(\mathbf{X}) = p_{\mathbf{d}}(\mathbf{X})$ and $\mu_{\mathbf{d}}(\mathbf{X})=m_{\mathbf{d}}(\mathbf{X})$. 

\subsubsection*{PDATT identification and proof Corollary~\ref{cor:ATT_identification}}
Finally, we show that ${\mathbb{E}\left[p_{\mathbf{d}}(\mathbf{X})\right]}^{-1}\mathbb{E}\left[(m_{\mathbf{d}}(\mathbf{X})-m_{\mathbf{d^\prime}}(\mathbf{X}))p_\mathbf{d}(\mathbf{X})\right] = \tau_{\mathbf{dd^\prime}}.$
With $\mathbf{d}^\prime=\mathbf{0}$, 
\begin{align}
	m_{\mathbf{d}}(\mathbf{X})-m_{\mathbf{d}^\prime}(\mathbf{X}) =& \mathbb{E}\left[Y_T(\mathbf{d})-Y_0(\mathbf{d})|\mathbf{D}=\mathbf{d},\mathbf{X}\right]-\mathbb{E}\left[Y_T(\mathbf{0})-Y_0(\mathbf{0})|\mathbf{D}=\mathbf{d}',\mathbf{X}\right] \nonumber \\
	=& \mathbb{E}\left[Y_T(\mathbf{d})-Y_0(\mathbf{0})|\mathbf{D}=\mathbf{d},\mathbf{X}\right]-\mathbb{E}\left[Y_T(\mathbf{0})-Y_0(\mathbf{0})|\mathbf{D}=\mathbf{d}',\mathbf{X}\right]\nonumber \\
	=& \mathbb{E}\left[Y_T(\mathbf{d})-Y_T(\mathbf{0})|\mathbf{D}=\mathbf{d},\mathbf{X}\right],
\end{align}
where the second line uses Assumption~\ref{ass:did_extn}.1, and the third line Assumption~\ref{ass:did_extn}.2. Hence,
\begin{align}
	\mathbb{E}\left[(m_{\mathbf{d}}(\mathbf{X})-m_{\mathbf{d}^\prime}(\mathbf{X}))\frac{p_{\mathbf{d}}(\mathbf{X})}{\mathbb{E}\left[p_{\mathbf{d}}(\mathbf{X})\right]}\right]=&
\mathbb{E}\left[\mathbb{E}\left[Y_T(\mathbf{d})-Y_T(\mathbf{0})|\mathbf{D}=\mathbf{d},\mathbf{X}\right]\frac{\mathbb{P}(\mathbf{D}=\mathbf{d}|\mathbf{X})}{\mathbb{P}(\mathbf{D}=\mathbf{d})}\right] \nonumber \\
	=& \int_\mathbf{X} \mathbb{E}\left[Y_T(\mathbf{d})-Y_T(\mathbf{0})|\mathbf{D}=\mathbf{d},\mathbf{X}\right] d \mathbb{P}(\mathbf{X}|\mathbf{D}=\mathbf{d}) \nonumber \\
	=& \mathbb{E}\left[Y_T(\mathbf{d})-Y_T(\mathbf{0})|\mathbf{D}=\mathbf{d}\right] = \tau_{\mathbf{dd^\prime}}.
\end{align}
This concludes the proof of Theorem~\ref{thm:robust}. The proof of Corollary~\ref{cor:ATT_identification_md} follows from \eqref{eq:w1dy} and \eqref{eq:w1mu} with 
$\phi_{\mathbf{d}_{-h}}(\mathbf{X})=q_{\mathbf{d}_{-h}}(\mathbf{X})$ and $\mu_{\mathbf{d}}(\mathbf{X})=m_{\mathbf{d}}(\mathbf{X})$, and from \eqref{eq:w1dy} and \eqref{eq:w2dy} with 
$\phi_{\mathbf{d}_{-h}}(\mathbf{X})=q_{\mathbf{d}_{-h}}(\mathbf{X})$ and 
$\pi_{\mathbf{d}}(\mathbf{X}) = p_{\mathbf{d}}(\mathbf{X})$.

\section{Semiparametric efficiency bound: Proof of Theorem~\ref{thm:seb}}\label{sec:proof_seb}
\begin{proof}
First, consider the density of: $(Y_2(1,1), Y_2(1,0), Y_2(0,1)$, $Y_2(0,0), Y_0(0,0), \mathbf{D}, \mathbf{X})$. This is given as $\bar{f}(y_2(1,1), y_2(1,0), y_2(0,1), y_2(0,0), y_0(0,0), \mathbf{d}, \mathbf{x})$
{\normalsize \begin{align*}
    = \bar{f}(y_2(1,1), y_2(1,0), y_2(0,1), y_2(0,0), y_0(0,0)|\mathbf{D}=(d_1, d_2), \mathbf{x})\cdot p_{d_1 d_2}(\mathbf{x}) \cdot f(\mathbf{x}),
\end{align*}}
where $ \bar{f}(y_2(1,1), y_2(1,0), y_2(0,1), y_2(0,0), y_0(0,0)|\mathbf{D}=(d_1, d_2), \mathbf{x})$ denotes the conditional density of $(Y_2(1,1), Y_2(1,0), Y_2(0,1), Y_2(0,0), Y_0(0,0))$ conditional on $\mathbf{D}= (d_1, d_2), \mathbf{X}=\mathbf{x}$ where $(d_1, d_2)\in \{0,1\}^2$ and $f(\mathbf{x})$  denotes the marginal density of $\mathbf{X}$. Now, $Y_2 = Y_2(D_1, D_2)$ and $Y_0 = Y_0(0,0)$. So, the density of $(Y_2, Y_0, \mathbf{D}, \mathbf{X})$ is given as
{\small \begin{align*}
    &f(y_2, y_0, \mathbf{d}, \mathbf{x})
     = \left\{ f_{11}(y_2, y_0|\mathbf{D}=(1, 1), \mathbf{x})\cdot p_{11}(\mathbf{x})\right\}^{d_1d_2}\times \left\{ f_{01}(y_2, y_0|\mathbf{D}=(0, 1), \mathbf{x})\cdot p_{01}(\mathbf{x})\right\}^{(1-d_1)d_2} \times\\
    & \left\{ f_{10}(y_2, y_0|\mathbf{D}=(1, 0), \mathbf{x})\cdot p_{10}(\mathbf{x})\right\}^{d_1(1-d_2)} \times\left\{ f_{00}(y_2, y_0|\mathbf{D}=(0, 0), \mathbf{x})\cdot p_{00}(\mathbf{x})\right\}^{(1-d_1)(1-d_2)} \times f(\mathbf{x}), 
\end{align*}}
where 
{\small \begin{align*}
 		f_{11}(\cdot, \cdot|\mathbf{D}=(1, 1), \mathbf{x}) &= \int \int\int\bar{f}(\cdot, y_2(1,0), y_2(0,1), y_2(0,0), \cdot|\mathbf{D}=(1, 1), \mathbf{x})dy_2(1,0)dy_2(0,1)dy_2(0,0); \\
 		f_{01}(\cdot, \cdot|\mathbf{D}=(0, 1), \mathbf{x}) &= \int \int\int\bar{f}(y_2(1,1), y_2(1,0), \cdot, y_2(0,0), \cdot|\mathbf{D}=(0, 1), \mathbf{x})dy_2(1,1)dy_2(1,0)dy_2(0,0); \\
 		f_{10}(\cdot, \cdot|\mathbf{D}=(1, 0), \mathbf{x}) &= \int \int\int\bar{f}(y_2(1,1), \cdot, y_2(0,1), y_2(0,0), \cdot|\mathbf{D}=(1, 0), \mathbf{x})dy_2(1,1)dy_2(0,1)dy_2(0,0); \\
 		f_{00}(\cdot, \cdot|\mathbf{D}=(0, 0), \mathbf{x}) &= \int \int\int\bar{f}(y_2(1,1), y_2(1,0), y_2(0,1), \cdot, \cdot|\mathbf{D}=(0, 0), \mathbf{x})dy_2(1,1)dy_2(1,0)dy_2(0,1). 
\end{align*}} 
The tangent space can be characterized by considering a regular parametric submodel
{\small \begin{align*}
 		&f_{\theta}(y_2, y_0, \mathbf{d}, \mathbf{x}) = \left\{ f_{11, \theta}(y_2, y_0|\mathbf{D}=(1, 1), \mathbf{x})\cdot p_{11, \theta}(\mathbf{x})\right\}^{d_1d_2}\times \left\{ f_{01, \theta}(y_2, y_0|\mathbf{D}=(0, 1), \mathbf{x})\cdot p_{01, \theta}(\mathbf{x})\right\}^{(1-d_1)d_2} \\
 		& \times \left\{ f_{10, \theta}(y_2, y_0|\mathbf{D}=(1, 0), \mathbf{x})\cdot p_{10, \theta}(\mathbf{x})\right\}^{d_1(1-d_2)} \times\left\{ f_{00, \theta}(y_2, y_0|\mathbf{D}=(0, 0), \mathbf{x})\cdot p_{00, \theta}(\mathbf{x})\right\}^{(1-d_1)(1-d_2)} \\
        &\times f_\theta(\mathbf{x}), 
\end{align*}} 
which equals $f(y_2, y_0, \mathbf{d}, \mathbf{x})$ at $\theta = \theta_0$. This implies
{\normalsize \begin{align*}
 		&logf_{\theta}(y_2, y_0, \mathbf{d}, \mathbf{x})= d_1d_2\cdot logf_{11, \theta}(y_2, y_0|\mathbf{D}=(1, 1), \mathbf{x})+(1-d_1)d_2\cdot logf_{01, \theta}(y_2, y_0|\mathbf{D}=(0, 1), \mathbf{x}) \\
 		& + d_1(1-d_2)\cdot logf_{10, \theta}(y_2, y_0|\mathbf{D}=(1, 0), \mathbf{x}) +(1-d_1)(1-d_2)\cdot logf_{00, \theta}(y_2, y_0|\mathbf{D}=(0, 0), \mathbf{x}) \\
 		& +d_1d_2\cdot logp_{11, \theta}(\mathbf{x}) +(1-d_1)d_2\cdot logp_{01, \theta}(\mathbf{x}) +d_1(1-d_2)\cdot logp_{10, \theta}(\mathbf{x})\\
 		&+(1-d_1)(1-d_2)\cdot logp_{00,\theta}(\mathbf{x})+logf_{\theta}(\mathbf{x}).
\end{align*}}
Then the score for this parametric submodel is given as
{\normalsize \begin{align}\label{eq:score_submodel}
 		&l_{\theta}(y_2, y_0, \mathbf{d}, \mathbf{x}) \equiv d_1d_2\cdot l_{11, \theta}(y_2, y_0|\mathbf{D}=(1,1),\mathbf{x})+(1-d_1)d_2\cdot l_{01, \theta}(y_2,y_0|\mathbf{D}=(0,1), \mathbf{x}) \nonumber \\
 		&+d_1(1-d_2)\cdot l_{10, \theta}(y_2,y_0|\mathbf{D}=(1,0), \mathbf{x})+(1-d_1)(1-d_2)\cdot l_{00, \theta}(y_2,y_0|\mathbf{D}=(0,0), \mathbf{x}) \nonumber \\
 		&+\frac{d_1d_2}{p_{11, \theta}(\mathbf{x})}\dot{p}_{11, \theta}(\mathbf{x})+\frac{(1-d_1)d_2}{p_{01, \theta}(\mathbf{x})}\dot{p}_{01, \theta}(\mathbf{x})+\frac{d_1(1-d_2)}{p_{10, \theta}(\mathbf{x})}\dot{p}_{10, \theta}(\mathbf{x})+\frac{(1-d_1)(1-d_2)}{p_{00, \theta}(\mathbf{x})}\dot{p}_{00, \theta}(\mathbf{x})+t_{\theta}(\mathbf{x}),
\end{align}}
where for each $\mathbf{d}\in \{0,1\}^2$, $l_{\mathbf{d}, \theta}(y_2, y_0|\mathbf{D}=\mathbf{d},\mathbf{x})= \frac{d}{d\theta} log f_{\mathbf{d}, \theta}(y_2, y_0|\mathbf{D}=\mathbf{d}, \mathbf{x})$, $\dot{p}_{\mathbf{d}, \theta} = \frac{d}{d\theta}p_{\mathbf{d}, \theta}(\mathbf{x})$, and $t_{\theta}(\mathbf{x}) = \frac{d}{d\theta} logf_{\theta}(\mathbf{x})$. So, the tangent subspace for this model is given as
{\normalsize \begin{align*}
 		\mathcal{T} &= \bigg\{d_1d_2\cdot l_{11}(y_2, y_0|\mathbf{D}=(1,1),\mathbf{x})+(1-d_1)d_2\cdot l_{01}(y_2,y_0|\mathbf{D}=(0,1), \mathbf{x}) \\
 		&+d_1(1-d_2)\cdot l_{10}(y_2,y_0|\mathbf{D}=(1,0), \mathbf{x})+(1-d_1)(1-d_2)\cdot l_{00}(y_2,y_0|\mathbf{D}=(0,0), \mathbf{x}) \\
 		&+a_{11}(\mathbf{x})\cdot d_1d_2+a_{01}(\mathbf{x})\cdot (1-d_1)d_2+a_{10}(\mathbf{x})\cdot d_1(1-d_2)+a_{00}(\mathbf{x})\cdot (1-d_1)(1-d_2)+t(\mathbf{x})\bigg\},
\end{align*}} 
such that for each $\mathbf{d} \in \{0,1\}^2$, $\int \int l_{\mathbf{d}}(y_2, y_0|\mathbf{D}=\mathbf{d},\mathbf{x})f_{\mathbf{d}}(y_2, y_0|\mathbf{D}=\mathbf{d}, \mathbf{x})dy_2 dy_0 = 0 \  \forall \ \mathbf{x}$, $\int t(\mathbf{x})f(\mathbf{x})d\mathbf{x} = 0$, 
 and $a_{\mathbf{d}}(\mathbf{x})$ is any square-integrable measurable function of $\mathbf{x}$. Under Assumption~\ref{ass:did}, $\tau_{\mathbf{dd^\prime}} =\mathbb{E}\left[	\mathbb{E}(Y_2- Y_0 |\mathbf{D}=\mathbf{d}, \mathbf{X}) - \mathbb{E}(Y_2-Y_0|\mathbf{D}=\mathbf{d^\prime}, \mathbf{X})|\mathbf{D}=\mathbf{d}\right]$. Next, we show that the target parameter, $\tau_{\mathbf{dd^\prime}}$, is path-dependent differentiable. For the parametric submodel under consideration, 
{\normalsize \begin{align*}
 		\tau_{\mathbf{dd^\prime}}(\theta) &= \frac{\int\int\int(y_2-y_0)f_{\mathbf{d}, \theta}(y_2, y_0|\mathbf{D}=\mathbf{d}, \mathbf{x})p_{\mathbf{d}, \theta}(\mathbf{x})f_{\theta}(\mathbf{x})dy_2dy_0d\mathbf{x}}{\int p_{\mathbf{d}, \theta}(\mathbf{x})f_{\theta}(\mathbf{x})d\mathbf{x}}\\
 		&-  \frac{\int\int\int(y_2-y_0)f_{\mathbf{d^\prime}, \theta}(y_2, y_0|\mathbf{D}=\mathbf{d^\prime}, \mathbf{x})p_{\mathbf{d}, \theta}(\mathbf{x})f_{\theta}(\mathbf{x})dy_2dy_0d\mathbf{x}}{\int p_{\mathbf{d}, \theta}(\mathbf{x})f_{\theta}(\mathbf{x})d\mathbf{x}}. 
\end{align*}} Then, 
{\normalsize \begin{align*}
 		\frac{\partial \tau_{\mathbf{dd^\prime}}(\theta_0)}{\partial \theta} &= \frac{\int\int\int(y_2-y_0)l_{\mathbf{d}}(y_2, y_0|\mathbf{D}=\mathbf{d}, \mathbf{x})f_{\mathbf{d}}(y_2, y_0|\mathbf{D}=\mathbf{d}, \mathbf{x})p_{\mathbf{d}}(\mathbf{x})f(\mathbf{x})dy_2dy_0d\mathbf{x}}{\mathbb{P}(\mathbf{D}=\mathbf{d})}\\
 		&- \frac{\int\int\int(y_2-y_0)l_{\mathbf{d^\prime}}(y_2, y_0|\mathbf{D}=\mathbf{d^\prime}, \mathbf{x})f_{\mathbf{d^\prime}}(y_2, y_0|\mathbf{D}=\mathbf{d^\prime}, \mathbf{x})p_{\mathbf{d}}(\mathbf{x})f(\mathbf{x})dy_2dy_0d\mathbf{x}}{\mathbb{P}(\mathbf{D}=\mathbf{d})} \\
 		&+\frac{\int (m_{\mathbf{d}}(\mathbf{x})-m_{\mathbf{d}^\prime}(\mathbf{x})-\tau_{\mathbf{dd}^\prime})\dot{p}_{\mathbf{d}}(\mathbf{x})f(\mathbf{x})d\mathbf{x}}{\mathbb{P}(\mathbf{D}=\mathbf{d})}+\frac{\int (m_{\mathbf{d}}(\mathbf{x})-m_{\mathbf{d}^\prime}(\mathbf{x})-\tau_{\mathbf{dd}^\prime})p_{\mathbf{d}}(\mathbf{x})t(\mathbf{x})f(\mathbf{x})d\mathbf{x}}{\mathbb{P}(\mathbf{D}=\mathbf{d})}.
\end{align*}}
Let $F_{\tau_{\mathbf{dd^\prime}}}(Y_2, Y_0, \mathbf{D}, \mathbf{X}) = \frac{\mathbbm{1}[\mathbf{D}=\mathbf{d}]}{\mathbb{P}(\mathbf{D}=\mathbf{d})}(\Delta Y-m_{\mathbf{d}}(\mathbf{X}))-\frac{\mathbbm{1}[\mathbf{D}=\mathbf{d}^\prime]\cdot p_{\mathbf{d}}(\mathbf{X})}{\mathbb{P}(\mathbf{D}=\mathbf{d})\cdot p_{\mathbf{d}^\prime}(\mathbf{X})}(\Delta Y-m_{\mathbf{d}^\prime}(\mathbf{X})) \\
+ \frac{(m_{\mathbf{d}}(\mathbf{X})-m_{\mathbf{d}^\prime}(\mathbf{X})-\tau_{\mathbf{dd}^\prime})(\mathbbm{1}[\mathbf{D}=\mathbf{d}]-p_{\mathbf{d}}(\mathbf{X}))}{\mathbb{P}(\mathbf{D}=\mathbf{d})}+\frac{(m_{\mathbf{d}}(\mathbf{X})-m_{\mathbf{d}^\prime}(\mathbf{X})-\tau_{\mathbf{dd}^\prime})\cdot p_{\mathbf{d}}(\mathbf{X})}{\mathbb{P}(\mathbf{D}=\mathbf{d})}.$ For the parametric submodel whose score is given by \eqref{eq:score_submodel}, we have that
{\normalsize \begin{equation*}
 		\frac{\partial \tau_{\mathbf{dd^\prime}}(\theta_0)}{\partial \theta} =\mathbb{E}[F_{\tau_{\mathbf{dd^\prime}}}(Y_2, Y_0, \mathbf{D}, \mathbf{X})\cdot l_{\theta_0}(Y_2, Y_0, \mathbf{D}, \mathbf{X})],
\end{equation*}} 
which proves that $\tau_{\mathbf{dd^\prime}}$ is path-dependent differentiable. Then the efficient influence function for full data, denoted by $F_{\tau_{\mathbf{dd^\prime}}}(Y_2, Y_0, \mathbf{D}, \mathbf{X})$,  is equal to
{\normalsize \begin{equation*} \frac{\mathbbm{1}[\mathbf{D}=\mathbf{d}]}{\mathbb{P}(\mathbf{D}=\mathbf{d})}(m_{\mathbf{d}}(\mathbf{X})-m_{\mathbf{d}^\prime}(\mathbf{X})-\tau_{\mathbf{dd}^\prime}) +\frac{\mathbbm{1}[\mathbf{D}=\mathbf{d}]}{\mathbb{P}(\mathbf{D}=\mathbf{d})}(\Delta Y-m_{\mathbf{d}}(\mathbf{X}))-\frac{\mathbbm{1}[\mathbf{D}=\mathbf{d}^\prime]\cdot p_{\mathbf{d}}(\mathbf{X})}{\mathbb{P}(\mathbf{D}=\mathbf{d})\cdot p_{\mathbf{d}^\prime}(\mathbf{X})}(\Delta Y-m_{\mathbf{d}^\prime}(\mathbf{X})).
\end{equation*}}	
Using Theorem 7.2 in Tsiatis, it follows that the efficient influence function for observed data, $\mathbf{W} = (\Delta Y, S, SD_1, D_2, \mathbf{X})$, is given by 
{\normalsize \begin{align}\label{if_obs}
 		&F_{\tau_{\mathbf{dd^\prime}}}(\mathbf{W}) = \frac{S}{q_{D_2}( \mathbf{X})}\left(F_{\tau_{\mathbf{dd^\prime}}}(Y_2, Y_0, \mathbf{D}, \mathbf{X})-\mathbb{E}\left[F_{\tau_{\mathbf{dd^\prime}}}(Y_2, Y_0, \mathbf{D}, \mathbf{X})|\Delta Y, SD_1, D_2, \mathbf{X}\right]\right) \nonumber \\
 		&+\mathbb{E}\left[F_{\tau_{\mathbf{dd^\prime}}}(Y_2, Y_0, \mathbf{D}, \mathbf{X})|\Delta Y, SD_1, D_2, \mathbf{X}\right]  	\text{ where } \\
		 &\mathbb{E}\left[F_{\tau_{\mathbf{dd^\prime}}}(Y_2, Y_0, \mathbf{D}, \mathbf{X})|\Delta Y, SD_1, D_2, \mathbf{X}\right] =\frac{p_{d_1|d_2}(\mathbf{X}) \mathbbm{1}[D_2=d_2]}{\mathbb{P}(\mathbf{D}=\mathbf{d})}\cdot  (m_{\mathbf{d}}(\mathbf{X})-m_{\mathbf{d^\prime}}(\mathbf{X})-\tau_{\mathbf{dd^\prime}}).
\end{align}}	 
\end{proof}
\section{Inference}\label{sec:conditions} 
Consider the following regularity conditions: (1) $\pi(\cdot; \bm{\gamma}_{\mathbf{d}})$, $\phi(\cdot; \bm{\delta}_{d_2})$, and $\mu(\cdot; \bm{\beta}_{\mathbf{d}})$ are continuous for each $\bm{\gamma}_{\mathbf{d}}\in \bm{\Gamma}, \bm{\delta}_{d_2}\in\bm{\Delta}$, and $\bm{\beta}_{\mathbf{d}}\in \bm{B}$; (2) $\bm{\gamma^\ast}_{\mathbf{d}} \in\bm{\Gamma}$, $\bm{\delta^\ast}_{d_2} \in \bm{\Delta}$, and $\bm{\beta^\ast}_{\mathbf{d}} \in \bm{B}$, where $\bm{\Gamma}$, $\bm{\Delta}$, and $\bm{B}$ are compact parameter spaces; (3) $\mathbb{E}[\underset{\bm{\gamma}_{\mathbf{d}}\in\bm{\Gamma}}{\textup{sup}}|\pi(\cdot; \bm{\gamma}_{\mathbf{d}})|]<\infty$, $\mathbb{E}[\underset{\bm{\delta}_{d_2}\in\bm{\Delta}}{\textup{sup}}|\phi(\cdot; \bm{\delta})|]<\infty$, and $\mathbb{E}[\underset{\bm{\beta}_{\mathbf{d}}\in\bm{B}}{\textup{sup}}|\mu(\cdot; \bm{\beta}_{\mathbf{d}})|]<\infty$; (4) $\pi(\cdot; \bm{\gamma}_{\mathbf{d}})$, $\phi(\cdot; \bm{\delta}_{d_2})$, and $\mu(\cdot; \bm{\beta}_{\mathbf{d}})$ are all twice continuously differentiable on $\textup{int}(\bm{\Gamma})$, $\textup{int}(\bm{\Delta})$, and $\textup{int}(\bm{B})$, respectively; (5) For each $\bm{\eta} = \bm{\beta}_{\mathbf{d}}, \bm{\beta}_{\mathbf{d^\prime}}, \bm{\gamma}_{\mathbf{d}}, \bm{\gamma}_{\mathbf{d^\prime}}, \bm{\delta}_{d_2}, \bm{\delta}_{d_2^\prime}$, the following asymptotic linear representation of the estimators of the nuisance parameters is assumed $\sqrt{n}\left(\bm{\widehat{\eta}}-\bm{\eta^\ast}\right) = \frac{1}{\sqrt{n}}\sum_{i=1}^{n}\mathbf{b}_{i\bm{\eta}}+o_p(1)$ where $\mathbf{b}_{i\bm{\eta}}$ is a mean-zero, finite-variance influence function whose form depends on the estimation method used. See Supplementary Appendix \ref*{sec:simsadd} for the influence function expressions associated with the commonly employed functional forms for the three models. Note that conditions (1)-(4) guarantee that $\bm{\widehat{\eta}} \overset{p}{\rightarrow}\bm{\eta^\ast}$ under uniform weak convergence.

\subsection{Proof Theorem~\ref{thm:asyvar}}\label{sec:proof_asyvar}
\begin{proof}
Part i) Consistency: As $n\rightarrow \infty$, by the continuous mapping theorem and the weak law of large numbers, $	\widehat{w}_1(\bm{\widehat{\delta}}_{d_2})\overset{p}{\rightarrow} w_1(\bm{\delta^\ast}_{d_2})$, $\widehat{w}_2(\bm{\widehat{\gamma}},\bm{\widehat{\delta}}_{d_2^\prime})\overset{p}{\rightarrow}w_2(\bm{\gamma^\ast},\bm{\delta^\ast}_{d_2^\prime})$, $\widehat{w}_3(\bm{\widehat{\gamma}}_{d_1|d_2})\overset{p}{\rightarrow}w_3(\bm{\gamma^\ast}_{d_1|d_2})$, and $\widehat{w}_4(\bm{\widehat{\gamma}}_{d_1|d_2}, \bm{\widehat{\delta}}_{d_2})\overset{p}{\rightarrow}w_4(\bm{\gamma^\ast}_{d_1|d_2}, \bm{\delta^\ast}_{d_2})$ which implies that $\widehat{\tau}_{\mathbf{dd^\prime}}^{\text{R}} \overset{p}{\rightarrow} \tau^\textup{R}_{\mathbf{dd^\prime}}$. Now, if any two of the three models are correctly specified, $\tau^\textup{R}_{\mathbf{dd^\prime}} = \tau_{\mathbf{dd^\prime}}$. 

Part ii) Asymptotic linear representation: We define some notation first. Let $\bm{\dot{\mu}}(\bm{\beta^\ast}_{\mathbf{d}}) \equiv \textup{d}\mu(\bm{\beta}_{\mathbf{d}})/\textup{d}\bm{\beta}_{\mathbf{d}}$ evaluated at the pseudo-true value $\bm{\beta^\ast}_{\mathbf{d}}$ for all $\mathbf{d}$, $\bm{\dot{\pi}}(\bm{\gamma^\ast}_{d_1|d_2}) \equiv \textup{d}\pi(\bm{\gamma}_{d_1|d_2})/\textup{d}\bm{\gamma}_{d_1|d_2}$ evaluated at the pseudo-true value  $\bm{\gamma^\ast}_{d_1|d_2}$ for all $\mathbf{d}$, and $\bm{\dot{\pi}}(\bm{\gamma^\ast}_{d_2}) \equiv \textup{d}\pi(\bm{\gamma}_{d_2})/\textup{d}\bm{\gamma}_{d_2}$ and $\bm{\dot{\phi}}(\bm{\delta^\ast}_{d_2}) \equiv \textup{d}\phi(\bm{\delta}_{d_2})/\textup{d}\bm{\delta}_{d_2}$ evaluated at the pseudo-true values $\bm{\gamma^\ast}_{d_2}$ and $\bm{\delta^\ast}_{d_2}$, respectively, for $d_2=0,1$. Next,  
{\normalsize \begin{align}\label{asy_expand}
		&\sqrt{n}(\widehat{\tau}^\textup{R}_{\mathbf{dd^\prime}}-\tau^{\textup{R}}_{\mathbf{dd^\prime}}) \nonumber \\
		&= \frac{1}{\sqrt{n}}\sum_{i=1}^{n}\bigg\{\left(\widehat{w}_{i1}(\bm{\widehat{\delta}}_{d_2})\Delta Y_i - \mathbb{E}[w_1(\bm{\delta^\ast}_{d_2})\Delta Y]\right) - \left(\widehat{w}_{i1}(\bm{\widehat{\delta}}_{d_2})\mu_{i}(\bm{\widehat{\beta}}_{\mathbf{d^\prime}})-\mathbb{E}[w_1(\bm{\delta^\ast}_{d_2})\mu(\bm{\beta^\ast}_{\mathbf{d^\prime}})]\right)\nonumber \\
		&-\left(\widehat{w}_{i2}(\bm{\widehat{\gamma}}, \bm{\widehat{\delta}}_{d_2^\prime})\Delta Y_i-\mathbb{E}[w_2(\bm{\gamma^\ast}, \bm{\delta^\ast}_{d_2^\prime})\Delta Y]\right)+\left(\widehat{w}_{i2}(\bm{\widehat{\gamma}}, \bm{\widehat{\delta}}_{d_2^\prime})\mu_{i}(\bm{\widehat{\beta}}_{\mathbf{d^\prime}})-\mathbb{E}[w_2(\bm{\gamma^\ast}, \bm{\delta^\ast}_{d_2^\prime})\mu(\bm{\beta^\ast}_{\mathbf{d^\prime}})]\right) \nonumber\\
		&+\left(\widehat{w}_{i3}(\bm{\widehat{\gamma}}_{d_1|d_2})\mu_{i}(\bm{\widehat{\beta}}_{\mathbf{d}})-\mathbb{E}[w_3(\bm{\gamma^\ast}_{d_1|d_2})\mu(\bm{\beta^\ast}_{\mathbf{d}})]\right)-\left(\widehat{w}_{i3}(\bm{\widehat{\gamma}}_{d_1|d_2})\mu_{i}(\bm{\widehat{\beta}}_{\mathbf{d^\prime}})-\mathbb{E}[w_3(\bm{\gamma^\ast}_{d_1|d_2})\mu(\bm{\beta^\ast}_{\mathbf{d^\prime}})]\right) \nonumber \\
		&- \left(\widehat{w}_{i4}(\bm{\widehat{\gamma}}_{d_1|d_2}, \bm{\widehat{\delta}}_{d_2})\mu_{i}(\bm{\widehat{\beta}}_{\mathbf{d}})-\mathbb{E}[w_4(\bm{\gamma^\ast}_{d_1|d_2}, \bm{\delta^\ast}_{d_2})\mu(\bm{\beta^\ast}_{\mathbf{d}})]\right) \nonumber \\
        & +\left(\widehat{w}_{i4}(\bm{\widehat{\gamma}}_{d_1|d_2}, \bm{\widehat{\delta}}_{d_2})\mu_{i}(\bm{\widehat{\beta}}_{\mathbf{d^\prime}})-\mathbb{E}[w_4(\bm{\gamma^\ast}_{d_1|d_2}, \bm{\delta^\ast}_{d_2})\mu(\bm{\beta^\ast}_{\mathbf{d^\prime}})]\right)\bigg\}.
\end{align}} 
Consider first, 
{\normalsize \begin{align}\label{t1_expand}
		&\frac{1}{\sqrt{n}}\sum_{i=1}^{n}\widehat{w}_{i1}(\bm{\widehat{\delta}}_{d_2})\Delta Y_i = 	\frac{1}{\sqrt{n}}\sum_{i=1}^{n}\frac{\frac{S_i\mathbbm{1}[\mathbf{D}_i=\mathbf{d}_i]}{\phi(\bm{\widehat{\delta}}_{d_2})}}{\mathbb{E}_n\left[\frac{S\mathbbm{1}[\mathbf{D}=\mathbf{d}]}{\phi(\bm{\widehat{\delta}}_{d_2})}\right]}\Delta Y_i \nonumber \\
		& = \frac{1}{\sqrt{n}}\sum_{i=1}^{n}\frac{\frac{S_i\mathbbm{1}[\mathbf{D}_i=\mathbf{d}_i]}{\phi(\bm{\widehat{\delta}}_{d_2})}}{\mathbb{E}\left[\frac{S\mathbbm{1}[\mathbf{D}=\mathbf{d}]}{\phi(\bm{\delta^\ast}_{d_2})}\right]}\Delta Y_i-\frac{\mathbb{E}\left[\frac{S\mathbbm{1}[\mathbf{D}=\mathbf{d}]}{\phi(\bm{\delta^\ast}_{d_2})}\Delta Y\right]}{\mathbb{E}\left[\frac{S\mathbbm{1}[\mathbf{D}=\mathbf{d}]}{\phi(\bm{\delta^\ast}_{d_2})}\right]^2}\times {\sqrt{n}\left(\mathbb{E}_n\left[\frac{S\mathbbm{1}[\mathbf{D}=\mathbf{d}]}{\phi(\bm{\widehat{\delta}}_{d_2})}\right]-\mathbb{E}\left[\frac{S\mathbbm{1}[\mathbf{D}=\mathbf{d}]}{\phi(\bm{\delta^\ast}_{d_2})}\right]\right)} \nonumber\\
        & \quad +o_p(1) \nonumber\\
		& = \frac{1}{\sqrt{n}}\sum_{i=1}^{n}\left\{\widetilde{w}_{i1}(\bm{\widehat{\delta}}_{d_2})\Delta Y_i-(\widetilde{w}_{i1}(\bm{\widehat{\delta}}_{d_2})-1)\mathbb{E}[w_{1}(\bm{\delta^\ast}_{d_2})\Delta Y]\right\}+o_p(1),
\end{align}} 
where $\widetilde{w}_1(\bm{\widehat{\delta}}_{d_2}) \equiv \frac{S\mathbbm{1}[\mathbf{D}=\mathbf{d}]}{\phi(\bm{\widehat{\delta}}_{d_2})}/\mathbbm{E}\left[\frac{S\mathbbm{1}[\mathbf{D}=\mathbf{d}]}{\phi(\bm{\delta^\ast}_{d_2})}\right]$. Then, the above implies that
{\normalsize\begin{equation*}
		\frac{1}{\sqrt{n}}\sum_{i=1}^{n}\left(\widehat{w}_{i1}(\bm{\widehat{\delta}}_{d_2})\Delta Y_i - \mathbb{E}[w_1(\bm{\delta^\ast}_{d_2})\Delta Y]\right)  = \frac{1}{\sqrt{n}}\sum_{i=1}^{n}\widetilde{w}_{i1}(\bm{\widehat{\delta}}_{d_2})\left\{\Delta Y_i - \mathbb{E}[w_1(\bm{\delta^\ast}_{d_2})\Delta Y]\right\} +o_p(1).
\end{equation*}} 
A second-order taylor expansion of the above around the pseudo-true $\bm{\delta^\ast}_{d_2}$ gives us
{\normalsize \begin{align}\label{t1_final}
		&\frac{1}{\sqrt{n}}\sum_{i=1}^{n}\left(\widehat{w}_{i1}(\bm{\widehat{\delta}}_{d_2})\Delta Y_i - \mathbb{E}[w_1(\bm{\delta^\ast}_{d_2})\Delta Y]\right) = \frac{1}{\sqrt{n}}\sum_{i=1}^{n}w_{i1}(\bm{\delta^\ast}_{d_2})\left(\Delta Y_i-\mathbb{E}[w_1(\bm{\delta^\ast}_{d_2})\Delta Y]\right) \nonumber \\
		&+ \sqrt{n}(\bm{\widehat{\delta}}_{d_2}-\bm{\delta^\ast}_{d_2})^\prime\cdot \frac{1}{n}\sum_{i=1}^{n}\bm{\dot{w}}_{i1}(\bm{\delta^\ast}_{d_2})(\Delta Y_i-\mathbb{E}[w_1(\bm{\delta^\ast}_{d_2})\Delta Y])+o_p(1) \nonumber \\
		& =  \frac{1}{\sqrt{n}}\sum_{i=1}^{n}\bigg\{w_{i1}(\bm{\delta^\ast}_{d_2})\left(\Delta Y_i-\mathbb{E}[w_1(\bm{\delta^\ast}_{d_2})\Delta Y]\right)+ \mathbf{b}_{i\bm{\delta}_{d_2}}^\prime\cdot  \mathbb{E}\left[\bm{\dot{w}}_1(\bm{\delta^\ast}_{d_2})(\Delta Y-\mathbb{E}[w_1(\bm{\delta^\ast}_{d_2})\Delta Y])\right]\bigg\} \nonumber \\
        &\qquad +o_p(1).
\end{align}}	
Expanding the remaining seven terms\footnote{See Supplementary Appendix \ref*{sec:inferencedetails} for the expansions.} in \eqref{asy_expand} using asymptotic arguments analogous to \eqref{t1_final}, and subsequently re-organizing them, yields the asymptotic linear representation presented below where $\sqrt{n}(\widehat{\tau}^\textup{R}_{\mathbf{dd^\prime}}-\tau^{\textup{R}}_{\mathbf{dd^\prime}})$ is equal to
{\normalsize  \begin{align}\label{if_r}
		& = \frac{1}{\sqrt{n}}\sum_{i=1}^{n}\bigg\{\psi_i+\mathbf{b}^\prime_{i\bm{\beta}_{\mathbf{d}}} \mathbb{E}(\Psi_{\bm{\beta}_{\mathbf{d}}}) -\mathbf{b}^\prime_{i\bm{\beta}_{\mathbf{d^\prime}}}  \mathbb{E}(\Psi_{\bm{\beta}_{\mathbf{d^\prime}}}) -\mathbf{b}^\prime_{i\bm{\gamma}_{d_1|d_2}}\mathbb{E}(\Psi_{\bm{\gamma}_{d_1|d_2}}) - \mathbf{b}^\prime_{i\bm{\gamma}_{d_2}} \mathbb{E}(\Psi_{\bm{\gamma}_{d_2}}) -\mathbf{b}^\prime_{i\bm{\gamma}_{\mathbf{d^\prime}}} \mathbb{E}(\Psi_{\bm{\gamma}_{\mathbf{d^\prime}}}) \nonumber \\
		&+\mathbf{b}^\prime_{i\bm{\delta}_{d_2}} \mathbb{E}(\Psi_{\bm{\delta}_{d_2}})  - \mathbf{b}^\prime_{i\bm{\delta}_{d^\prime_2}} \mathbb{E}(\Psi_{\bm{\delta}_{d^\prime_2}})\bigg\}+o_p(1) \equiv \frac{1}{\sqrt{n}}\sum_{i=1}^{n}\xi_i(\bm{\beta^\ast}, \bm{\gamma^\ast}, \bm{\delta^\ast})+o_p(1),
\end{align}} where $\mathbf{b}_{i\bm{\eta}}$ is the influence function of $\bm{\eta}=\bm{\beta}_{\mathbf{d}},\bm{\beta}_{\mathbf{d^\prime}}, \bm{\gamma}_{d_1|d_2}, \bm{\gamma}_{d_1^\prime|d_2^\prime}, \bm{\gamma}_{d_2}, \bm{\delta}_{d_2}$, $\bm{\delta}_{d_2^\prime}$, and
{\normalsize\begin{align*}
	\psi&\equiv  w_1(\bm{\delta^\ast}_{d_2})\bigg((\Delta Y-\mu(\bm{\beta^\ast}_{\mathbf{d^\prime}}))-\mathbb{E}[w_1(\bm{\delta^\ast}_{d_2})(\Delta Y-\mu(\bm{\beta^\ast}_{\mathbf{d^\prime}}))]\bigg)\\
    &-w_2(\bm{\gamma^\ast}, \bm{\delta^\ast}_{d_2^\prime})\bigg((\Delta Y-\mu(\bm{\beta^\ast}_{\mathbf{d^\prime}})) - \mathbb{E}[w_2(\bm{\gamma^\ast}, \bm{\delta^\ast}_{d_2^\prime})(\Delta Y-\mu(\bm{\beta^\ast}_{\mathbf{d^\prime}}))]\bigg) \\
    &+ w_3(\bm{\gamma^\ast}_{d_1|d_2})\bigg((\mu(\bm{\beta^\ast}_{\mathbf{d}})-\mu(\bm{\beta^\ast}_{\mathbf{d^\prime}}))-\mathbb{E}[w_3(\bm{\gamma^\ast}_{d_1|d_2})(\mu(\bm{\beta^\ast}_{\mathbf{d}})-\mu(\bm{\beta^\ast}_{\mathbf{d^\prime}}))]\bigg) \\
	&-w_4(\bm{\gamma^\ast}_{d_1|d_2}, \bm{\delta^\ast}_{d_2})\bigg((\mu(\bm{\beta^\ast}_{\mathbf{d}})-\mu(\bm{\beta^\ast}_{\mathbf{d^\prime}})) - \mathbb{E}[w_4(\bm{\gamma^\ast}_{d_1|d_2}, \bm{\delta^\ast}_{d_2})(\mu_{\mathbf{d}}(\bm{\beta^\ast})-\mu(\bm{\beta^\ast}_{\mathbf{d^\prime}}))]\bigg); \\
	\Psi_{\bm{\beta}_{\mathbf{d}}}&\equiv\bigg(w_3(\bm{\gamma^\ast}_{d_1|d_2})-w_4(\bm{\gamma^\ast}_{d_1|d_2}, \bm{\delta^\ast}_{d_2})\bigg)\bm{\dot{\mu}}(\bm{\beta^\ast}_{\mathbf{d}}); \\
	\Psi_{\bm{\beta}_{\mathbf{d^\prime}}} & \equiv\bigg(w_1(\bm{\delta^\ast}_{d_2})-w_2(\bm{\gamma^\ast}, \bm{\delta^\ast}_{d_2^\prime}) + w_3(\bm{\gamma^\ast}_{d_1|d_2})-w_4(\bm{\gamma^\ast}_{d_1|d_2}, \bm{\delta^\ast}_{d_2})\bigg)\bm{\dot{\mu}}(\bm{\beta^\ast}_{\mathbf{d^\prime}}); \\
	\Psi_{\bm{\gamma}_{d_1|d_2}}& \equiv \bm{\dot{w}}_{2,\bm{\gamma}_{d_1|d_2}}(\bm{\gamma^\ast}, \bm{\delta^\ast}_{d_2^\prime})\bigg((\Delta Y - \mu(\bm{\beta^\ast}_{\mathbf{d^\prime}})) - \mathbb{E}[w_2(\bm{\gamma^\ast}, \bm{\delta^\ast}_{d_2^\prime})(\Delta Y -\mu(\bm{\beta^\ast}_{\mathbf{d^\prime}}) )]\bigg)\nonumber \\
	&-\bm{\dot{w}}_3(\bm{\gamma^\ast}_{d_1|d_2})\bigg((\mu(\bm{\beta^\ast}_{\mathbf{d}})-\mu(\bm{\beta^\ast}_{\mathbf{d^\prime}}))-\mathbb{E}[w_3(\bm{\gamma^\ast}_{d_1|d_2})(\mu(\bm{\beta^\ast}_{\mathbf{d}})-\mu(\bm{\beta^\ast}_{\mathbf{d^\prime}}))]\bigg)\\
	&+\bm{\dot{w}}_{4,\bm{\gamma}_{d_1|d_2}}(\bm{\gamma^\ast}_{d_1|d_2}, \bm{\delta^\ast}_{d_2})\bigg((\mu(\bm{\beta^\ast}_{\mathbf{d}})-\mu(\bm{\beta^\ast}_{\mathbf{d^\prime}}))- \mathbb{E}[w_4(\bm{\gamma^\ast}_{d_1|d_2}, \bm{\delta^\ast}_{d_2})(\mu(\bm{\beta^\ast}_{\mathbf{d}})-\mu(\bm{\beta^\ast}_{\mathbf{d^\prime}}))]\bigg);\\
	\Psi_{\bm{\gamma}_{d_2}}&\equiv\bm{\dot{w}}_{2,\bm{\gamma}_{d_2}}(\bm{\gamma^\ast}, \bm{\delta^\ast}_{d_2^\prime})\bigg((\Delta Y-\mu(\bm{\beta^\ast_{\mathbf{d^\prime}}}))-\mathbb{E}\left[w_2(\bm{\gamma^\ast}, \bm{\delta^\ast}_{d_2^\prime})\left(\Delta Y - \mu(\bm{\beta^\ast}_{\mathbf{d^\prime}})\right)\right]\bigg);
        \end{align*}}	
{\normalsize\begin{align*} 
	\Psi_{\bm{\gamma}_{\mathbf{d^\prime}}}& \equiv\bm{\dot{w}}_{2,\bm{\gamma}_{\mathbf{d^\prime}}}(\bm{\gamma^\ast}, \bm{\delta^\ast}_{d_2^\prime})\bigg((\Delta Y-\mu(\bm{\beta^\ast_{\mathbf{d^\prime}}}))-\mathbb{E}\left[w_2(\bm{\gamma^\ast},\bm{\delta^\ast}_{d_2^\prime})\left(\Delta Y - \mu(\bm{\beta^\ast}_{\mathbf{d^\prime}})\right)\right]\bigg);\\
	\Psi_{\bm{\delta}_{d_2}}& \equiv \bm{\dot{w}}_1(\bm{\delta^\ast}_{d_2})\bigg((\Delta Y-\mu(\bm{\beta^\ast}_{\mathbf{d^\prime}})) - \mathbb{E}[w_1(\bm{\delta^\ast}_{d_2})(\Delta Y-\mu(\bm{\beta^\ast}_{\mathbf{d^\prime}}))]\bigg)\\
    &-\bm{\dot{w}}_{4,\bm{\delta}_{d_2}}(\bm{\gamma^\ast}_{d_1|d_2},\bm{\delta^\ast}_{d_2})\bigg((\mu(\bm{\beta^\ast}_{\mathbf{d}})-\mu(\bm{\beta^\ast}_{\mathbf{d^\prime}}))-\mathbb{E}[w_4(\bm{\gamma^\ast}_{d_1|d_2}, \bm{\delta^\ast}_{d_2})(\mu(\bm{\beta^\ast}_{\mathbf{d}})-\mu(\bm{\beta^\ast}_{\mathbf{d^\prime}}))]\bigg); \\
	\Psi_{\bm{\delta}_{d^\prime_2}}& \equiv\bm{\dot{w}}_{2,\bm{\delta}_{d_2^\prime}}(\bm{\gamma^\ast}, \bm{\delta^\ast}_{d_2^\prime})\bigg((\Delta Y-\mu(\bm{\beta^\ast}_{\mathbf{d^\prime}})) - \mathbb{E}[w_2(\bm{\gamma^\ast}, \bm{\delta^\ast}_{d_2^\prime})(\Delta Y-\mu(\bm{\beta^\ast}_{\mathbf{d^\prime}}))]\bigg),
 	\end{align*}}		where 
    {\normalsize \begin{align*}
		\bm{\dot{w}}_1(\bm{\delta^\ast}_{d_2})  \equiv \frac{\partial}{\partial \bm{\delta}_{d_2}}\widetilde{w}_1(\bm{\delta^\ast}_{d_2}) 
		&= - w_1(\bm{\delta^\ast}_{d_2})\cdot \frac{\bm{\dot{\phi}}(\bm{\delta^\ast}_{d_2})}{\phi(\bm{\delta^\ast}_{d_2})};\\
	 \bm{\dot{w}}_{2,\bm{\gamma}_{d_1|d_2}}(\bm{\gamma^\ast}, \bm{\delta^\ast}_{d_2^\prime}) \equiv \frac{\partial}{\partial \bm{\gamma}_{d_1|d_2}}\widetilde{w}_2(\bm{\gamma^\ast}, \bm{\delta^\ast}_{d_2^\prime}) 
		& = w_2(\bm{\gamma^\ast}, \bm{\delta^\ast}_{d_2^\prime})\cdot \frac{\bm{\dot{\pi}}(\bm{\gamma^\ast}_{d_1|d_2})}{\pi(\bm{\gamma^\ast}_{d_1|d_2})};\\
		\bm{\dot{w}}_{2,\bm{\gamma}_{d_2}}(\bm{\gamma}^\ast, \bm{\delta^\ast}_{d_2^\prime}) \equiv \frac{\partial}{\partial \bm{\gamma}_{d_2}}\widetilde{w}_2(\bm{\gamma^\ast},\bm{\delta^\ast}_{d_2^\prime}) 
		& = w_2(\bm{\gamma^\ast}, \bm{\delta^\ast}_{d_2^\prime})\cdot \frac{\bm{\dot{\pi}}(\bm{\gamma^\ast}_{d_2})}{\pi(\bm{\gamma^\ast}_{d_2})};\\
		\bm{\dot{w}}_{2,\bm{\gamma}_{\mathbf{d^\prime}}}(\bm{\gamma^\ast}, \bm{\delta^\ast}_{d_2^\prime}) \equiv \frac{\partial}{\partial \bm{\gamma}_{\mathbf{d^\prime}}}\widetilde{w}_2(\bm{\gamma^\ast},\bm{\delta^\ast}_{d_2^\prime}) 
		& = -w_2(\bm{\gamma^\ast}, \bm{\delta^\ast}_{d_2^\prime})\cdot \frac{\bm{\dot{\pi}}(\bm{\gamma^\ast}_{\mathbf{d^\prime}})}{\pi(\bm{\gamma^\ast}_{\mathbf{d^\prime}})}; \\
		\bm{\dot{w}}_{2,\bm{\delta}_{d_2^\prime}}(\bm{\gamma^\ast}, \bm{\delta^\ast}_{d_2^\prime}) \equiv \frac{\partial}{\partial \bm{\delta}_{d_2^\prime}}\widetilde{w}_2(\bm{\gamma^\ast},\bm{\delta^\ast}_{d_2^\prime}) 
		& = -w_2(\bm{\gamma^\ast}, \bm{\delta^\ast}_{d_2^\prime})\cdot \frac{\bm{\dot{\phi}}(\bm{\delta^\ast}_{d_2^\prime})}{\phi(\bm{\delta^\ast}_{d_2^\prime})}; \\
		\bm{\dot{w}}_3(\bm{\gamma^\ast}_{d_1|d_2}) \equiv \frac{\partial}{\partial \bm{\gamma}_{d_1|d_2}}\widetilde{w}_3(\bm{\gamma^\ast}_{d_1|d_2}) 
		& =w_3(\bm{\gamma^\ast}_{d_1|d_2})\cdot\frac{\bm{\dot{\pi}}(\bm{\gamma^\ast}_{d_1|d_2})}{\pi(\bm{\gamma^\ast}_{d_1|d_2})}; \\
		\bm{\dot{w}}_{4,\bm{\gamma}_{d_1|d_2}}(\bm{\gamma^\ast}_{d_1|d_2}, \bm{\delta^\ast}_{d_2})\equiv \frac{\partial}{\partial \bm{\gamma}_{d_1|d_2}}\widetilde{w}_4(\bm{\gamma^\ast}_{d_1|d_2},\bm{\delta^\ast}_{d_2})
		& =w_4(\bm{\gamma^\ast}_{d_1|d_2}, \bm{\delta^\ast}_{d_2})\cdot\frac{\bm{\dot{\pi}}(\bm{\gamma^\ast}_{d_1|d_2})}{\pi(\bm{\gamma^\ast}_{d_1|d_2})}; \\
		\bm{\dot{w}}_{4,\bm{\delta}_{d_2}}(\bm{\gamma^\ast}_{d_1|d_2}, \bm{\delta^\ast}_{d_2})\equiv \frac{\partial}{\partial \bm{\delta}_{d_2}}\widetilde{w}_4(\bm{\gamma^\ast}_{d_1|d_2},\bm{\delta^\ast}_{d_2}) 
		& = -w_4(\bm{\gamma^\ast}_{d_1|d_2}, \bm{\delta^\ast}_{d_2})\cdot \frac{\bm{\dot{\phi}}(\bm{\delta^\ast}_{d_2})}{\phi(\bm{\delta^\ast}_{d_2})}.
\end{align*}}
Finally, asymptotic normality follows from the Lindberg-Levy central limit theorem.
\end{proof}

\subsection{Proof Corollary~\ref{cor:efficient}}\label{sec:proof_cor_efficient}
\begin{proof}
    To prove the robust estimator achieves the efficiency bound, we will first show that when all three models are correctly specified, i.e. $\mu= m$, $\pi= p$, and $\phi = q$, then $\mathbb{E}(\Psi_{\bm{\eta}}) = \mathbf{0}$ for each $\bm{\eta}=\bm{\beta}_{\mathbf{d}}$, $\bm{\beta}_{\mathbf{d^\prime}}$, $\bm{\gamma}_{d_1|d_2}$, $\bm{\gamma}_{d_1^\prime|d_2^\prime}$, $\bm{\gamma}_{d_2}$, $\bm{\delta}_{d_2}$, and $\bm{\delta}_{d_2^\prime}$. 

To see this, consider first $\mathbb{E}(\Psi_{\bm{\beta}_{\mathbf{d}}})= \mathbb{E}\left[\left(w_3(\bm{\gamma^\ast}_{d_1|d_2})-w_4(\bm{\gamma^\ast}_{d_1|d_2}, \bm{\delta^\ast}_{d_2})\right)\bm{\dot{\mu}}(\bm{\beta^\ast}_{\mathbf{d}})\right]$. By LIE, one can show easily that this term is zero since
{\normalsize \begin{align}\label{psi_beta_d}
	\mathbb{E}[w_3(\bm{\gamma^\ast}_{d_1|d_2})\bm{\dot{\mu}}(\bm{\beta^\ast}_{\mathbf{d}})]=\mathbb{E}[w_4(\bm{\gamma^\ast}_{d_1|d_2}, \bm{\delta^\ast}_{d_2})\bm{\dot{\mu}}(\bm{\beta^\ast}_{\mathbf{d}})]& =\mathbb{P}(\mathbf{D}=\mathbf{d})^{-1}\cdot\mathbb{E}\left[\pi(\bm{\gamma^\ast}_{\mathbf{d}})\bm{\dot{\mu}}(\bm{\beta^\ast}_{\mathbf{d}})\right]. 		
\end{align}} 
Next, consider, $\mathbb{E}(\Psi_{\bm{\beta}_{\mathbf{d^\prime}}})= \mathbb{E}\left[\left(w_1(\bm{\delta^\ast}_{d_2}) - w_2(\bm{\gamma^\ast}, \bm{\delta^\ast}_{d_2^\prime}) + w_3(\bm{\gamma^\ast}_{d_1|d_2})-w_4(\bm{\gamma^\ast}_{d_1|d_2}, \bm{\delta^\ast}_{d_2})\right)\bm{\dot{\mu}}(\bm{\beta^\ast}_{\mathbf{d^\prime}})\right]$.
Again, with LIE, one can easily show that 
{\normalsize \begin{align}\label{psi_beta_d'}
		\mathbb{E}[w_1(\bm{\delta^\ast}_{d_2})\bm{\dot{\mu}}(\bm{\beta^\ast}_{\mathbf{d^\prime}})] =\mathbb{E}[ w_2(\bm{\gamma^\ast}, \bm{\delta^\ast}_{d_2^\prime})\bm{\dot{\mu}}(\bm{\beta^\ast}_{\mathbf{d^\prime}})] &= \mathbb{P}(\mathbf{D}=\mathbf{d})^{-1}\mathbb{E}[\pi(\bm{\gamma^\ast}_{\mathbf{d}})\bm{\dot{\mu}}(\bm{\beta^\ast}_{\mathbf{d^\prime}})] \text{ and } \nonumber \\
		\mathbb{E}[w_3(\bm{\gamma^\ast}_{d_1|d_2})\bm{\dot{\mu}}(\bm{\beta^\ast}_{\mathbf{d^\prime}})]=\mathbb{E}[w_4(\bm{\gamma^\ast}_{d_1|d_2}, \bm{\delta^\ast}_{d_2})\bm{\dot{\mu}}(\bm{\beta^\ast}_{\mathbf{d^\prime}})]&=\mathbb{P}(\mathbf{D}=\mathbf{d})^{-1}\mathbb{E}[\pi(\bm{\gamma^\ast}_{\mathbf{d}})\bm{\dot{\mu}}(\bm{\beta^\ast}_{\mathbf{d^\prime}})].
\end{align}} which implies that $\mathbb{E}(\Psi_{\bm{\beta}_{\mathbf{d^\prime}}}) = \mathbf{0}$.
Next, consider $\mathbb{E}(\Psi_{\bm{\gamma}_{d_1|d_2}})$.
First, note that when all three models are correctly specified, 
{\normalsize\begin{align}
		\mathbb{E}[w_2(\bm{\gamma}^\ast, \bm{\delta^\ast}_{d_2^\prime})(\Delta Y - \mu(\bm{\beta^\ast}_{\mathbf{d^\prime}}))]& = 0 \label{psi_gamma_t1}, \\
		\mathbb{E}[w_3(\bm{\gamma^\ast}_{d_1|d_2})\left(\mu(\bm{\beta^\ast}_{\mathbf{d}})-\mu(\bm{\beta^\ast}_{\mathbf{d^\prime}})\right)] &= \mathbb{E}\left[w_4(\bm{\gamma^\ast}_{d_1|d_2}, \bm{\delta^\ast}_{d_2})(\mu(\bm{\beta^\ast}_{\mathbf{d}})-\mu(\bm{\beta^\ast}_{\mathbf{d^\prime}}))\right] \nonumber \\
		& = \mathbb{P}(\mathbf{D}=\mathbf{d})^{-1}\mathbb{E}[\pi(\bm{\gamma^\ast}_{\mathbf{d}}) (\mu(\bm{\beta^\ast}_{\mathbf{d}})-\mu(\bm{\beta^\ast}_{\mathbf{d^\prime}}))] = \tau_{\mathbf{dd^\prime}}, \label{psi_gamma1_t23} 
\end{align}}
which implies that
{\normalsize \begin{align}\label{psi_gamma1}
	\mathbb{E}(\Psi_{\bm{\gamma}_{d_1|d_2}}) &=  \mathbb{E}[\bm{\dot{w}}_{2,\bm{\gamma}_{d_1|d_2}}(\bm{\gamma^\ast}, \bm{\delta^\ast}_{d_2^\prime})(\Delta Y - \mu(\bm{\beta^\ast}_{\mathbf{d^\prime}}))] - \mathbb{E}[\bm{\dot{w}}_3(\bm{\gamma^\ast}_{d_1|d_2})\left(\mu(\bm{\beta^\ast}_{\mathbf{d}}) - \mu(\bm{\beta^\ast}_{\mathbf{d^\prime}}) - \tau_{\mathbf{dd^\prime}}\right)] \nonumber  \\
	&+\mathbb{E}[\bm{\dot{w}}_{4,\bm{\gamma}_{d_1|d_2}}(\bm{\gamma^\ast}_{d_1|d_2}, \bm{\delta^\ast}_{d_2})\left(\mu(\bm{\beta^\ast}_{\mathbf{d}}) - \mu(\bm{\beta^\ast}_{\mathbf{d^\prime}}) - \tau_{\mathbf{dd^\prime}}\right)].
\end{align}}
Now, the first term of $\eqref{psi_gamma1}$ is zero	since (by LIE)
{\normalsize \begin{align*} 
		\mathbb{E}\left[\frac{S\cdot 	\mathbbm{1}[\mathbf{D}=\mathbf{d^\prime}]\pi(\bm{\gamma^\ast}_{\mathbf{d}})}{\phi(\bm{\delta^\ast}_{d_2^\prime})\pi(\bm{\gamma^\ast}_{\mathbf{d^\prime}})}\right] = \mathbb{P}(\mathbf{D}=\mathbf{d}) \text{ and }
		\mathbb{E}\bigg[\frac{S\cdot \mathbbm{1}[\mathbf{D}=\mathbf{d^\prime}]\bm{\dot{\pi}}(\bm{\gamma^\ast}_{d_1|d_2})\pi(\bm{\gamma^\ast}_{d_2})}{\phi(\bm{\delta^\ast}_{d_2^\prime})\pi(\bm{\gamma^\ast}_{\mathbf{d^\prime}})}\left(\Delta Y-\mu(\bm{\beta^\ast}_{\mathbf{d^\prime}})\right)\bigg]= \mathbf{0}.
\end{align*}} 
Finally, we can also show that
{\normalsize \begin{equation} \label{psi_gamma_t2_2}
	 \mathbb{E}\left[\bm{\dot{w}}_3(\bm{\gamma^\ast}_{d_1|d_2})\left(\mu(\bm{\beta^\ast}_{\mathbf{d}})-\mu(\bm{\beta^\ast}_{\mathbf{d^\prime}})-\tau_{\mathbf{dd^\prime}}\right)\right] =\mathbb{E}\left[\bm{\dot{w}}_{4,\bm{\gamma}_{d_1|d_2}}(\bm{\gamma^\ast}_{d_1|d_2}, \bm{\delta^\ast}_{d_2})\left(\mu(\bm{\beta^\ast}_{\mathbf{d}})-\mu(\bm{\beta^\ast}_{\mathbf{d^\prime}})-\tau_{\mathbf{dd^\prime}}\right)\right]. 
\end{equation}}
since by LIEs
{\normalsize \begin{align*}
		\mathbb{E}[\pi(\bm{\gamma^\ast}_{d_1|d_2})\mathbbm{1}[D_2=d_2]] = \mathbb{E}\left[\frac{S\cdot \mathbbm{1}[D_2=d_2]\pi(\bm{\gamma^\ast}_{d_1|d_2})}{\phi(\bm{\delta^\ast}_{d_2})}\right] &= \mathbb{P}(\mathbf{D}=\mathbf{d}); \\
		\mathbb{E}[\mathbbm{1}[D_2=d_2]\bm{\dot{\pi}}(\bm{\gamma^\ast}_{d_1|d_2})(\mu(\bm{\beta^\ast}_{\mathbf{d}}) - \mu(\bm{\beta^\ast}_{\mathbf{d^\prime}})-\tau_{\mathbf{dd^\prime}})]& = \mathbb{E}[\bm{\dot{\pi}}(\bm{\gamma^\ast}_{d_1|d_2}) \ \pi(\bm{\gamma^\ast}_{d_2})(\mu(\bm{\beta^\ast}_{\mathbf{d}}) - \mu(\bm{\beta^\ast}_{\mathbf{d^\prime}})-\tau_{\mathbf{dd^\prime}})]; \\
		\mathbb{E}\left[\frac{S\cdot \mathbbm{1}[D_2=d_2]\bm{\dot{\pi}}(\bm{\gamma^\ast}_{d_1|d_2})}{\phi(\bm{\delta^\ast}_{d_2})}(\mu(\bm{\beta^\ast}_{\mathbf{d}})-\mu(\bm{\beta^\ast}_{\mathbf{d^\prime}})-\tau_{\mathbf{dd^\prime}})\right]  &= \mathbb{E}\left[\bm{\dot{\pi}}(\bm{\gamma^\ast}_{d_1|d_2})\pi(\bm{\gamma^\ast}_{d_2})(\mu(\bm{\beta^\ast}_{\mathbf{d}})-\mu(\bm{\beta^\ast}_{\mathbf{d^\prime}})-\tau_{\mathbf{dd^\prime}})\right]. 
\end{align*}}
This proves that $\mathbb{E}(\Psi_{\bm{\gamma}_{d_1|d_2}}) = \mathbf{0}$. In a similar vein, consider $\mathbb{E}(\Psi_{\bm{\gamma}_{d_2}})$ which simplifies to $\mathbb{E}[\bm{\dot{w}}_{2,\bm{\gamma}_{d_2}}(\bm{\gamma^\ast}, \bm{\delta^\ast}_{d_2^\prime})(\Delta Y-\mu(\bm{\beta^\ast_{\mathbf{d^\prime}}}))]$ since $\mathbb{E}[w_2(\bm{\gamma^\ast}, \bm{\delta^\ast}_{d_2^\prime})\left(\Delta Y - \mu(\bm{\beta^\ast}_{\mathbf{d^\prime}})\right)] = 0$. Now, one can show that $\mathbb{E}[\bm{\dot{w}}_{2,\bm{\gamma}_{d_2}}(\bm{\gamma^\ast}, \bm{\delta^\ast}_{d_2^\prime})(\Delta Y-\mu(\bm{\beta^\ast_{\mathbf{d^\prime}}}))] = \mathbf{0}$  by LIE because
\begin{equation*}
	\mathbb{E}\bigg[\frac{S\cdot \mathbbm{1}[\mathbf{D}=\mathbf{d^\prime}]\pi(\bm{\gamma^\ast}_{d_1|d_2})\bm{\dot{\pi}}(\bm{\gamma^\ast}_{d_2})}{\phi(\bm{\delta^\ast}_{d_2^\prime})\pi(\bm{\gamma^\ast}_{\mathbf{d^\prime}})}\left(\Delta Y-\mu(\bm{\beta^\ast}_{\mathbf{d^\prime}})\right)\bigg] = \mathbf{0}.
\end{equation*}
Next, consider $\mathbb{E}(\Psi_{\bm{\gamma}_{\mathbf{d^\prime}}})$. Following similar arguments as above, the right hand side can be simplified such that $\mathbb{E}(\Psi_{\bm{\gamma}_{\mathbf{d^\prime}}})= \mathbb{E}[\bm{\dot{w}}_{2,\bm{\gamma}_{\mathbf{d^\prime}}}(\bm{\gamma^\ast}, \bm{\delta^\ast}_{d_2^\prime})(\Delta Y - \mu(\bm{\beta^\ast}_{\mathbf{d^\prime}}))]$ which itself equals
{\normalsize \begin{align}\label{Psi_gamma_d'}
		&=-\mathbb{E}\left[\frac{S\cdot \mathbbm{1}[\mathbf{D}=\mathbf{d^\prime}]\pi(\bm{\gamma^\ast}_{\mathbf{d}})\bm{\dot{\pi}}(\bm{\gamma^\ast}_{\mathbf{d^\prime}})}{\phi(\bm{\delta^\ast}_{d_2^\prime})\pi(\bm{\gamma^\ast}_{\mathbf{d^\prime}})^2}(\Delta Y - \mu(\bm{\beta^\ast}_{\mathbf{d^\prime}}))\right]\bigg/\mathbb{P}(\mathbf{D}=\mathbf{d})= \mathbf{0}. \ \ \ \ \text{ (by LIE)}
\end{align}}
For $\mathbb{E}(\Psi_{\bm{\delta}_{d_2}})$, one can show that under correct specification of the models, successively applying LIE gives us 
{\normalsize \begin{align}\label{psi_delta_t1_1}
		\mathbb{E}[w_1(\bm{\delta^\ast}_{d_2})(\Delta Y - \mu(\bm{\beta^\ast}_{\mathbf{d^\prime}}))] &= \mathbb{P}(\mathbf{D}=\mathbf{d})^{-1}\cdot \mathbb{E}[(\mu(\bm{\beta^\ast}_{\mathbf{d}})-\mu(\bm{\beta^\ast}_{\mathbf{d^\prime}})\pi(\bm{\gamma^\ast}_{\mathbf{d}})] =\tau_{\mathbf{dd^\prime}}	\text{ and } \\
		\mathbb{E}[w_4(\bm{\gamma^\ast}_{d_1|d_2}, \bm{\delta^\ast}_{d_2})(\mu(\bm{\beta^\ast}_{\mathbf{d}})-\mu(\bm{\beta^\ast}_{\mathbf{d^\prime}}))] &=\mathbb{P}(\mathbf{D}=\mathbf{d})^{-1}\cdot \mathbb{E}[(\mu(\bm{\beta^\ast}_{\mathbf{d}})-\mu(\bm{\beta^\ast}_{\mathbf{d^\prime}})\pi(\bm{\gamma^\ast}_{\mathbf{d}})] =\tau_{\mathbf{dd^\prime}}.
\end{align}} This implies that we can write
{\normalsize \begin{equation*}
	\mathbb{E}(\Psi_{\bm{\delta}_{d_2}})	= \mathbb{E}\big[\bm{\dot{w}}_1(\bm{\delta^\ast}_{d_2})(\Delta Y- \mu(\bm{\beta^\ast}_{\mathbf{d^\prime}})-\tau_{\mathbf{dd^\prime}})-\bm{\dot{w}}_{4,\bm{\delta}_{d_2}}(\bm{\gamma^\ast}_{d_1|d_2}, \bm{\delta^\ast}_{d_2})(\mu(\bm{\beta^\ast}_{\mathbf{d}})-\mu(\bm{\beta^\ast}_{\mathbf{d^\prime}})-\tau_{\mathbf{dd^\prime}})\big].
\end{equation*}}
It can again be shown through applications of LIE that 
\begin{equation}
	\mathbb{E}[\bm{\dot{w}}_1(\bm{\delta^\ast}_{d_2})(\Delta Y- 	\mu(\bm{\beta^\ast}_{\mathbf{d^\prime}})-\tau_{\mathbf{dd^\prime}})] = \mathbb{E}[\bm{\dot{w}}_{4,\bm{\delta}_{d_2}}(\bm{\gamma^\ast}_{d_1|d_2}, \bm{\delta^\ast}_{d_2})(\mu(\bm{\beta^\ast}_{\mathbf{d}})-\mu(\bm{\beta^\ast}_{\mathbf{d^\prime}})-\tau_{\mathbf{dd^\prime}})],	
\end{equation} since
{\small \begin{align*}
		\mathbb{E}\left[\frac{S\cdot \mathbbm{1}[\mathbf{D}=\mathbf{d}]}{\phi(\bm{\delta^\ast}_{d_2})}\right] = 	\mathbb{E}\left[\frac{S\cdot \mathbbm{1}[D_2=d_2]\pi(\bm{\gamma^\ast}_{d_1|d_2})}{\phi(\bm{\delta^\ast}_{d_2})}\right]& = \mathbb{P}(\mathbf{D}=\mathbf{d});\\
		\mathbb{E}\left[\frac{S\cdot \mathbbm{1}[\mathbf{D}=\mathbf{d}]\bm{\dot{\phi}}(\bm{\delta^\ast}_{d_2})}{\phi^2(\bm{\delta^\ast}_{d_2})}(\Delta Y- \mu(\bm{\beta^\ast}_{\mathbf{d^\prime}})-\tau_{\mathbf{dd^\prime}})\right]& = \mathbb{E}\left[\frac{\bm{\dot{\phi}}(\bm{\delta^\ast}_{d_2})}{\phi(\bm{\delta^\ast}_{d_2})}\pi(\bm{\gamma^\ast}_{\mathbf{d}})(\mu_{\mathbf{d}}(\bm{\beta^\ast})-\mu(\bm{\beta^\ast}_{\mathbf{d^\prime}})-\tau_{\mathbf{dd^\prime}})\right]; \\		
		\mathbb{E}\left[\frac{S\cdot 	\mathbbm{1}[D_2=d_2]\pi(\bm{\gamma^\ast}_{d_1|d_2})\bm{\dot{\phi}}(\bm{\delta^\ast}_{d_2})}{\phi^2(\bm{\delta^\ast}_{d_2})}(\mu(\bm{\beta^\ast}_{\mathbf{d}})-\mu(\bm{\beta^\ast}_{\mathbf{d^\prime}})-\tau_{\mathbf{dd^\prime}})\right] & = \mathbb{E}\left[\frac{\bm{\dot{\phi}}(\bm{\delta^\ast}_{d_2})}{\phi(\bm{\delta^\ast}_{d_2})}\pi(\bm{\gamma^\ast}_{\mathbf{d}})(\mu(\bm{\beta^\ast}_{\mathbf{d}})-\mu(\bm{\beta^\ast}_{\mathbf{d^\prime}})-\tau_{\mathbf{dd^\prime}})\right].
\end{align*}} Finally,  
$\mathbb{E}[\bm{\dot{w}}_{2,\bm{\delta}_{d_2^\prime}}(\bm{\gamma^\ast}, \bm{\delta^\ast}_{d_2^\prime})(\Delta Y - \mu(\bm{\beta^\ast}_{\mathbf{d^\prime}}))] = \mathbf{0}$ since $\mathbb{E}\left[\frac{S\cdot\mathbbm{1}[\mathbf{D}=\mathbf{d^\prime}]\pi(\bm{\gamma^\ast}_{\mathbf{d}})\bm{\dot{\phi}}(\bm{\delta^\ast}_{d_2^\prime})}{\phi^2(\bm{\delta^\ast}_{d_2^\prime})\pi(\bm{\gamma^\ast}_{\mathbf{d^\prime}})}(\Delta Y-\mu(\bm{\beta^\ast}_{\mathbf{d^\prime}}))\right] = \mathbf{0}$. This  proves that $\mathbb{E}(\Psi_{\bm{\delta}_{d^\prime_2}}) = \mathbf{0}$. Therefore,  when all three models are correctly specified, the influence function of the robust estimator $\widehat{\tau}^\textup{R}_{\mathbf{dd^\prime}}$ simplifies to
{\normalsize \begin{align}\label{equivalence}
		\psi& = w_1(\bm{\delta^\ast}_{d_2})\left(\Delta Y-\mu(\bm{\beta^\ast}_{\mathbf{d^\prime}})-\tau_{\mathbf{dd^\prime}}\right)-w_2(\bm{\gamma^\ast}, \bm{\delta^\ast}_{d_2^\prime})\big(\Delta Y-\mu(\bm{\beta^\ast}_{\mathbf{d^\prime}})\big) +\left(w_3(\bm{\gamma^\ast}_{d_1|d_2})-w_4(\bm{\gamma^\ast}_{d_1|d_2}, \bm{\delta^\ast}_{d_2})\right) \nonumber \\
		&\times \left(\mu(\bm{\beta^\ast}_{\mathbf{d}})-\mu(\bm{\beta^\ast}_{\mathbf{d^\prime}})-\tau_{\mathbf{dd^\prime}}\right) \nonumber \\
		& = w_1(S, \mathbf{D}, \mathbf{X})\left(\Delta Y-m_{\mathbf{d^\prime}}(\mathbf{X})-\tau_{\mathbf{dd^\prime}}\right)- w_2(S, \mathbf{D}, \mathbf{X})(\Delta Y-m_{\mathbf{d^\prime}}(\mathbf{X})) \nonumber \\
		&+\left(w_3(D_{2}, \mathbf{X}) - w_4(S, D_2, \mathbf{X})\right)\left(m_{\mathbf{d}}(\mathbf{X})-m_{\mathbf{d^\prime}}(\mathbf{X})-\tau_{\mathbf{dd^\prime}}\right), 
\end{align}}
which equals the efficient influence function, $F_{\tau_{\mathbf{dd^\prime}}}(\mathbf{W})$, for the target parameter $\tau_{\mathbf{dd^\prime}}$. 
\end{proof}

\subsection{Estimation routine}\label{sec:routine}
Theorem~\ref{thm:asyvar} accommodates any set of generic parametric models for the outcome means, propensity scores, and missing treatment probabilities, for which $\sqrt{n}$-consistent estimators of their pseudo-true values are available. In practice, specific parametric models and estimators have to be selected. Consider the most commonly used choices: $\phi(\bm{\delta}) = \Lambda(\mathbf{X}\bm{\delta})$, $\pi(\bm{{\gamma}}) = \Lambda(\mathbf{X}\bm{\gamma})$, and $\mu(\bm{{\beta}}) = \mathbf{X}\bm{\beta}$ where $\Lambda(\cdot)$ denotes the inverse logit function. 
The following procedure outlines the estimation steps for our robust method: 
\begin{procedure}[Estimation with $\hat{\tau}^{\textup{R}}_{\mathbf{dd^\prime}}$] \ 
  \begin{enumerate}
    \item Estimate $\bm{\delta}_{d_2}$ by maximizing the log-likelihood function 
    {\normalsize \begin{equation*}
       \sum_{i=1}^N S_i\textup{log}[\Lambda(\mathbf{X}_i\bm{\delta}_{d_2})]+(1-S_i)\textup{log}[1-\Lambda(\mathbf{X}_i\bm{\delta}_{d_2})] \text{ if } D_{i2} = d_{i2}. 
    \end{equation*}}  Obtain the predicted probability $\phi(\bm{\hat{\delta}}_{d_2}) = \Lambda(\mathbf{X}\bm{\hat{\delta}}_{d_2})$ for each $d_2=0,1$.
    \item Estimate $\bm{\gamma}_{d_1|d_2}$, $\bm{\gamma}_{d_1^\prime|d_2^\prime}$, and $\bm{\gamma}_{d_2}$, by maximizing the log-likelihood functions given by 
    \begin{itemize}
        \item[2a)] $\sum_{i=1}^N \mathbbm{1}[D_{i1}=d_{i1}]\textup{log}[\Lambda(\mathbf{X}_i\bm{\gamma}_{d_1|d_2})]+(1-\mathbbm{1}[D_{i1}=d_{i1}])\textup{log}[1-\Lambda(\mathbf{X}_i\bm{\gamma}_{d_1|d_2})]$  if  $S_i=1$ and $D_{i2} = d_{i2}$; 
        \item[2b)] $\sum_{i=1}^N \mathbbm{1}[D_{i1}=d^\prime_{i1}]\textup{log}[\Lambda(\mathbf{X}_i\bm{\gamma}_{d^\prime_1|d^\prime_2})]+(1-\mathbbm{1}[D_{i1}=d^\prime_{i1}])\textup{log}[1-\Lambda(\mathbf{X}_i\bm{\gamma}_{d^\prime_1|d^\prime_2})]$ if $S_i=1$ and $D_{i2} = d^\prime_{i2}$;
        \item[2c)] $\sum_{i=1}^N D_{i2}\textup{log}[\Lambda(\mathbf{X}_i\bm{\gamma}_{d_2})]+(1-D_{i2}])\textup{log}[1-\Lambda(\mathbf{X}_i\bm{\gamma}_{d_2})]$.
    \end{itemize}
    respectively. Obtain the predicted probabilities $\pi(\bm{\hat{\gamma}}_{d_1|d_2}) = \Lambda(\mathbf{X}\bm{\hat{\gamma}}_{d_1|d_2})$, $\pi(\bm{\hat{\gamma}}_{d^\prime_1|d^\prime_2}) = \Lambda(\mathbf{X}\bm{\hat{\gamma}}_{d^\prime_1|d^\prime_2})$, and $\pi(\bm{\hat{\gamma}}_{d_2}) = \Lambda(\mathbf{X}\bm{\hat{\gamma}}_{d_2})$ which gives us the propensity score estimates, $\pi(\bm{\hat{\gamma}}_{\mathbf{d}}) = \Lambda(\mathbf{X}\bm{\hat{\gamma}}_{d_1|d_2})\times \Lambda(\mathbf{X}\bm{\hat{\gamma}}_{d_2})$ and $\pi(\bm{\hat{\gamma}}_{\mathbf{d^\prime}}) = \Lambda(\mathbf{X}\bm{\hat{\gamma}}_{d^\prime_1|d^\prime_2})\times(1-\Lambda(\mathbf{X}\bm{\hat{\gamma}}_{d_2}))$.
	\item Estimate $\bm{\widehat{\beta}}_{\mathbf{d}}$ and $\bm{\widehat{\beta}}_{\mathbf{d^\prime}}$ using least squares, where estimation is conditioned on $\{S=1, D_1=d_1, D_2=d_2\}$ and $\{S=1, D_1=d_1^\prime, D_2=d_2^\prime\}$ samples, respectively. Obtain the predicted values, $\mu(\bm{\hat{\beta}}_{\mathbf{d}}) = \mathbf{X}\bm{\hat{\beta}}_{\mathbf{d}}$ and $\mu(\bm{\hat{\beta}}_{\mathbf{d^\prime}}) = \mathbf{X}\bm{\hat{\beta}}_{\mathbf{d^\prime}}$.
	\item Use the predicted values from the previous steps (1-3) to estimate the weights as in \eqref{eq:weightSat}. Use these weights to compute $\widehat{\tau}^\textup{R}_{\mathbf{dd^\prime}}$ as in \eqref{eq:tau_hat}.
	\item Construct asymptotically valid confidence intervals with CI$=\widehat{\tau}_{\mathbf{dd^\prime}}^{\textup{R}} \pm z_{\alpha/2}\sqrt{\widehat{\mathbb{V}}[\widehat{\tau}_{\mathbf{dd^\prime}}^{\textup{R}}]}$ where $\widehat{\mathbb{V}}[\widehat{\tau}^\textup{R}_{\mathbf{dd^\prime}}] = \widehat{\Omega}$ is a consistent estimator of the asymptotic variance specified in Theorem \ref{thm:asyvar}.
\end{enumerate}  
\end{procedure}
The routine for estimating $\hat{\tau}_{\mathbf{dd^\prime}}$ is easy to implement, computationally tractable, and can be accomplished in two steps. Alternative choices for the working models also fit in our framework, but may result in more computationally involved estimation steps. For instance, probit models for the propensity scores or missing treatment models require numerical integration. 

\clearpage
\renewcommand{\footnotelayout}{\setstretch{1.0}}
\setcounter{section}{0}
\renewcommand{\thetable}{S\arabic{table}}
\renewcommand{\thefigure}{S\arabic{figure}}
\renewcommand{\thesection}{S\Alph{section}}
\renewcommand{\thesubsection}{S\Alph{section}.\arabic{subsection}}
\renewcommand{\theequation}{S\arabic{equation}}

\numberwithin{equation}{section}
\numberwithin{table}{section}
\numberwithin{figure}{section}
\numberwithin{lemma}{section}
\numberwithin{proposition}{section}
\numberwithin{assumption}{section}
\setcounter{page}{1}
\title{Supplementary Appendix for \\ ``Identification of dynamic treatment effects when treatment histories are partially observed''}
\date{}
\author{Akanksha Negi and Didier Nibbering}
\maketitle

\section{Alternative identifying assumptions}

\subsection{Alternative parallel trend assumption}\label{sec:cpt_alt}
There are two types of parallel trends assumptions that combined with the no-anticipation assumption can identify $\tau_{\mathbf{dd^\prime}}$. In the main text, we invoke Assumption \ref{ass:did}.2 that is commonly employed in the DID literature whenever conditional methods are being used/proposed. This assumption postulates that outcomes would have evolved in parallel between the treated and control groups \textit{in the absence of the treatment}. This is assumed to hold for each subpopulation of $\mathbf{X}$. This is sufficient for identifying $\tau_{\mathbf{dd^\prime}}$ for each $\mathbf{d}\in \left\{(1,1), (0,1), (1,0)\right\}$ and $\mathbf{d^\prime}=(0,0)$. However, comparisons involving $\mathbf{d^\prime}= (1,1)$ can be used to identify treatment effects $\tau_{\mathbf{dd^\prime}}$ for each $\mathbf{d} \in  \{(1,0), (0,1)\}$ and require an analogous version of Assumption \ref{ass:did}.2, which is rarely invoked (see \citet{hull2018estimating} for an exception).
\begin{assumption}[Conditional parallel trends]\label{ass:cpt2} \ 
        $\mathbb{E}\left[Y_2(\mathbf{1})-Y_0(\mathbf{1})|\mathbf{D}=\mathbf{d}, \mathbf{X}\right] = \mathbb{E}[Y_2(\mathbf{1})-Y_0(\mathbf{1})|\mathbf{X}]$ for each $\mathbf{d}$.
\end{assumption}
Then, $\mathbb{E}[\Delta Y|\mathbf{D}=\mathbf{d}, \mathbf{X}]-\mathbb{E}[\Delta Y|\mathbf{D}=\mathbf{d^\prime}, \mathbf{X}]$ is equal to
{\normalsize \begin{align}\label{eq:m2taux}
        & = \mathbb{E}[Y_2-Y_0|\mathbf{D}=\mathbf{d}, \mathbf{X}]- \mathbb{E}[Y_2-Y_0|\mathbf{D}=\mathbf{d^\prime}, \mathbf{X}] \nonumber\\
        & = \mathbb{E}[Y_2(\mathbf{d})-Y_0(\mathbf{d})|\mathbf{D}=\mathbf{d}, \mathbf{X}]- \mathbb{E}[Y_2(\mathbf{d^\prime})-Y_0(\mathbf{d^\prime})|\mathbf{D}=\mathbf{d^\prime}, \mathbf{X}] \nonumber\\
        & =   \mathbb{E}[Y_2(\mathbf{d})-Y_2(\mathbf{d^\prime})|\mathbf{D}=\mathbf{d}, \mathbf{X}] +\mathbb{E}[Y_2(\mathbf{d^\prime})-Y_0(\mathbf{d^\prime})|\mathbf{D}=\mathbf{d}, \mathbf{X}] \nonumber\\
        &- \mathbb{E}[Y_2(\mathbf{d^\prime})-Y_0(\mathbf{d^\prime})|\mathbf{D}=\mathbf{d^\prime}, \mathbf{X}] \nonumber\\
        & =  \mathbb{E}[Y_2(\mathbf{d})-Y_2(\mathbf{d^\prime})|\mathbf{D}=\mathbf{d}, \mathbf{X}],
\end{align}} 
where the third equality follows from Assumption \ref{ass:did}.1 and the fourth equality follows from Assumption \ref{ass:did}.2 if we use $\mathbf{d^\prime} = (0,0)$ as the comparison group, or Assumption \ref{ass:cpt2} if we use $\mathbf{d^\prime} = (1,1)$ as the comparison group.

\subsection{Alternative missing at random assumption}\label{sec:weakmar}
The conditional independence of $S$ on $\Delta Y$ in Assumption~\ref{ass:mar} may be considered too strong for certain empirical settings. We show that under a weaker version of Assumption \ref{ass:mar}, $\tau_{\mathbf{dd^\prime}}$ is identified using an inverse probability weighted estimand.  

\begin{assumption}[Missingness assumptions]\label{ass:mar2} \ 
 	\begin{enumerate}
 		\item $ S \perp D_1| (D_2, \Delta Y, \mathbf{X})$.
 		\item $0< \mathbb{P}(S=1|D_2,\mathbf{X}, \Delta Y) \equiv q(D_2, \mathbf{X}, \Delta Y)  \leq 1$.
 	\end{enumerate}
\end{assumption} 

Assumption~\ref{ass:mar2} does not allow for the identification of $\mathbb{E}[\Delta Y|\mathbf{D}=\mathbf{d},\mathbf{X}]$ when there are elements in $\mathbf{D}$ missing. Hence, commonly used estimands cannot identify ATTs under this assumption. However, when the propensity score $\mathbb{P}(\mathbf{D}=\mathbf{d}|\mathbf{X})$ and missing data model $q(D_2, \mathbf{X}, \Delta Y)$ can be correctly identified from the data, an adapted inverse probability weighted estimand does identify the PDATTs.

\begin{lemma}[Inverse probability weighted estimand]\label{corr:propscore}\, \\
 	Under Assumptions~\ref{ass:did} and \ref{ass:mar2}, it holds for each $d$ and $\mathbf{d^\prime}=(0,0)$ that
 	\begin{align}
 		{\mathbb{E}\left[p_\mathbf{d}(\mathbf{X})\right]}^{-1}\mathbb{E}\left[\frac{S}{q(D_2,\Delta Y,\mathbf{X})}\left({\mathbbm{1}[\mathbf{D}=\mathbf{d}]}-{\frac{p_{\mathbf{d}}(\mathbf{X})}{p_{\mathbf{d^\prime}}(\mathbf{X})}\mathbbm{1}[\mathbf{D}=\mathbf{d}']}\right) \Delta Y \right]=& \tau_{\mathbf{dd^\prime}}, 
 	\end{align}
    where $p_\mathbf{d}(\mathbf{X})=\mathbb{P}(\mathbf{D}=\mathbf{d}|\mathbf{X})$.
\end{lemma}	 
The estimand in Lemma~\ref{corr:propscore} is similar to the one in Lemma~\ref{cor:ATT_identification_md}.2, with a missing data model that allows missingness to depend on $\Delta Y$. 

\begin{proof}
 \begin{align}
 	{\mathbb{E}\left[p_\mathbf{d}(\mathbf{X})\right]}^{-1}\mathbb{E}\left[\frac{S}{q(D_2,\Delta Y,\mathbf{X})}\left({\mathbbm{1}[\mathbf{D}=\mathbf{d}]}-{\frac{p_{\mathbf{d}}(\mathbf{X})}{p_{\mathbf{d^\prime}}(\mathbf{X})}\mathbbm{1}[\mathbf{D}=\mathbf{d^\prime}]}\right) \Delta Y \right]=& \tau_{\mathbf{dd^\prime}}.  
\end{align}

First, we show that the propensity score $p_\mathbf{d}(\mathbf{X})$ is identified from the data:
{\normalsize \begin{align}
 	p_{\mathbf{d}}(\mathbf{X}) =& \sum_{\Delta Y}\mathbb{P}(\mathbf{D}=\mathbf{d}|\Delta Y,\mathbf{X})\cdot \mathbb{P}(\Delta Y|\mathbf{X}) \nonumber \\
 	=& \sum_{\Delta Y}\mathbb{P}(D_1=d_1|D_2=d_2,\Delta Y,\mathbf{X})\cdot \mathbb{P}(D_2=d_2|\Delta Y,\mathbf{X})\cdot \mathbb{P}(\Delta Y|\mathbf{X}) \nonumber\\
 	=&\sum_{\Delta Y}\frac{\mathbb{P}(S=1|D_2=d_2,\Delta Y,\mathbf{X})\mathbb{P}(D_1=d_1|D_2=d_2,\Delta Y,\mathbf{X})}{\mathbb{P}(S=1|D_2=d_2,\Delta Y,\mathbf{X})} \cdot \mathbb{P}(D_2=d_2|\Delta Y,\mathbf{X})\cdot \mathbb{P}(\Delta Y|\mathbf{X}) \nonumber \\
 	{=}&   
 	\sum_{\Delta Y}\frac{\mathbb{P}(S\mathbbm{1}[D_1=d_1]|D_2=d_2,\Delta Y,\mathbf{X})}{\mathbb{P}(S=1|D_2=d_2,\Delta Y,\mathbf{X})}  \cdot \mathbb{P}(D_2=d_2|\Delta Y,\mathbf{X})\cdot \mathbb{P}(\Delta Y|\mathbf{X})\nonumber \\=&
 	\mathbb{E}\left[ \frac{S\mathbbm{1}[\mathbf{D}=\mathbf{d}]}{\mathbb{P}(S=1|D_2=d_2,\Delta Y,\mathbf{X})} |\mathbf{X} \right],
\end{align}}
where the fourth line uses $S \perp D_1| (D_2, \Delta Y, \mathbf{X})$.

Second, we show that the estimand equals $\tau_{\mathbf{dd^\prime}}$:
{\normalsize \begin{align}
 	\mathbb{E}\left[\frac{S}{q(D_2,\Delta Y,\mathbf{X})}\mathbbm{1}[\mathbf{D}=\mathbf{d}]\Delta Y \right] =&  
 	\mathbb{E}\left\{{\mathbb{E}\left[\frac{S}{q(D_2,\Delta Y,\mathbf{X})}\mathbbm{1}[\mathbf{D}=\mathbf{d}]\Delta Y|\mathbf{X}\right]}  \right\} \nonumber  \\
 	=& \mathbb{E}\bigg\{\sum_{\Delta Y} \Delta Y \mathbb{E}\left[\frac{S}{q(D_2,\Delta Y,\mathbf{X})}\mathbbm{1}[\mathbf{D}=\mathbf{d}]|\Delta Y, \mathbf{X}\right] \cdot \mathbb{P}(\Delta Y|\mathbf{X}) \bigg\} \nonumber \\
 	=& \mathbb{E}\bigg\{\sum_{\Delta Y} \Delta Y\mathbb{E}\bigg[\frac{S}{q(D_2,\Delta Y,\mathbf{X})}\mathbbm{1}[D_1=d_1]|D_2=d_2,\Delta Y,\mathbf{X}\bigg] \nonumber \\
 		& \cdot \mathbb{P}(D_2=d_2|\Delta Y,\mathbf{X})\cdot \mathbb{P}(\Delta Y|\mathbf{X}) \bigg\} \nonumber \\
 	{=}&    \mathbb{E}\bigg\{\sum_{\Delta Y} \Delta Y\frac{q(D_2=d_2,\Delta Y,\mathbf{X})}{q(D_2=d_2,\Delta Y,\mathbf{X})} \nonumber \\
 	& \cdot \mathbb{P}(D_1=d_1|D_2=d_2,\Delta Y,\mathbf{X})\cdot \mathbb{P}(D_2=d_2|\Delta Y,\mathbf{X})\cdot  \mathbb{P}(\Delta Y|\mathbf{X}) \bigg\} \nonumber \\
 	=& \mathbb{E}\left[m_{\mathbf{d}}(\mathbf{X})p_\mathbf{d}(\mathbf{X})\right],
\end{align}}
where the fourth line uses $S \perp D_1| (D_2, \Delta Y, \mathbf{X})$ and $m_{\mathbf{d}}(\mathbf{X}) \equiv \mathbb{E}[\Delta Y|\mathbf{D}=\mathbf{d},\mathbf{X}]$. Similarly 
\begin{align}
 	\mathbb{E}\left[\frac{S}{q(D_2,\Delta Y,\mathbf{X})}\frac{p_\mathbf{d}(\mathbf{X})}{p_{\mathbf{d^\prime}}(\mathbf{X})} \mathbbm{1}[\mathbf{D}=\mathbf{d^\prime}]\Delta Y \right]=
 	\mathbb{E}\left[m_{\mathbf{d^\prime}}(\mathbf{X})p_\mathbf{d}(\mathbf{X}) \right].
\end{align}
It now follows that the estimand equals
\begin{align}\label{eq:mp2tau}
 	\mathbb{E}\left[(m_{\mathbf{d}}(\mathbf{X})-m_{\mathbf{d^\prime}}(\mathbf{X}))\frac{p_\mathbf{d}(\mathbf{X})}{\mathbb{E}\left[p_{\mathbf{d}}(\mathbf{X})\right]}\right]=&
 	\mathbb{E}\left[\mathbb{E}\left[Y_2(\mathbf{d})-Y_2(\mathbf{d^\prime})|\mathbf{D}=\mathbf{d},\mathbf{X}\right]\frac{\mathbb{P}(\mathbf{D}=\mathbf{d}|\mathbf{X})}{\mathbb{P}(\mathbf{D}=\mathbf{d})}\right] \nonumber \\
 	=& \int_\mathbf{X} \mathbb{E}\left[Y_2(\mathbf{d})-Y_2(\mathbf{d^\prime})|\mathbf{D}=\mathbf{d},\mathbf{X}\right] d \mathbb{P}(\mathbf{X}|\mathbf{D}=\mathbf{d}) \nonumber \\
 	=& \mathbb{E}\left[Y_2(\mathbf{d})-Y_2(\mathbf{d^\prime})|\mathbf{D}=\mathbf{d}\right] \nonumber \\
 	=& \tau_{\mathbf{dd^\prime}},
\end{align}
where the first line uses \eqref{eq:m2taux} to write $m_{\mathbf{d}}(\mathbf{X})-m_{\mathbf{d^\prime}}(\mathbf{X})=\mathbb{E}\left[Y_2(\mathbf{d})-Y_2(\mathbf{d^\prime})|\mathbf{D}=\mathbf{d},\mathbf{X}\right]$.
\end{proof}

\section{Partial identification of PDATTs}\label{sec:partial_DID}
\begin{proposition}[Partial-identification of PDATT]\label{prop:partial}
 	Under Assumption~\ref{ass:did}, 
    and monotone treatment responses $\mathbb{E}[Y_2(1,0)-Y_2(0,0)|D_1=1,D_2=0,\mathbf{X}]\geq0$ and a lower bound $Y_{\min}\leq Y$ it holds that
 	{ \begin{align*}
 		&\mathbbm{E}[\Delta Y|D_2=1,\mathbf{X}]-\mathbbm{E}[\Delta Y|D_2=0,\mathbf{X}] \\
 		&\leq \mathbb{E}[Y_2(1,1)-Y_2(0,0)|D_1=1, D_2=1,\mathbf{X}]\cdot \mathbbm{P}(D_1=1|D_2=1,\mathbf{X})\\
        &+\mathbb{E}[Y_2(0,1)-Y_2(0,0)|D_1=0, D_2=1,\mathbf{X}]\cdot \mathbbm{P}(D_1=0|D_2=1,\mathbf{X}) \\
        &\leq \mathbbm{E}[\Delta Y|D_2=1,\mathbf{X}]-\mathbbm{E}[\Delta Y|D_2=0,\mathbf{X}]+\mathbb{E}[Y_2|D_2=0,\mathbf{X}]-Y_{\min}.
 	\end{align*}}
\end{proposition}
Proposition~\ref{prop:partial} provides bounds that identify $\mathbbm{E}[Y_2(1,1)-Y_2(0,0)|D_1=1, D_2=1, \mathbf{X}]$ if there are no late-adopters, under monotone treatment responses and bounded response (\citet{molinari2010missing}). Even under these strict assumptions, these bounds may be wide if late-adopters are present. 

\begin{proof} Assume monotone treatment responses $\mathbb{E}[Y_2(1,0)-Y_2(0,0)|D_1=1,D_2=0,\mathbf{X}]\geq0$ to construct a lower bound on $\mathbb{E}[Y_2(1,1)-Y_2(0,0)|D_2=1,\mathbf{X}]$:
{\normalsize \begin{align}
 	&\mathbbm{E}[\Delta Y|D_2=1,\mathbf{X}]-\mathbbm{E}[\Delta Y|D_2=0,\mathbf{X}] \nonumber \\
    &= \mathbb{E}[Y_2(1,1)-Y_2(0,0)|D_1=1, D_2=1,\mathbf{X}]\cdot \mathbbm{P}(D_1=1|D_2=1,\mathbf{X}) \nonumber \\
 	&+\mathbb{E}[Y_2(0,1)-Y_2(0,0)|D_1=0, D_2=1,\mathbf{X}]\cdot \mathbbm{P}(D_1=0|D_2=1,\mathbf{X})\nonumber \\
 	&- \mathbb{E}[Y_2(1,0)-Y_2(0,0)|D_1=1, D_2=0,\mathbf{X}]\cdot \mathbbm{P}(D_1=1|D_2=0,\mathbf{X}) \nonumber \\
 	&\leq \mathbb{E}[Y_2(1,1)-Y_2(0,0)|D_1=1, D_2=1,\mathbf{X}]\cdot \mathbbm{P}(D_1=1|D_2=1,\mathbf{X})\nonumber \\
 	&+ \mathbb{E}[Y_2(0,1)-Y_2(0,0)|D_1=0, D_2=1,\mathbf{X}]\cdot \mathbbm{P}(D_1=0|D_2=1,\mathbf{X}).
\end{align}}
Note that $\mathbb{E}[Y_2|D_2=0,\mathbf{X}]=\mathbb{E}[Y_2(0,0)|D_1=0,D_2=0,\mathbf{X}]\mathbbm{P}(D_1=0|D_2=0,\mathbf{X})+\mathbb{E}[Y_2(1,0)|D_1=1,D_2=0,\mathbf{X}]\mathbbm{P}(D_1=1|D_2=0,\mathbf{X})$. Assume a lower bound $Y_{\min}\leq Y$ to construct an upper bound on $\mathbb{E}[Y_2(1,1)-Y_2(0,0)|D_2=1,\mathbf{X}]$:
{\normalsize\begin{align}
 	&\mathbbm{E}[\Delta Y|D_2=1,\mathbf{X}]-\mathbbm{E}[\Delta Y|D_2=0,\mathbf{X}]+\mathbb{E}[Y_2|D_2=0,\mathbf{X}]-Y_{\min} \nonumber \\
    &=\mathbb{E}[Y_2(1,1)-Y_2(0,0)|D_1=1, D_2=1,\mathbf{X}]\cdot \mathbbm{P}(D_1=1|D_2=1,\mathbf{X})  \nonumber \\
 	&+\mathbb{E}[Y_2(0,1)-Y_2(0,0)|D_1=0, D_2=1,\mathbf{X}]\cdot \mathbbm{P}(D_1=0|D_2=1,\mathbf{X}) \nonumber\\
    &+\mathbb{E}[Y_2(0,0)-Y_{\min}|D_2=0] \nonumber \\
 	&\geq\mathbb{E}[Y_2(1,1)-Y_2(0,0)|D_1=1, D_2=1,\mathbf{X}]\cdot \mathbbm{P}(D_1=1|D_2=1,\mathbf{X}) \nonumber\\
    &+ \mathbb{E}[Y_2(0,1)-Y_2(0,0)|D_1=0, D_2=1,\mathbf{X}] \cdot \mathbbm{P}(D_1=0|D_2=1,\mathbf{X}).
\end{align}}
\end{proof}
\section{Asymptotic expansions of terms in \ref{asy_expand}}\label{sec:inferencedetails}
We can expand the other seven terms in the asymptotic distribution of 
{\small  \begin{align}\label{t2_final}
		&\frac{1}{\sqrt{n}}\sum_{i=1}^{n}\left(\widehat{w}_{i1}(\bm{\widehat{\delta}}_{d_2})\mu_i(\bm{\widehat{\beta}}_{\mathbf{d^\prime}})-\mathbb{E}[w_1(\bm{\delta^\ast}_{d_2})\mu(\bm{\beta^\ast}_{\mathbf{d^\prime}})]\right)  = \frac{1}{\sqrt{n}}\sum_{i=1}^{n}\bigg\{w_{i1}(\bm{\delta^\ast}_{d_2})(\mu_i(\bm{\beta^\ast}_{\mathbf{d^\prime}})-\mathbb{E}[w_1(\bm{\delta^\ast}_{d_2})\mu(\bm{\beta^\ast}_{\mathbf{d^\prime}})]) \nonumber \\
		&+\mathbf{b}_{i\bm{\delta}_{d_2}}^\prime\cdot \mathbb{E}[\bm{\dot{w}}_1(\bm{\delta^\ast}_{d_2})(\mu(\bm{\beta^\ast}_{\mathbf{d^\prime}})-\mathbb{E}[w_1(\bm{\delta^\ast}_{d_2})\mu(\bm{\beta^\ast}_{\mathbf{d^\prime}})])] +\mathbf{b}^\prime_{i\bm{\beta}_{\mathbf{d^\prime}}}\cdot\mathbb{E}[w_1(\bm{\delta^\ast}_{d_2})\bm{\dot{\mu}}(\bm{\beta^\ast}_{\mathbf{d^\prime}})]\bigg\}+o_p(1); 
	\end{align}}
	{\footnotesize \begin{align}\label{t3_final}
		&\frac{1}{\sqrt{n}}\sum_{i=1}^{n}\left(\widehat{w}_{i2}(\bm{\widehat{\gamma}}, \bm{\widehat{\delta}}_{d_2^\prime})\Delta Y_i-\mathbb{E}[w_2(\bm{\gamma^\ast}, \bm{\delta^\ast}_{d_2^\prime})\Delta Y]\right)  = \frac{1}{\sqrt{n}}\sum_{i=1}^{n}\bigg\{w_{i2}(\bm{\gamma^\ast}, \bm{\delta^\ast}_{d_2^\prime})(\Delta Y_i-\mathbb{E}[w_2(\bm{\gamma^\ast}, \bm{\delta^\ast}_{d_2^\prime})\Delta Y])\nonumber \\
		&+\mathbf{b}^\prime_{i\bm{\gamma}_{d_1|d_2}}\cdot \mathbb{E}[\bm{\dot{w}}_{2,\bm{\gamma}_{d_1|d_2}}(\bm{\gamma^\ast},\bm{\delta^\ast}_{d_2^\prime})(\Delta Y-\mathbb{E}[w_2(\bm{\gamma^\ast}, \bm{\delta^\ast}_{d_2^\prime})\Delta Y])]+\mathbf{b}^\prime_{i\bm{\gamma}_{d_2}}\cdot \mathbb{E}[\bm{\dot{w}}_{2,\bm{\gamma}_{d_2}}(\bm{\gamma^\ast},\bm{\delta^\ast}_{d_2^\prime})(\Delta Y-\mathbb{E}[w_2(\bm{\gamma^\ast}, \bm{\delta^\ast}_{d_2^\prime})\Delta Y])] \nonumber \\ &+\mathbf{b}^\prime_{i\bm{\gamma}_{\mathbf{d^\prime}}}\cdot \mathbb{E}[\bm{\dot{w}}_{2,\bm{\gamma}_{\mathbf{d^\prime}}}(\bm{\gamma^\ast},\bm{\delta^\ast}_{d_2^\prime})(\Delta Y-\mathbb{E}[w_2(\bm{\gamma^\ast}, \bm{\delta^\ast}_{d_2^\prime})\Delta Y])] + \mathbf{b}^\prime_{i\bm{\delta}_{d_2^\prime}}\cdot \mathbb{E}[\bm{\dot{w}}_{2,\bm{\delta}_{d_2^\prime}}(\bm{\gamma^\ast}, \bm{\delta^\ast}_{d_2^\prime})(\Delta Y-\mathbb{E}[w_2(\bm{\gamma^\ast}, \bm{\delta^\ast}_{d_2^\prime})\Delta Y])]\bigg\}\nonumber \\
        &+o_p(1);
		\end{align}}
		{\footnotesize \begin{align}\label{t4_final}
			&\frac{1}{\sqrt{n}}\sum_{i=1}^{n}\left(\widehat{w}_{i2}(\bm{\widehat{\gamma}},\bm{\widehat{\delta}}_{d_2^\prime})\mu_i(\bm{\widehat{\beta}}_{\mathbf{d^\prime}})-\mathbb{E}[w_2(\bm{\gamma^\ast},\bm{\delta^\ast}_{d_2^\prime})\mu(\bm{\beta^\ast}_{\mathbf{d^\prime}})]\right) = \frac{1}{\sqrt{n}}\sum_{i=1}^{n}\bigg\{w_{i2}(\bm{\gamma^\ast}, \bm{\delta^\ast}_{d_2^\prime})(\mu_i(\bm{\beta^\ast}_{\mathbf{d^\prime}})-\mathbb{E}[w_2(\bm{\gamma^\ast},\bm{\delta^\ast}_{d_2^\prime})\mu(\bm{\beta^\ast}_{\mathbf{d^\prime}})])\nonumber \\
            &+\mathbf{b}^\prime_{i\bm{\gamma}_{d_1|d_2}}\cdot \mathbb{E}[\bm{\dot{w}}_{2,\bm{\gamma}_{d_1|d_2}}(\bm{\gamma^\ast}, \bm{\delta^\ast}_{d_2^\prime})(\mu(\bm{\beta^\ast}_{\mathbf{d^\prime}})-\mathbb{E}[w_2(\bm{\gamma^\ast}, \bm{\delta^\ast}_{d_2^\prime})\mu(\bm{\beta^\ast}_{\mathbf{d^\prime}})])] \nonumber \\
			&+\mathbf{b}^\prime_{i\bm{\gamma}_{d_2}}\cdot \mathbb{E}[\bm{\dot{w}}_{2,\bm{\gamma}_{d_2}}(\bm{\gamma^\ast}, \bm{\delta^\ast}_{d_2^\prime})(\mu(\bm{\beta^\ast}_{\mathbf{d^\prime}})-\mathbb{E}[w_2(\bm{\gamma^\ast}, \bm{\delta^\ast}_{d_2^\prime})\mu(\bm{\beta^\ast}_{\mathbf{d^\prime}})])] +\mathbf{b}^\prime_{i\bm{\gamma}_{\mathbf{d^\prime}}}\cdot \mathbb{E}[\bm{\dot{w}}_{2,\bm{\gamma}_{\mathbf{d^\prime}}}(\bm{\gamma^\ast}, \bm{\delta^\ast}_{d_2^\prime})(\mu(\bm{\beta^\ast}_{\mathbf{d^\prime}}) \nonumber \\
            &-\mathbb{E}[w_2(\bm{\gamma^\ast}, \bm{\delta^\ast}_{d_2^\prime})\mu(\bm{\beta^\ast}_{\mathbf{d^\prime}})])] + \mathbf{b}^\prime_{i\bm{\delta}_{d_2^\prime}}\cdot \mathbb{E}[\bm{\dot{w}}_{2,\bm{\delta}_{d_2^\prime}}(\bm{\gamma^\ast}, \bm{\delta^\ast}_{d_2^\prime})(\mu(\bm{\beta^\ast}_{\mathbf{d^\prime}})-\mathbb{E}[w_2(\bm{\gamma^\ast},\bm{\delta^\ast}_{d_2^\prime})\mu(\bm{\beta^\ast}_{\mathbf{d^\prime}})])] \nonumber \\
            &+\mathbf{b}^\prime_{i\bm{\beta}_{\mathbf{d^\prime}}}\cdot \mathbb{E}[w_2(\bm{\gamma^\ast}, \bm{\delta^\ast}_{d_2^\prime})\bm{\dot{\mu}}(\bm{\beta^\ast}_{\mathbf{d^\prime}})]\bigg\}+o_p(1);	
		\end{align}}
		{\footnotesize \begin{align} \label{t5_final} 
				&\frac{1}{\sqrt{n}}\sum_{i=1}^{n}\left(\widehat{w}_{i3}(\bm{\widehat{\gamma}}_{d_1|d_2})\mu_{i}(\bm{\widehat{\beta}}_{\mathbf{d}})-\mathbb{E}[w_3(\bm{\gamma^\ast}_{d_1|d_2})\mu(\bm{\beta^\ast}_{\mathbf{d}})]\right)= \frac{1}{\sqrt{n}}\sum_{i=1}^{n}\bigg\{w_{i3}(\bm{\gamma^\ast}_{d_1|d_2})(\mu_i(\bm{\beta^\ast}_{\mathbf{d}})-\mathbb{E}[w_3(\bm{\gamma^\ast}_{d_1|d_2})\mu(\bm{\beta^\ast}_{\mathbf{d}})]) \nonumber \\
				&+\mathbf{b}^\prime_{i\bm{\gamma}_{d_1|d_2}}\cdot\mathbb{E}[\bm{\dot{w}}_3(\bm{\gamma^\ast}_{d_1|d_2})(\mu(\bm{\beta^\ast}_{\mathbf{d}})-\mathbb{E}[w_3(\bm{\gamma^\ast}_{d_1|d_2})\mu(\bm{\beta^\ast}_{\mathbf{d}})])]+\mathbf{b}^\prime_{i\bm{\beta}_{\mathbf{d}}}\cdot\mathbb{E}[w_3(\bm{\gamma^\ast}_{d_1|d_2})\bm{\dot{\mu}}(\bm{\beta^\ast}_{\mathbf{d}})]\bigg\}+o_p(1);
		\end{align}}
		{\small \begin{align} \label{t6_final}
			&\frac{1}{\sqrt{n}}\sum_{i=1}^{n}\left(\widehat{w}_{i3}(\bm{\widehat{\gamma}}_{d_1|d_2})\mu_i(\bm{\widehat{\beta}}_{\mathbf{d^\prime}})-\mathbb{E}[w_3(\bm{\gamma^\ast}_{d_1|d_2})\mu(\bm{\beta^\ast}_{\mathbf{d^\prime}})]\right) =\frac{1}{\sqrt{n}}\sum_{i=1}^{n}\bigg\{w_{i3}(\bm{\gamma^\ast}_{d_1|d_2})(\mu_i(\bm{\beta^\ast}_{\mathbf{d^\prime}}) \nonumber \\
            &-\mathbb{E}[w_3(\bm{\gamma^\ast}_{d_1|d_2})\mu(\bm{\beta^\ast}_{\mathbf{d^\prime}})]) 
			+\mathbf{b}^\prime_{i\bm{\gamma}_{d_1|d_2}}\cdot\mathbb{E}[\bm{\dot{w}}_3(\bm{\gamma^\ast}_{d_1|d_2})(\mu(\bm{\beta^\ast}_{\mathbf{d^\prime}})-\mathbb{E}[w_3(\bm{\gamma^\ast}_{d_1|d_2})\mu(\bm{\beta^\ast}_{\mathbf{d^\prime}})])]\nonumber \\
            &+\mathbf{b}^\prime_{i\bm{\beta}_{\mathbf{d^\prime}}}\cdot\mathbb{E}[w_3(\bm{\gamma^\ast}_{d_1|d_2})\bm{\dot{\mu}}(\bm{\beta^\ast}_{\mathbf{d^\prime}})]\bigg\}+o_p(1);
	\end{align}}
	{\small \begin{align} \label{t7_final} 
			&\frac{1}{\sqrt{n}}\sum_{i=1}^{n}\left(\widehat{w}_{i4}(\bm{\widehat{\gamma}}_{d_1|d_2}, \bm{\widehat{\delta}}_{d_2})\mu_i(\bm{\widehat{\beta}}_{\mathbf{d}})-\mathbb{E}[w_4(\bm{\gamma^\ast}_{d_1|d_2},\bm{\delta^\ast}_{d_2})\mu(\bm{\beta^\ast}_{\mathbf{d}})]\right) = \frac{1}{\sqrt{n}}\sum_{i=1}^{n}\bigg\{w_{i4}(\bm{\gamma^\ast}_{d_1|d_2}, \bm{\delta^\ast}_{d_2})(\mu_i(\bm{\beta^\ast}_{\mathbf{d}})\nonumber \\
			&-\mathbb{E}[w_4(\bm{\gamma^\ast}_{d_1|d_2}, \bm{\delta^\ast}_{d_2})\mu(\bm{\beta^\ast}_{\mathbf{d}})]) +\mathbf{b}^\prime_{i\bm{\gamma}_{d_1|d_2}}\cdot\mathbb{E}[\bm{\dot{w}}_{4,\bm{\gamma}_{d_1|d_2}}(\bm{\gamma^\ast}_{d_1|d_2}, \bm{\delta^\ast}_{d_2})(\mu(\bm{\beta^\ast}_{\mathbf{d}})-\mathbb{E}[w_4(\bm{\gamma^\ast}_{d_1|d_2}, \bm{\delta^\ast}_{d_2})\mu(\bm{\beta^\ast}_{\mathbf{d}})])] \nonumber \\
			&+\mathbf{b}^\prime_{i\bm{\delta}_{d_2}}\cdot \mathbb{E}[\bm{\dot{w}}_{4,\bm{\delta}_{d_2}}(\bm{\gamma^\ast}_{d_1|d_2}, \bm{\delta^\ast}_{d_2})(\mu(\bm{\beta^\ast}_{\mathbf{d}})-\mathbb{E}[w_4(\bm{\gamma^\ast}_{d_1|d_2}, \bm{\delta^\ast}_{d_2})\mu(\bm{\beta^\ast}_{\mathbf{d}})])] +\mathbf{b}^\prime_{i\bm{\beta}_{\mathbf{d}}}\cdot\mathbb{E}[w_4(\bm{\gamma^\ast}_{d_1|d_2}, \bm{\delta^\ast}_{d_2})\bm{\dot{\mu}}(\bm{\beta^\ast}_{\mathbf{d}})]\bigg\}\nonumber \\
            &+o_p(1);
		\end{align}}
	{\small \begin{align} \label{t8_final}
			&\frac{1}{\sqrt{n}}\sum_{i=1}^{n}\left(\widehat{w}_{i4}(\bm{\widehat{\gamma}}_{d_1|d_2},\bm{\widehat{\delta}}_{d_2})\mu_i(\bm{\widehat{\beta}}_{\mathbf{d^\prime}})-\mathbb{E}[w_4(\bm{\gamma^\ast}_{d_1|d_2}, \bm{\delta^\ast}_{d_2})\mu(\bm{\beta^\ast}_{\mathbf{d^\prime}})]\right) 
			= \frac{1}{\sqrt{n}}\sum_{i=1}^{n}\bigg\{w_{i4}(\bm{\gamma^\ast}_{d_1|d_2}, \bm{\delta^\ast}_{d_2})(\mu_i(\bm{\beta^\ast}_{\mathbf{d^\prime}}) \nonumber \\
			&-\mathbb{E}[w_4(\bm{\gamma^\ast}_{d_1|d_2}, \bm{\delta^\ast}_{d_2})\mu(\bm{\beta^\ast}_{\mathbf{d}^\prime})])+\mathbf{b}^\prime_{i\bm{\gamma}_{d_1|d_2}}\cdot\mathbb{E}[\bm{\dot{w}}_{4,\bm{\gamma}_{d_1|d_2}}(\bm{\gamma^\ast}_{d_1|d_2}, \bm{\delta^\ast}_{d_2})(\mu(\bm{\beta^\ast}_{\mathbf{d^\prime}})-\mathbb{E}[w_4(\bm{\gamma^\ast}_{d_1|d_2}, \bm{\delta^\ast}_{d_2})\mu(\bm{\beta^\ast}_{\mathbf{d^\prime}})])] \nonumber \\ &+\mathbf{b}^\prime_{i\bm{\delta}_{d_2}}\cdot\mathbb{E}[\bm{\dot{w}}_{4,\bm{\delta}_{d_2}}(\bm{\gamma^\ast}_{d_1|d_2}, \bm{\delta^\ast}_{d_2})(\mu(\bm{\beta^\ast}_{\mathbf{d^\prime}})-\mathbb{E}[w_4(\bm{\gamma^\ast}_{d_1|d_2}, \bm{\delta^\ast}_{d_2})\mu(\bm{\beta^\ast}_{\mathbf{d^\prime}})])] +\mathbf{b}^\prime_{i\bm{\beta}_{\mathbf{d^\prime}}}\cdot\mathbb{E}[w_4(\bm{\gamma^\ast}_{d_1|d_2}, \bm{\delta^\ast}_{d_2})\bm{\dot{\mu}}(\bm{\beta^\ast}_{\mathbf{d^\prime}})]\bigg\}\nonumber \\
            &+o_p(1).
	\end{align}}

\section{Inference-robust alternatives}\label{sec:inferencerobust}
We aim to find estimators that do not affect the asymptotic behavior of the robust estimator, even when one of the models is misspecified. To find such estimators, we develop some notation. Note that with parametric working models, each weight in \eqref{eq:weights} can be written as $w_j(\bm\theta)=f_j(\bm\theta)/\mathbb{E}[f_j(\bm\theta)]$, the estimated weights in \eqref{eq:weightSat} as $\widehat{w}_j(\bm\theta)=f_j(\bm\theta)/\mathbb{E}_n[f_j(\bm\theta)]$, and we also define $\widetilde{w}_j(\bm\theta)=f_j(\bm\theta)/\mathbb{E}_n[f_j(\bm{\theta^\ast})]$, where $\bm\theta=(\bm\beta,\bm\gamma,\bm\delta)$ and $\bm{\theta^\ast}=(\bm{\beta^\ast},\bm{\gamma^\ast},\bm{\delta^\ast})$. Define
{\small \begin{align}
	\psi(\bm{W},\bm\theta) =
	& \widetilde{w}_1(\bm{\delta}_{d_2})\bigg(\Delta Y-\mu(\bm{\beta}_{\mathbf{d^\prime}})-\mathbb{E}[w_1(\bm{\delta^\ast}_{d_2})(\Delta Y-\mu(\bm{\beta^\ast}_{\mathbf{d^\prime}}))]\bigg)- \nonumber \\
	&\widetilde{w}_2(\bm{\gamma}, \bm{\delta}_{d_2^\prime})\bigg(\Delta Y-\mu(\bm{\beta}_{\mathbf{d^\prime}}) - \mathbb{E}[w_2(\bm{\gamma^\ast}, \bm{\delta^\ast}_{d_2^\prime})(\Delta Y-\mu(\bm{\beta^\ast}_{\mathbf{d^\prime}}))]\bigg) + \nonumber\\ 
	&\widetilde{w}_3(\bm{\gamma}_{d_1|d_2})\bigg(\mu(\bm{\beta}_{\mathbf{d}})-\mu(\bm{\beta}_{\mathbf{d^\prime}})-\mathbb{E}[w_3(\bm{\gamma^\ast}_{d_1|d_2})(\mu(\bm{\beta^\ast}_{\mathbf{d}})-\mu(\bm{\beta^\ast}_{\mathbf{d^\prime}}))]\bigg)- \nonumber \\
	&\widetilde{w}_4(\bm{\gamma}_{d_1|d_2}, \bm{\delta}_{d_2})\bigg(\mu(\bm{\beta}_{\mathbf{d}})-\mu(\bm{\beta}_{\mathbf{d^\prime}}) - \mathbb{E}[w_4(\bm{\gamma^\ast}_{d_1|d_2}, \bm{\delta^\ast}_{d_2})(\mu_{\mathbf{d}}(\bm{\beta^\ast})-\mu(\bm{\beta^\ast}_{\mathbf{d^\prime}}))]\bigg).
\end{align}}
The following result follows directly from the proof of Theorem~\ref{thm:asyvar}:

\begin{corollary}[Estimation effect of the working models]\label{cor:improved}\, \\
	Under Assumptions \ref{ass:did}-\ref{ass:rs}, conditions 1-5 in Supplementary Appendix \ref*{sec:conditions}, and provided that either 
	$\mu(\bm{\beta}_\mathbf{d}^\ast)=m_{\mathbf{d}}(\mathbf{X})$ and $\pi(\bm{\gamma^\ast}_{\mathbf{d}})=p_{\mathbf{d}}(\mathbf{X})$;
	$\phi(\bm{\delta^\ast}_{d_2})=q_{d_2}(\mathbf{X})$ and 
	$\pi(\bm{\gamma^\ast}_{\mathbf{d}})=p_{\mathbf{d}}(\mathbf{X})$; or
	$\phi(\bm{\delta}_{d_2}^\ast)=q_{d_2}(\mathbf{X})$ and 
	$\mu(\bm{\beta}_\mathbf{d}^\ast)=m_{\mathbf{d}}(\mathbf{X})$
	, 
	as $n\rightarrow \infty$, 
	$$\xi(\bm{W},\bm{\theta^\ast}) = \psi(\bm{W},\bm{\theta^\ast}) \text{ and } \Omega  = \mathbb{E}[\psi(\mathbf{W},\bm{\theta^\ast})^2],$$
	with $\bm\theta^\ast$ a solution to $\mathbb{E}[\partial\psi(\bm{W},\bm\theta^\ast)/\partial\bm\theta]=\bm0$.
\end{corollary}
Corollary~\ref{cor:improved} shows that the asymptotic variance of $\widehat{\tau}^\textup{R}_{\mathbf{dd^\prime}}$ does not depend on the estimators of the working models when their parameters follow from $\mathbb{E}[\partial\psi(\bm{W},\bm\theta^\ast)/\partial\bm\theta]=\bm0$. Hence, this result suggests that we can solve $\mathbb{E}_n[\partial\psi(\bm{W},\bm{\hat{\theta}})/\partial\bm{\theta}]=\bm0$ for $\bm{\hat{\theta}}$ to obtain estimates for $\bm\theta$. This procedure is similar to the improved estimation proposed in \citet{sant2020doubly}. \citet{vermeulen2015bias} consider robust estimators with all working models misspecified, and use a procedure similar to the one described here to reduce the squared asymptotic bias of $\widehat{\tau}^\textup{R}_{\mathbf{dd^\prime}}$. Since Corollary~\ref{cor:improved} assumes that at least two of the three working models are correct, this bias is zero under our assumptions. 

Intuitively, when the asymptotic variance of $\widehat{\tau}^\textup{R}_{\mathbf{dd^\prime}}$ does not depend on the estimators of the working models, this robust estimator is more efficient compared to a robust estimator that relies on first-stage estimators without this property. Indeed, we find that this estimation strategy improves over standard maximum likelihood estimation in simulations. However, there are no theoretical guarantees: different estimators, $\bm{\hat{\theta}}$, may have different pseudo-true values, $\bm{\theta^\ast}$, with a different variance expression for $\psi(\bm{W},\bm{\theta^\ast})$, or the variance of the estimand itself may be smaller at estimated instead of true parameter values.

For the sake of illustration, consider the commonly used working models, $\mu_{\mathbf{d}}(\mathbf{X}) = \mathbf{X}\bm{\beta}_{\mathbf{d}}$, $\phi(\mathbf{X}; \bm{\delta}_{d_2}) = \Lambda(\mathbf{X}\bm{\delta}_{d_2})$, $\pi(\mathbf{X}, \bm{\gamma}_{d_1|d_2}) =\Lambda(\mathbf{X}\bm{\gamma}_{d_1|d_2})$ and $\pi(\mathbf{X}, \bm{\gamma}_{d_2}) = \Lambda(\mathbf{X}\bm{\gamma}_{d_2})$. We have seven parameters to estimate: $\bm{\beta}_{\mathbf{d}}, \bm{\beta}_{\mathbf{d^\prime}}, \bm{\delta}_{d_2}, \bm{\delta}_{d_2^\prime}, \bm{\gamma}_{d_1|d_2}, \bm{\gamma}_{d_1^\prime|d_2^\prime}$, and $\bm{\gamma}_{d_2}$. 

The first approach is a stepwise algorithm outlined below.
\begin{algorithm}[H]
\setstretch{1.5}
{
    \begin{algorithmic}[1]
    \Procedure{Estimate model parameters}{}
        \State{$\bm{\hat{\gamma}}_{d_1|d_2} = \arg\max_{\bm{\gamma}_{d_1|d_2}} \mathbb{E}_n\left[\mathbbm{1}[D_1=d_1]\mathbf{X}\bm{\gamma}_{d_1|d_2}-\ln(1+\exp(\mathbf{X}\bm{\gamma}_{d_1|d_2})) |S=1,D_2=d_2\right]$}
        \State{$\bm{\hat{\gamma}}_{d_1'|d_2'} = \arg\max_{\bm{\gamma}_{d_1'|d_2'}} \mathbb{E}_n\left[\mathbbm{1}[D_1=d_1']\mathbf{X}\bm{\gamma}_{d_1'|d_2'}-\ln(1+\exp(\mathbf{X}\bm{\gamma}_{d_1'|d_2'})) |S=1,D_2=d_2'\right]$}
        \State{$\bm{\hat{\gamma}}_{d_2} = \arg\max_{\bm{\gamma}_{d_2}} \mathbb{E}_n\left[\mathbbm{1}[D_2=d_2]\mathbf{X}\bm{\gamma}_{d_2}-\ln(1+\exp(\mathbf{X}\bm{\gamma}_{d_2}))\right]$}   \State{$\bm{\hat{\delta}}_{d_2}=\arg\max_{\bm{\delta}_{d_2}} \mathbb{E}_n\left[\pi(\bm{\hat{\gamma}}_{d_1|d_2})\left((S-1)\mathbf{X}\bm{\delta}_{d_2} -S\exp(-\mathbf{X}\bm{\delta}_{d_2}) \right)|D_2=d_2\right]$}
        \State{\small $\bm{\hat{\delta}}_{d_2^\prime}=\arg\max_{\bm{\delta}_{d_2'}} \mathbb{E}_n\left[
        \frac{\pi(\bm{\hat{\gamma}}_{d_1|d_2})}{\pi(\bm{\hat{\gamma}}_{d_1'|d_2'})}\frac{\pi(\bm{\hat{\gamma}}_{d_2})}{(1-\pi(\bm{\hat{\gamma}}_{d_2}))}\mathbbm{1}[\mathbf{D}=\mathbf{d'}]\left(\mathbf{X}\bm{\delta}_{d_2'}-\exp(-\mathbf{X}\bm{\delta}_{d_2'}) \right)
        -\frac{\mathbbm{1}[\mathbf{D}=\mathbf{d}]}{\phi(\bm{\hat{\delta}}_{d_2})}\mathbf{X}\bm{\delta}_{d_2'}|S=1\right]$}
        \State{\small $\bm{\hat{\beta}_{d'}}=\arg\min_{\bm{\beta_{d'}}} \mathbb{E}_n\left[\left(\Delta Y -\mathbf{\breve{X}_{d'}}\bm{\hat{\beta}}_{\mathbf{d^\prime}}\right)^2\right]$} where $\mathbf{\breve{X}_{d'}}=(\mathbf{X},\widehat{w}_2\mathbf{X},\hat{\pi}_{d_1|d_2}\widehat{w}_2\mathbf{X},\hat{\pi}_{d_1'|d_2'}\widehat{w}_2\mathbf{X},\widehat{w}_1\mathbf{X},\hat{\phi}_{d_2}\widehat{w}_1\mathbf{X})$.
        \State{$\bm{\hat{\beta}_{d}}=\arg\min_{\bm{\beta_{d}}} \mathbb{E}_n\left[\left(\mathbf{\breve{X}_d}\bm{\beta_{d}}-\mu(\bm{\hat{\beta}}_{\mathbf{d'}}) \right)^2\right]$ where $\mathbf{\breve{X}_{d}} = (\mathbf{X},\hat{\pi}_{d_1|d_2}(\widehat{w}_3-\widehat{w}_4)\mathbf{X},\widehat{w}_3\mathbf{X},\widehat{w}_4\mathbf{X},\hat{\phi}_{d_2}\widehat{w}_4\mathbf{X})$.}
    \EndProcedure
    \Procedure{predicted values}{}
    \State{$\pi(\bm{\gamma}_{d_1|d_2}) = \frac{\exp(\mathbf{X}\bm{\gamma}_{d_1|d_2})}{1+\exp(\mathbf{X}\bm{\gamma}_{d_1|d_2})}$, $\pi(\bm{{\gamma}}_{d_1'|d_2'}) = \frac{\exp(\mathbf{X}\bm{\gamma}_{d_1'|d_2'})}{1+\exp(\mathbf{X}\bm{\gamma}_{d_1'|d_2'})}$, $\pi(\bm{\gamma}_{d_2}) = \frac{\exp(\mathbf{X}\bm{\gamma}_{d_2})}{1+\exp(\mathbf{X}\bm{\gamma}_{d_2})}$}
        \State{$\phi(\bm{\delta}_{d_2}) = \frac{\exp(\mathbf{X}\bm{\delta}_{d_2})}{1+\exp(\mathbf{X}\bm{\delta}_{d_2})}$, $\phi(\bm{\delta}_{d_2'}) = \frac{\exp(\mathbf{X}\bm{\delta}_{d_2'})}{1+\exp(\mathbf{X}\bm{\delta}_{d_2'})}$}, 
        \State{$\mu(\bm{\beta}_{\mathbf{d}}) = \mathbf{\breve{X}_{d}}\bm{\beta}_{\mathbf{d}}$, $\mu(\bm{{\beta}_{\mathbf{d^\prime}}}) = \mathbf{\breve{X}_{d'}}\bm{\beta}_{\mathbf{d^\prime}}.$}
    \EndProcedure
    \Procedure{Estimated weights}{}
        \State{$\widehat{w}_1(\bm{\widehat{\delta}}_{d_2}) = {\frac{S}{\phi(\bm{\widehat{\delta}}_{d_2})}\mathbbm{1}[\mathbf{D}=\mathbf{d}]}/{\mathbb{E}_n\left[\frac{S}{\phi(\bm{\widehat{\delta}}_{d_2})}\mathbbm{1}[\mathbf{D}=\mathbf{d}]\right]}$}
        \State{$\widehat{w}_2(\bm{\widehat{\gamma}},\bm{\widehat{\delta}}_{d_2'}) ={\frac{S}{\phi(\bm{\widehat{\delta}}_{d_2'})}\frac{\pi(\bm{\widehat{\gamma}}_{d_1|d_2})}{\pi(\bm{\widehat{\gamma}}_{d_1'|d_2'})}\frac{\pi(\bm{\widehat{\gamma}}_{d_2})}{(1-\pi(\bm{\widehat{\gamma}}_{d_2}))}\mathbbm{1}[\mathbf{D}=\mathbf{d}']}/{\mathbb{E}_n\left[\frac{S}{\phi(\bm{\widehat{\delta}}_{d_2'})}\frac{\pi(\bm{\widehat{\gamma}}_{d_1|d_2})}{\pi(\bm{\widehat{\gamma}}_{d_1'|d_2'})}\frac{\pi(\bm{\widehat{\gamma}}_{d_2})}{(1-\pi(\bm{\widehat{\gamma}}_{d_2}))}\mathbbm{1}[\mathbf{D}=\mathbf{d}']\right]}$}
        \State{$\widehat{w}_3(\bm{\widehat{\gamma}}_{d_1|d_2}) = {\pi(\bm{\widehat{\gamma}}_{d_1|d_2})\mathbbm{1}[D_2=d_2]}/{\mathbb{E}_n\left[\pi(\bm{\widehat{\gamma}}_{d_1|d_2})\mathbbm{1}[D_2=d_2]\right]}$}
        \State{$\widehat{w}_4(\bm{\widehat{\gamma}}_{d_1|d_2},\bm{\widehat{\delta}}_{d_2}) = {\frac{S}{\phi(\bm{\widehat{\delta}}_{d_2})}\pi(\bm{\widehat{\gamma}}_{d_1|d_2})\mathbbm{1}[D_2=d_2]}/{\mathbb{E}_n\left[\frac{S}{\phi(\bm{\widehat{\delta}}_{d_2})}\pi(\bm{\widehat{\gamma}}_{d_1|d_2})\mathbbm{1}[D_2=d_2]\right]}$}
        \EndProcedure
        \Procedure{Estimate PDATT, its variance, and confidence interval}{}
        \State{$\widehat{\tau}_{\mathbf{dd^\prime}}^{\textup{R}}=\mathbb{E}_n[  
	 	(\widehat{w}_1(\bm{\widehat{\delta}}_{d_2})-\widehat{w}_2(\bm{\widehat{\gamma}},\bm{\widehat{\delta}}_{d_2'}))(\Delta Y -\mu(\bm{\widehat{\beta}}_{\mathbf{d^\prime}}))+(\widehat{w}_3(\bm{\widehat{\gamma}}_{d_1|d_2})-\widehat{w}_4(\bm{\widehat{\gamma}}_{d_1|d_2},\bm{\widehat{\delta}}_{d_2}))( \mu(\bm{\widehat{\beta}}_{\mathbf{d}})-\mu(\bm{\widehat{\beta}}_{\mathbf{d^\prime}}))]$}
        \State{$\widehat{\mathbb{V}}[\widehat{\tau}_{\mathbf{dd^\prime}}^{\textup{R}}]=  \mathbb{E}_n[( 
     \widehat{w}_1(\bm{\widehat{\delta}}_{d_2})(\Delta Y-\mu(\bm{\widehat{\beta}}_{\mathbf{d^\prime}})-\widehat{\tau}_{\mathbf{dd^\prime}}) 
			- \widehat{w}_2(\bm{\widehat{\gamma}},\bm{\widehat{\delta}}_{d_2'})  (\Delta Y -\mu(\bm{\widehat{\beta}}_{\mathbf{d^\prime}})) + 
   (\widehat{w}_3(\bm{\widehat{\gamma}}_{d_1|d_2})-\widehat{w}_4(\bm{\widehat{\gamma}}_{d_1|d_2},\bm{\widehat{\delta}}_{d_2}))(\mu(\bm{\widehat{\beta}}_{\mathbf{d}})-\mu(\bm{\widehat{\beta}}_{\mathbf{d^\prime}})-\widehat{\tau}_{\mathbf{dd^\prime}}))^2 ] $}
   \State{CI$=\widehat{\tau}_{\mathbf{dd^\prime}}^{\textup{R}} \pm z_{\alpha/2}\sqrt{\widehat{\mathbb{V}}[\widehat{\tau}_{\mathbf{dd^\prime}}^{\textup{R}}]}$}
   \EndProcedure
    \end{algorithmic}}
    \caption{Inference-robust estimator for PDATT}
    \label{alg:inference}
\end{algorithm}

In the second approach, we stack the first-order conditions that need to be satisfied by the first-stage parameter estimates into a joint GMM problem. Then, consider the following first-order conditions:
{\small \begin{align*}
		\mathbb{E}(\Psi_{\bm{\beta}_{\mathbf{d}}}) &= \mathbb{E}\left[\left(w_3(\bm{\gamma^\ast}_{d_1|d_2})-w_4(\bm{\gamma^\ast}_{d_1|d_2}, \bm{\delta^\ast}_{d_2})\right)\mathbf{X^\prime}\right] = \mathbf{0}; \\
		\mathbb{E}(\Psi_{\bm{\beta}_{\mathbf{d^\prime}}}) &= \mathbb{E}\left[\left(w_1(\bm{\delta^\ast}_{d_2})-w_2(\bm{\gamma^\ast}, \bm{\delta^\ast}_{d_2^\prime})+w_3(\bm{\gamma^\ast}_{d_1|d_2})-w_4(\bm{\gamma^\ast}_{d_1|d_2}, \bm{\delta^\ast}_{d_2})\right)\mathbf{X^\prime}\right] = \mathbf{0}; \\
		\mathbb{E}(\Psi_{\bm{\gamma}_{d_1|d_2}}) & = \mathbb{E}\bigg[w_2(\bm{\gamma^\ast}, \bm{\delta^\ast}_{d_2^\prime})\frac{\bm{\dot{\pi}}(\bm{\gamma^\ast}_{d_1|d_2})}{\pi(\bm{\gamma^\ast}_{d_1|d_2})}\bigg(\Delta Y - \mu(\bm{\beta^\ast}_{\mathbf{d^\prime}}) - \mathbb{E}[w_2(\bm{\gamma^\ast}, \bm{\delta^\ast}_{d_2^\prime})(\Delta Y - \bm{\mu}(\bm{\beta^\ast}_{\mathbf{d^\prime}}))]\bigg)\bigg]\\
		& - \mathbb{E}\bigg[w_3(\bm{\gamma^\ast}_{d_1|d_2})\frac{\bm{\dot{\pi}}(\bm{\gamma^\ast}_{d_1|d_2})}{\pi(\bm{\gamma^\ast}_{d_1|d_2})}\bigg(\mu(\bm{\beta^\ast}_{\mathbf{d}})-\mu(\bm{\beta^\ast}_{\mathbf{d^\prime}})-\mathbb{E}[w_3(\bm{\gamma^\ast}_{d_1|d_2})(\mu(\bm{\beta^\ast}_{\mathbf{d}}) -\mu(\bm{\beta^\ast}_{\mathbf{d^\prime}}) )]\bigg)\bigg]\\
		& +\mathbb{E}\bigg[w_4(\bm{\gamma^\ast}_{d_1|d_2}, \bm{\delta^\ast}_{d_2})\frac{\bm{\dot{\pi}}(\bm{\gamma^\ast}_{d_1|d_2})}{\pi(\bm{\gamma^\ast}_{d_1|d_2})}\bigg(\mu(\bm{\beta^\ast}_{\mathbf{d}})-\mu(\bm{\beta^\ast}_{\mathbf{d^\prime}}) - \mathbb{E}[w_4(\bm{\gamma^\ast}_{d_1|d_2}, \bm{\delta^\ast}_{d_2})(\mu(\bm{\beta^\ast}_{\mathbf{d}}) - \mu(\bm{\beta^\ast}_{\mathbf{d^\prime}}))]\bigg)\bigg] =  \mathbf{0};\\
		\mathbb{E}(\Psi_{\bm{\gamma}_{d_2}})& = \mathbb{E}\bigg[w_2(\bm{\gamma^\ast}, \bm{\delta^\ast}_{d_2^\prime})\frac{\bm{\dot{\pi}}(\bm{\gamma^\ast}_{d_2})}{\pi(\bm{\gamma^\ast}_{d_2})}\bigg(\Delta Y - \mu(\bm{\beta^\ast}_{\mathbf{d^\prime}}) - \mathbb{E}[w_2(\bm{\gamma^\ast}, \bm{\delta^\ast}_{d_2^\prime})(\Delta Y - \mu(\bm{\beta^\ast}_{\mathbf{d^\prime}}))]\bigg)\bigg] = \mathbf{0};\\
		\mathbb{E}(\Psi_{\bm{\gamma}_{\mathbf{d^\prime}}}) & = -\mathbb{E}\bigg[w_2(\bm{\gamma^\ast}, \bm{\delta^\ast}_{d_2^\prime})\frac{\bm{\dot{\pi}}(\bm{\gamma^\ast}_{\mathbf{d^\prime}})}{\pi(\bm{\gamma^\ast}_{\mathbf{d^\prime}})}\bigg(\Delta Y - \mu(\bm{\beta^\ast}_{\mathbf{d^\prime}}) - \mathbb{E}[w_2(\bm{\gamma^\ast}, \bm{\delta^\ast}_{d_2^\prime})(\Delta Y - \mu(\bm{\beta^\ast}_{\mathbf{d^\prime}}))]\bigg)\bigg] = \mathbf{0}; \\
		\mathbb{E}(\Psi_{\bm{\delta}_{d_2}})& = -\mathbb{E}\bigg[w_1(\bm{\delta^\ast}_{d_2})\frac{\bm{\dot{\phi}}(\bm{\delta^\ast}_{d_2})}{\phi(\bm{\delta^\ast}_{d_2})}\bigg(\Delta Y - \mu(\bm{\beta^\ast}_{\mathbf{d^\prime}}) - \mathbb{E}[w_1(\bm{\delta^\ast}_{d_2})(\Delta Y - \mu(\bm{\beta^\ast}_{\mathbf{d^\prime}}))]\bigg)\bigg] \\
		&- \mathbb{E}\bigg[w_4(\bm{\gamma^\ast}_{d_1|d_2}, \bm{\delta^\ast}_{d_2})\frac{\bm{\dot{\phi}}(\bm{\delta^\ast}_{d_2})}{\phi(\bm{\delta^\ast}_{d_2})}\bigg(\mu(\bm{\beta^\ast}_{\mathbf{d}}) - \mu(\bm{\beta^\ast}_{\mathbf{d^\prime}}) - \mathbb{E}[w_4(\bm{\gamma^\ast}_{d_1|d_2}, \bm{\delta^\ast}_{d_2})(\mu(\bm{\beta^\ast}_{\mathbf{d}}) - \mu(\bm{\beta^\ast}_{\mathbf{d^\prime}}))]\bigg)\bigg]  = \mathbf{0};\\
		\mathbb{E}(\Psi_{\bm{\delta}_{d^\prime_2}})& = - \mathbb{E}\bigg[w_2(\bm{\gamma^\ast}, \bm{\delta^\ast}_{d_2^\prime})\frac{\bm{\dot{\phi}}(\bm{\delta^\ast}_{d_2^\prime})}{\phi(\bm{\delta^\ast}_{d_2^\prime})}\bigg(\Delta Y - \mu(\bm{\beta^\ast}_{\mathbf{d^\prime}}) - \mathbb{E}[w_2(\bm{\gamma^\ast}, \bm{\delta^\ast}_{d_2^\prime})(\Delta Y - \mu(\bm{\beta^\ast}_{\mathbf{d^\prime}}))]\bigg)\bigg]  = \mathbf{0}.
\end{align*}}
Plugging-in the respective derivatives, we get
{\small \begin{align*}
		\mathbb{E}(\Psi_{\bm{\beta}_{\mathbf{d}}}) &= \mathbb{E}\left[\left(w_3(\bm{\gamma^\ast}_{d_1|d_2})-w_4(\bm{\gamma^\ast}_{d_1|d_2}, \bm{\delta^\ast}_{d_2})\right)\mathbf{X^\prime}\right] = \mathbf{0}; \\
		\mathbb{E}(\Psi_{\bm{\beta}_{\mathbf{d^\prime}}}) &= \mathbb{E}\left[\left(w_1(\bm{\delta^\ast}_{d_2})-w_2(\bm{\gamma^\ast}, \bm{\delta^\ast}_{d_2^\prime})+w_3(\bm{\gamma^\ast}_{d_1|d_2})-w_4(\bm{\gamma^\ast}_{d_1|d_2}, \bm{\delta^\ast}_{d_2})\right)\mathbf{X^\prime}\right] = \mathbf{0}; \\
		\mathbb{E}(\Psi_{\bm{\gamma}_{d_1|d_2}}) & = \mathbb{E}\bigg[w_2(\bm{\gamma^\ast}, \bm{\delta^\ast}_{d_2^\prime})(1-\Lambda(\mathbf{X}\bm{\gamma^\ast}_{d_1|d_2}))\mathbf{X^\prime}\bigg(\Delta Y - \mu(\bm{\beta^\ast}_{\mathbf{d^\prime}}) - \mathbb{E}[w_2(\bm{\gamma^\ast}, \bm{\delta^\ast}_{d_2^\prime})(\Delta Y - \bm{\mu}(\bm{\beta^\ast}_{\mathbf{d^\prime}}))]\bigg)\bigg]\\
		& - \mathbb{E}\bigg[w_3(\bm{\gamma^\ast}_{d_1|d_2})(1-\Lambda(\mathbf{X}\bm{\gamma^\ast}_{d_1|d_2}))\mathbf{X^\prime}\bigg(\mu(\bm{\beta^\ast}_{\mathbf{d}})-\mu(\bm{\beta^\ast}_{\mathbf{d^\prime}})-\mathbb{E}[w_3(\bm{\gamma^\ast}_{d_1|d_2})(\mu(\bm{\beta^\ast}_{\mathbf{d}}) -\mu(\bm{\beta^\ast}_{\mathbf{d^\prime}}) )]\bigg)\bigg]\\
		& +\mathbb{E}\bigg[w_4(\bm{\gamma^\ast}_{d_1|d_2}, \bm{\delta^\ast}_{d_2})(1-\Lambda(\mathbf{X}\bm{\gamma^\ast}_{d_1|d_2}))\mathbf{X^\prime}\bigg(\mu(\bm{\beta^\ast}_{\mathbf{d}})-\mu(\bm{\beta^\ast}_{\mathbf{d^\prime}}) - \mathbb{E}[w_4(\bm{\gamma^\ast}_{d_1|d_2}, \bm{\delta^\ast}_{d_2})(\mu(\bm{\beta^\ast}_{\mathbf{d}}) - \mu(\bm{\beta^\ast}_{\mathbf{d^\prime}}))]\bigg)\bigg] \\
        &=  \mathbf{0};\\
		\mathbb{E}(\Psi_{\bm{\gamma}_{d_2}})& = \mathbb{E}\bigg[w_2(\bm{\gamma^\ast}, \bm{\delta^\ast}_{d_2^\prime})(1-\Lambda(\mathbf{X}\bm{\gamma^\ast}_{d_2}))\mathbf{X^\prime}\bigg(\Delta Y - \mu(\bm{\beta^\ast}_{\mathbf{d^\prime}}) - \mathbb{E}[w_2(\bm{\gamma^\ast}, \bm{\delta^\ast}_{d_2^\prime})(\Delta Y - \mu(\bm{\beta^\ast}_{\mathbf{d^\prime}}))]\bigg)\bigg] = \mathbf{0};\\
		\mathbb{E}(\Psi_{\bm{\gamma}_{\mathbf{d^\prime}}}) & = -\mathbb{E}\bigg[w_2(\bm{\gamma^\ast}, \bm{\delta^\ast}_{d_2^\prime})\frac{\bm{\dot{\pi}}(\bm{\gamma^\ast}_{\mathbf{d^\prime}})}{\pi(\bm{\gamma^\ast}_{\mathbf{d^\prime}})}\bigg(\Delta Y - \mu(\bm{\beta^\ast}_{\mathbf{d^\prime}}) - \mathbb{E}[w_2(\bm{\gamma^\ast}, \bm{\delta^\ast}_{d_2^\prime})(\Delta Y - \mu(\bm{\beta^\ast}_{\mathbf{d^\prime}}))]\bigg)\bigg] = \mathbf{0}; \\
		\mathbb{E}(\Psi_{\bm{\delta}_{d_2}})& = -\mathbb{E}\bigg[w_1(\bm{\delta^\ast}_{d_2})(1-\Lambda(\mathbf{X}\bm{\delta^\ast}_{d_2}))\mathbf{X^\prime}\bigg(\Delta Y - \mu(\bm{\beta^\ast}_{\mathbf{d^\prime}}) - \mathbb{E}[w_1(\bm{\delta^\ast}_{d_2})(\Delta Y - \mu(\bm{\beta^\ast}_{\mathbf{d^\prime}}))]\bigg)\bigg] \\
		&- \mathbb{E}\bigg[w_4(\bm{\gamma^\ast}_{d_1|d_2}, \bm{\delta^\ast}_{d_2})(1-\Lambda(\mathbf{X}\bm{\delta^\ast}_{d_2}))\mathbf{X^\prime}\bigg(\mu(\bm{\beta^\ast}_{\mathbf{d}}) - \mu(\bm{\beta^\ast}_{\mathbf{d^\prime}}) - \mathbb{E}[w_4(\bm{\gamma^\ast}_{d_1|d_2}, \bm{\delta^\ast}_{d_2})(\mu(\bm{\beta^\ast}_{\mathbf{d}}) - \mu(\bm{\beta^\ast}_{\mathbf{d^\prime}}))]\bigg)\bigg] \\
        &= \mathbf{0};
            \end{align*}}
    {\small \begin{align*}    
		\mathbb{E}(\Psi_{\bm{\delta}_{d^\prime_2}})& = - \mathbb{E}\bigg[w_2(\bm{\gamma^\ast}, \bm{\delta^\ast}_{d_2^\prime})(1-\Lambda(\mathbf{X}\bm{\delta^\ast}_{d^\prime_2}))\mathbf{X^\prime}\bigg(\Delta Y - \mu(\bm{\beta^\ast}_{\mathbf{d^\prime}}) - \mathbb{E}[w_2(\bm{\gamma^\ast}, \bm{\delta^\ast}_{d_2^\prime})(\Delta Y - \mu(\bm{\beta^\ast}_{\mathbf{d^\prime}}))]\bigg)\bigg]  = \mathbf{0}.
\end{align*}}
Then, the GMM estimator that minimizes the following objective function remains insensitive to the choice of the first-stage models being a logit (for the probabilities) and linear regression (for the conditional mean of outcome). In other words, estimation of the parameters indexing these models has no effect on the asymptotic variance of the resulting robust estimator, which solves $\bm{\hat{\theta}}_{\textup{GMM}} =  \arg\min_{\bm{\theta}}\mathbbm{E}_n[\bm{\Psi}(\bm{\theta})'\bm{\Sigma}^{-1} \ \bm{\Psi}(\bm{\theta})]$, where $\bm{\Sigma}$ is the asymptotic variance-covariance matrix of the vector of moments conditions, $\bm{\Psi}$.

\section{Numerical experiments: additional details and results}\label{sec:simsadd}

\subsection{Implementation details DR, IPW, and OR estimators}\label{sec:simsdet}
Because we are assuming a logit working model for the propensity scores and missing treatment probability, $\pi(\cdot) = \phi(\cdot)= \Lambda(\cdot)$ where $\Lambda$ is the inverse logit function. Then, $\bm{\dot{\pi}}(\cdot) = \bm{\dot{\phi}} = \Lambda(\cdot) \cdot (1-\Lambda(\cdot))$. The outcome (or conditional mean) model is assumed to be linear which means that $\mu(\mathbf{X}, \bm{\beta}_{\mathbf{d}}) = \mathbf{X}\bm{\beta}_{\mathbf{d}}$. Given these choices, for all values of $\mathbf{d}=  (d_1, d_2)$ we have
{\small\begin{align*}
		\mathbf{b}_{i\bm{\delta}_{d_2}} &=	\left\{\mathbb{E}\left[\mathbbm{1}[D_2=d_2]\mathbf{X}^\prime\mathbf{X}\lambda(\mathbf{X}\bm{\delta^\ast}_{d_2})\right]\right\}^{-1}\left\{\mathbbm{1}[D_{2i}=d_{2i}]\mathbf{X}_i^\prime(S_i-\Lambda(\mathbf{X}_i\bm{\delta^\ast}))\right\}; \\
		\mathbf{b}_{i\bm{\beta}_{\mathbf{d}}} &= \{\mathbb{E}[S\mathbbm{1}[\mathbf{D}=\mathbf{d}]\mathbf{X}^\prime\mathbf{X}]\}^{-1}\left\{S_i\mathbbm{1}[\mathbf{D}_i=\mathbf{d}_i]\mathbf{X}_i^\prime \left(\Delta Y_i - \mathbf{X}_i\bm{\beta^\ast}_{\mathbf{d}}\right)\right\}; \\
		\mathbf{b}_{i\bm{\gamma}_{d_1|d_2}}& = \left\{\mathbb{E}\left[S\mathbbm{1}[D_2=d_2]\mathbf{X}^\prime\mathbf{X}\lambda(\mathbf{X}\bm{\gamma^\ast}_{d_1|d_2})\right]\right\}^{-1}\left\{S_i\mathbbm{1}[D_{2i}=d_{2i}]\mathbf{X}_i^\prime(D_{1i}-\Lambda(\mathbf{X}_i\bm{\gamma^\ast}_{d_1|d_2}))\right\}; 
        \end{align*}}
        {\small \begin{align*}
		\mathbf{b}_{i\bm{\gamma}_{d_2}}& = \left\{\mathbb{E}\left[\mathbf{X}^\prime\mathbf{X}\lambda(\mathbf{X}\bm{\gamma^\ast}_{d_2})\right]\right\}^{-1}\left\{\mathbf{X}_i^\prime(D_{2i}-\Lambda(\mathbf{X}_i\bm{\gamma^\ast}_{d_2}))\right\}.
\end{align*}} 
We also consider the OR, IPW, and DR estimators discussed in Section \ref{sec:adjusted}. To compare with the robust estimator, we consider their normalized versions which are given as 
{\small \begin{align*}
		\sqrt{n}\left(\widehat{\tau}^\textup{OR}_{\mathbf{dd^\prime}}-\tau^{\textup{OR}}_{\mathbf{dd^\prime}}\right) &= \frac{1}{\sqrt{n}}\sum_{i=1}^{n}\bigg\{\psi^{\textup{OR}}_i -\mathbf{b}^\prime_{i\bm{\beta}_{\mathbf{d^\prime}}} \mathbb{E}(\Psi^{\textup{OR}}_{\bm{\beta}_{\mathbf{d^\prime}}}) +\mathbf{b}^\prime_{i\bm{\delta}_{d_2}} \mathbb{E}(\Psi^{\textup{OR}}_{\bm{\delta}_{d_2}})\bigg\} + o_p(1); \\
		\sqrt{n}\left(\widehat{\tau}^\textup{IPW}_{\mathbf{dd^\prime}}-\tau^{\textup{IPW}}_{\mathbf{dd^\prime}}\right) &= \frac{1}{\sqrt{n}}\sum_{i=1}^{n}\bigg\{\psi^{\textup{IPW}}_i-\mathbf{b}^\prime_{i\bm{\gamma}_{\mathbf{d}}} \mathbb{E}(\Psi^{\textup{IPW}}_{\bm{\gamma}_{\mathbf{d}}})-\mathbf{b}^\prime_{i\bm{\gamma}_{\mathbf{d^\prime}}} \mathbb{E}(\Psi^{\textup{IPW}}_{\bm{\gamma}_{\mathbf{d^\prime}}})+\mathbf{b}^\prime_{i\bm{\delta}_{d_2}} \mathbb{E}(\Psi^{\textup{IPW}}_{\bm{\delta}_{d_2}})-\mathbf{b}^\prime_{i\bm{\delta}_{d^\prime_2}} \mathbb{E}(\Psi^{\textup{IPW}}_{\bm{\delta}_{d^\prime_2}})\bigg\} \\
        &+ o_p(1); \\
		\sqrt{n}\left(\widehat{\tau}^\textup{DR}_{\mathbf{dd^\prime}}-\tau^{\textup{DR}}_{\mathbf{dd^\prime}}\right) &= \frac{1}{\sqrt{n}}\sum_{i=1}^{n}\bigg\{\psi^{\textup{DR}}_i-\mathbf{b}^\prime_{i\bm{\beta}_{\mathbf{d^\prime}}}  \mathbb{E}(\Psi^{\textup{DR}}_{\bm{\beta}_{\mathbf{d^\prime}}}) -\mathbf{b}^\prime_{i\bm{\gamma}_{\mathbf{d}}} \mathbb{E}(\Psi^{\textup{DR}}_{\bm{\gamma}_{\mathbf{d}}}) - \mathbf{b}^\prime_{i\bm{\gamma}_{\mathbf{d^\prime}}} \mathbb{E}(\Psi^{\textup{DR}}_{\bm{\gamma}_{\mathbf{d^\prime}}}) +\mathbf{b}^\prime_{i\bm{\delta}_{d_2}}  \mathbb{E}(\Psi^{\textup{DR}}_{\bm{\delta}_{d_2}})\nonumber \\
        &-\mathbf{b}^\prime_{i\bm{\delta}_{d^\prime_2}}  \mathbb{E}(\Psi^{\textup{DR}}_{\bm{\delta}_{d^\prime_2}})\bigg\}+o_p(1),
\end{align*}}
where 
{\small \begin{align*}
		\psi^{\textup{OR}}& = w_1(\bm{\delta^\ast}_{d_2})\bigg((\Delta Y - \mu(\bm{\beta^\ast}_{\mathbf{d^\prime}}))-\mathbb{E}[w_1(\bm{\delta^\ast}_{d_2})(\Delta Y - \mu(\bm{\beta^\ast}_{\mathbf{d^\prime}}))]\bigg);	\\
		\Psi^{\textup{OR}}_{\bm{\beta}_{\mathbf{d^\prime}}}& =w_1(\bm{\delta^\ast}_{d_2})\bm{\dot{\mu}}(\bm{\beta^\ast}_{\mathbf{d^\prime}}); \\
		\Psi^{\textup{OR}}_{\bm{\delta}_{d_2}} & = \bm{\dot{w}}_1(\bm{\delta^\ast}_{d_2})\bigg((\Delta Y - \mu(\bm{\beta^\ast}_{\mathbf{d^\prime}}))-\mathbb{E}[w_1(\bm{\delta^\ast}_{d_2})(\Delta Y - \mu(\bm{\beta^\ast}_{\mathbf{d^\prime}}))]\bigg); \\
		\psi^{\textup{IPW}}& = w_1(\bm{\delta^\ast}_{d_2})\bigg(\Delta Y -\mathbb{E}[w_1(\bm{\delta^\ast}_{d_2})\Delta Y] \bigg)- w_2(\bm{\gamma^\ast}, \bm{\delta^\ast}_{d_2^\prime})\bigg(\Delta Y - \mathbb{E}[w_2(\bm{\gamma^\ast}, \bm{\delta^\ast}_{d_2^\prime})\Delta Y]\bigg);\\
		\Psi^{\textup{IPW}}_{\bm{\gamma}_{\mathbf{d}}} & = \bm{\dot{w}}_{2,\bm{\gamma}_{\mathbf{d}}}(\bm{\gamma^\ast}, \bm{\delta^\ast}_{d_2^\prime})\bigg(\Delta Y - \mathbb{E}[w_2(\bm{\gamma^\ast}, \bm{\delta^\ast}_{d_2^\prime})\Delta Y]\bigg);\\
		\Psi^{\textup{IPW}}_{\bm{\gamma}_{\mathbf{d^\prime}}} & =\bm{\dot{w}}_{2,\bm{\gamma}_{\mathbf{d^\prime}}}(\bm{\gamma^\ast}, \bm{\delta^\ast}_{d_2^\prime})\bigg(\Delta Y - \mathbb{E}[w_2(\bm{\gamma^\ast}, \bm{\delta^\ast}_{d_2^\prime})\Delta Y]\bigg); \\
		\Psi^{\textup{IPW}}_{\bm{\delta}_{d_2}}& = \bm{\dot{w}}_1(\bm{\delta^\ast}_{d_2})\bigg(\Delta Y - \mathbb{E}[w_1(\bm{\delta^\ast}_{d_2})\Delta Y]\bigg); \\
		\Psi^{\textup{IPW}}_{\bm{\delta}_{d^\prime_2}}& = \bm{\dot{w}}_{2,\bm{\delta}_{d_2^\prime}}(\bm{\gamma^\ast}, \bm{\delta^\ast}_{d_2^\prime})\bigg(\Delta Y - \mathbb{E}[w_2(\bm{\gamma^\ast}, \bm{\delta^\ast}_{d_2^\prime})\Delta Y]\bigg);\\
		\psi^{\textup{DR}}& = w_1(\bm{\delta^\ast}_{d_2})\bigg((\Delta Y - \mu(\bm{\beta^\ast}_{\mathbf{d^\prime}}))-\mathbb{E}[w_1(\bm{\delta^\ast}_{d_2})(\Delta Y - \mu(\bm{\beta^\ast}_{\mathbf{d^\prime}}))]\bigg) - w_2(\bm{\gamma^\ast}, \bm{\delta^\ast}_{d_2^\prime})\bigg((\Delta Y - \mu(\bm{\beta^\ast}_{\mathbf{d^\prime}})) \\
		&- \mathbb{E}[w_2(\bm{\gamma^\ast}, \bm{\delta^\ast}_{d_2^\prime})(\Delta Y - \mu(\bm{\beta^\ast}_{\mathbf{d^\prime}}))]\bigg); \\
		\Psi^{\textup{DR}}_{\bm{\beta}_{\mathbf{d^\prime}}}& = \bigg(w_1(\bm{\delta^\ast}_{d_2}) - w_2(\bm{\gamma^\ast}, \bm{\delta^\ast}_{d_2^\prime})\bigg)\bm{\dot{\mu}}(\bm{\beta^\ast}_{\mathbf{d^\prime}}); \\
		\Psi^{\textup{DR}}_{\bm{\gamma}_{\mathbf{d}}}& = \bm{\dot{w}}_{2,\bm{\gamma}_{\mathbf{d}}}(\bm{\gamma^\ast}, \bm{\delta^\ast}_{d_2^\prime})\bigg((\Delta Y - \mu(\bm{\beta^\ast}_{\mathbf{d^\prime}})) - \mathbb{E}[w_2(\bm{\gamma^\ast}, \bm{\delta^\ast}_{d_2^\prime})(\Delta Y - \mu(\bm{\beta^\ast}_{\mathbf{d^\prime}}))]\bigg);\\
		\Psi^{\textup{DR}}_{\bm{\gamma}_{\mathbf{d^\prime}}}& = \bm{\dot{w}}_{2,\bm{\gamma}_{\mathbf{d^\prime}}}(\bm{\gamma^\ast}, \bm{\delta^\ast}_{d_2^\prime})\bigg((\Delta Y - \mu(\bm{\beta^\ast}_{\mathbf{d^\prime}})) - \mathbb{E}[w_2(\bm{\gamma^\ast}, \bm{\delta^\ast}_{d_2^\prime})(\Delta Y - \mu(\bm{\beta^\ast}_{\mathbf{d^\prime}}))]\bigg); \\
		\Psi^{\textup{DR}}_{\bm{\delta}_{d_2}}& = \bm{\dot{w}}_1(\bm{\delta^\ast}_{d_2})\bigg((\Delta Y - \mu(\bm{\beta^\ast}_{\mathbf{d^\prime}})) - \mathbb{E}[w_1(\bm{\delta^\ast}_{d_2})(\Delta Y - \mu(\bm{\beta^\ast}_{\mathbf{d^\prime}}))]\bigg); 
        \end{align*}}
        {\small \begin{align*}
		\Psi^{\textup{DR}}_{\bm{\delta}_{d^\prime_2}}& =\bm{\dot{w}}_{2,\bm{\delta}_{d_2^\prime}}(\bm{\gamma^\ast}, \bm{\delta^\ast}_{d_2^\prime})\bigg((\Delta Y - \mu(\bm{\beta^\ast}_{\mathbf{d^\prime}})) - \mathbb{E}[w_2(\bm{\gamma^\ast}, \bm{\delta^\ast}_{d_2^\prime})(\Delta Y - \mu(\bm{\beta^\ast}_{\mathbf{d^\prime}}))]\bigg).
\end{align*}}

\subsection{Additional results: Monte Carlo experiments}\label{sec:simsresults}
\begin{table}[H]
	\centering
	\begin{threeparttable}
	\caption{Bias and coverage missingness-adjusted estimators}
	\label{tab:biascov}%
    \begin{tabular}{clrrrrrrrr}
    \toprule
          &       & \multicolumn{4}{c}{Bias}      & \multicolumn{4}{c}{Coverage} \\
          \cmidrule(lr){3-6} \cmidrule(lr){7-10}
    \multicolumn{1}{l}{Incorrect model} & PDATT & \multicolumn{1}{l}{R} & \multicolumn{1}{l}{DR} & \multicolumn{1}{l}{IPW} & \multicolumn{1}{l}{OR} & \multicolumn{1}{l}{R} & \multicolumn{1}{l}{DR} & \multicolumn{1}{l}{IPW} & \multicolumn{1}{l}{OR} \\
    \midrule
    \multirow{3}[0]{*}{M} & 11-00 & 0.000 & 0.019 & -0.036 & 0.018 & 0.951 & 0.941 & 0.933 & 0.941 \\
          & 10-00 & 0.000 & -0.028 & -0.028 & -0.028 & 0.954 & 0.921 & 0.924 & 0.917 \\
          & 01-00 & 0.000 & 0.036 & -0.026 & 0.036 & 0.955 & 0.913 & 0.952 & 0.912 \\
          \cmidrule(lr){3-10}
    \multirow{3}[0]{*}{P} & 11-00 & 0.000 & 0.000 & -0.116 & 0.000 & 0.944 & 0.944 & 0.781 & 0.945 \\
          & 10-00 & 0.000 & 0.000 & -0.052 & 0.000 & 0.955 & 0.955 & 0.814 & 0.954 \\
          & 01-00 & 0.002 & 0.002 & -0.036 & 0.002 & 0.948 & 0.945 & 0.928 & 0.953 \\
          \cmidrule(lr){3-10}
    \multirow{3}[0]{*}{O} & 11-00 & 0.001 & 0.001 & 0.002 & -0.098 & 0.943 & 0.942 & 0.945 & 0.612 \\
          & 10-00 & 0.001 & 0.001 & 0.000 & 0.025 & 0.944 & 0.942 & 0.945 & 0.923 \\
          & 01-00 & 0.000 & 0.000 & 0.000 & -0.066 & 0.934 & 0.935 & 0.945 & 0.833 \\
          \cmidrule(lr){3-10}
    \multirow{3}[0]{*}{None} & 11-00 & 0.001 & 0.001 & 0.003 & 0.000 & 0.944 & 0.944 & 0.929 & 0.952 \\
          & 10-00 & 0.001 & 0.002 & 0.002 & 0.001 & 0.954 & 0.954 & 0.951 & 0.954 \\
          & 01-00 & 0.001 & 0.001 & 0.004 & 0.001 & 0.945 & 0.943 & 0.931 & 0.948 \\
          \midrule
    \multirow{3}[0]{*}{M, P} & 11-00 & 0.013 & 0.009 & -0.157 & 0.009 & 0.944 & 0.946 & 0.290 & 0.937 \\
          & 10-00 & 0.015 & 0.010 & 0.005 & 0.010 & 0.944 & 0.953 & 0.953 & 0.953 \\
          & 01-00 & 0.000 & 0.032 & 0.003 & 0.032 & 0.954 & 0.907 & 0.953 & 0.901 \\
          \cmidrule(lr){3-10}
    \multirow{3}[0]{*}{M,O} & 11-00 & 0.089 & 0.109 & 0.089 & -0.030 & 0.780 & 0.669 & 0.741 & 0.919 \\
          & 10-00 & 0.039 & 0.005 & 0.005 & 0.007 & 0.891 & 0.952 & 0.953 & 0.958 \\
          & 01-00 & 0.181 & 0.185 & 0.146 & 0.065 & 0.385 & 0.357 & 0.530 & 0.805 \\
          \cmidrule(lr){3-10}
    \multirow{3}[0]{*}{P,O} & 11-00 & 0.204 & 0.204 & 0.147 & 0.099 & 0.311 & 0.309 & 0.536 & 0.597 \\
          & 10-00 & 0.098 & 0.098 & 0.053 & 0.073 & 0.482 & 0.474 & 0.801 & 0.671 \\
          & 01-00 & 0.212 & 0.212 & 0.178 & 0.171 & 0.093 & 0.098 & 0.269 & 0.140 \\
          \cmidrule(lr){3-10}
    \multirow{3}[0]{*}{All} & 11-00 & 0.296 & 0.294 & 0.252 & 0.231 & 0.001 & 0.001 & 0.022 & 0.004 \\
          & 10-00 & 0.161 & 0.156 & 0.154 & 0.146 & 0.054 & 0.065 & 0.075 & 0.094 \\
          & 01-00 & 0.366 & 0.372 & 0.361 & 0.362 & 0.000 & 0.000 & 0.000 & 0.000 \\
          \toprule
    \end{tabular}%
		\begin{tablenotes}[flushleft]
			\footnotesize
			\item \textit{Notes:} This table shows the bias and coverage of different missingness-adjusted estimators (R, DR, IPW, OR) for the PDATTs ${\tau}_{(11)(00)}$, ${\tau}_{(10)(00)}$, and ${\tau}_{(01)(00)}$. The first panel corresponds to the four experiments in which either only the missing data model (M), only the propensity score (P), only the outcome regression (O), or none of the models are misspecified (None). These results are also displayed in Figure~\ref{fig:bias}, where test size equals 1 minus the coverage. The second panel corresponds to the four experiments in which two or more models are misspecified.
		\end{tablenotes}
	\end{threeparttable}%
\end{table}%

\begin{table}[H]
	\centering
	\begin{threeparttable}
		\caption{Bias and coverage complete-case estimators}
		\label{tab:cc}%
    \begin{tabular}{clrrrrrr}
    \toprule
          &       & \multicolumn{3}{c}{Bias} & \multicolumn{3}{c}{Coverage} \\
          \cmidrule(lr){3-5} \cmidrule(lr){6-8}
    \multicolumn{1}{l}{Incorrect model} & PDATT & \multicolumn{1}{l}{DR} & \multicolumn{1}{l}{IPW} & \multicolumn{1}{l}{OR} & \multicolumn{1}{l}{DR} & \multicolumn{1}{l}{IPW} & \multicolumn{1}{l}{OR} \\
    \midrule
    \multirow{3}[0]{*}{M} & 11-00 & -0.075 & -0.070 & -0.076 & 0.780 & 0.853 & 0.720 \\
          & 10-00 & 0.082 & 0.081 & 0.082 & 0.671 & 0.681 & 0.636 \\
          & 01-00 & 0.060 & 0.061 & 0.059 & 0.843 & 0.821 & 0.823 \\
          \cmidrule(lr){3-8}
    \multirow{3}[0]{*}{P} & 11-00 & -0.073 & -0.160 & -0.072 & 0.840 & 0.616 & 0.754 \\
          & 10-00 & 0.139 & 0.143 & 0.139 & 0.120 & 0.106 & 0.115 \\
          & 01-00 & 0.243 & 0.196 & 0.243 & 0.042 & 0.238 & 0.005 \\
          \cmidrule(lr){3-8}
    \multirow{3}[0]{*}{O} & 11-00 & -0.090 & -0.083 & -0.152 & 0.725 & 0.830 & 0.261 \\
          & 10-00 & 0.125 & 0.124 & 0.166 & 0.348 & 0.358 & 0.084 \\
          & 01-00 & 0.010 & 0.029 & -0.069 & 0.942 & 0.928 & 0.785 \\
          \cmidrule(lr){3-8}
    \multirow{3}[0]{*}{None} & 11-00 & -0.073 & -0.045 & -0.073 & 0.788 & 0.925 & 0.734 \\
          & 10-00 & 0.143 & 0.143 & 0.143 & 0.200 & 0.219 & 0.157 \\
          & 01-00 & 0.205 & 0.217 & 0.205 & 0.143 & 0.183 & 0.051 \\
          \midrule
    \multirow{3}[0]{*}{M, P} & 11-00 & -0.068 & -0.212 & -0.067 & 0.788 & 0.116 & 0.731 \\
          & 10-00 & 0.079 & 0.100 & 0.079 & 0.588 & 0.422 & 0.580 \\
          & 01-00 & 0.108 & 0.084 & 0.109 & 0.466 & 0.686 & 0.444 \\
          \cmidrule(lr){3-8}
    \multirow{3}[0]{*}{M,O} & 11-00 & 0.005 & 0.026 & -0.126 & 0.941 & 0.919 & 0.424 \\
          & 10-00 & 0.103 & 0.103 & 0.133 & 0.540 & 0.553 & 0.285 \\
          & 01-00 & 0.222 & 0.225 & 0.083 & 0.154 & 0.131 & 0.709 \\
          \cmidrule(lr){3-8}
    \multirow{3}[0]{*}{P,O} & 11-00 & 0.115 & 0.079 & 0.065 & 0.742 & 0.796 & 0.811 \\
          & 10-00 & 0.202 & 0.199 & 0.203 & 0.004 & 0.005 & 0.004 \\
          & 01-00 & 0.195 & 0.173 & 0.192 & 0.172 & 0.322 & 0.067 \\
          \cmidrule(lr){3-8}
    \multirow{3}[0]{*}{All} & 11-00 & 0.250 & 0.221 & 0.180 & 0.028 & 0.124 & 0.076 \\
          & 10-00 & 0.224 & 0.233 & 0.206 & 0.003 & 0.002 & 0.006 \\
          & 01-00 & 0.390 & 0.388 & 0.391 & 0.000 & 0.000 & 0.000 \\
          \bottomrule
    \end{tabular}%
		\begin{tablenotes}[flushleft]
			\footnotesize
			\item \textit{Notes:} This table shows the bias and coverage of different complete case estimators (CC-DR, CC-IPW, CC-OR) for the PDATTs ${\tau}_{(11)(00)}$, ${\tau}_{(10)(00)}$, and ${\tau}_{(01)(00)}$. The first panel corresponds to the four experiments in which either only the missing data model (M), only the propensity score (P), only the outcome regression (O), or none of the models are misspecified (None). The second panel corresponds to the four experiments in which two or more models are misspecified.
		\end{tablenotes}
	\end{threeparttable}%
\end{table}%

\begin{table}[H]
	\centering
	\begin{threeparttable}
		\caption{Asymptotic variance missingness-adjusted estimators}
		\label{tab:asyvar}%
    \begin{tabular}{clrrrrr}
    \toprule
          &       &       & \multicolumn{4}{c}{Asymptotic variance} \\ \cmidrule(lr){4-7}
    \multicolumn{1}{l}{Incorrect model} & PDATT & \multicolumn{1}{l}{SEB} & \multicolumn{1}{l}{R} & \multicolumn{1}{l}{DR} & \multicolumn{1}{l}{IPW} & \multicolumn{1}{l}{OR} \\ \midrule
    \multirow{3}[0]{*}{M} & 11-00 & 35.219 & 37.777 & 37.343 & 50.763 & 26.509 \\
          & 10-00 & 27.580 & 25.689 & 25.736 & 26.648 & 24.216 \\
          & 01-00 & 49.783 & 50.518 & 49.905 & 66.202 & 32.152 \\
          \cmidrule(lr){3-7}
    \multirow{3}[0]{*}{P} & 11-00 & 57.115 & 57.898 & 57.862 & 86.586 & 29.412 \\
          & 10-00 & 26.115 & 21.505 & 21.284 & 22.759 & 20.766 \\
          & 01-00 & 52.421 & 37.433 & 39.576 & 49.701 & 32.309 \\
          \cmidrule(lr){3-7}
    \multirow{3}[0]{*}{O} & 11-00 & 49.132 & 63.760 & 63.766 & 63.716 & 33.606 \\
          & 10-00 & 26.405 & 29.653 & 29.804 & 28.431 & 27.687 \\
          & 01-00 & 69.098 & 84.985 & 85.770 & 81.449 & 41.790 \\
          \cmidrule(lr){3-7}
    \multirow{3}[0]{*}{None} & 11-00 & 51.099 & 49.404 & 49.404 & 66.035 & 29.526 \\
          & 10-00 & 26.594 & 26.606 & 26.802 & 27.794 & 24.511 \\
          & 01-00 & 66.058 & 63.900 & 66.508 & 84.983 & 40.782 \\
          \midrule
    \multirow{3}[0]{*}{M, P} & 11-00 & 58.774 & 29.436 & 29.236 & 39.910 & 23.650 \\
          & 10-00 & 26.071 & 19.306 & 19.254 & 20.205 & 19.156 \\
          & 01-00 & 65.492 & 27.841 & 27.913 & 33.560 & 25.729 \\
          \cmidrule(lr){3-7}
    \multirow{3}[0]{*}{M,O} & 11-00 & 35.149 & 51.667 & 51.049 & 50.764 & 29.758 \\
          & 10-00 & 27.277 & 28.509 & 28.384 & 28.045 & 27.261 \\
          & 01-00 & 50.897 & 71.076 & 69.880 & 65.923 & 34.572 \\
          \cmidrule(lr){3-7}
    \multirow{3}[0]{*}{P,O} & 11-00 & 54.012 & 78.425 & 78.520 & 81.508 & 33.172 \\
          & 10-00 & 25.823 & 24.185 & 23.764 & 22.828 & 22.910 \\
          & 01-00 & 50.923 & 41.826 & 42.238 & 47.309 & 31.432 \\
          \cmidrule(lr){3-7}
    \multirow{3}[0]{*}{All} & 11-00 & 56.496 & 35.752 & 35.747 & 38.027 & 26.300 \\
          & 10-00 & 25.993 & 21.512 & 21.396 & 21.606 & 20.906 \\
          & 01-00 & 61.038 & 29.826 & 29.817 & 31.485 & 26.399 \\
          \bottomrule
    \end{tabular}%
		\begin{tablenotes}[flushleft]
			\footnotesize
			\item \textit{Notes:} This table shows the semi-parametric efficiency bound (SEB) and the asymptotic variance of different missingness-adjusted estimators (R, DR, IPW, OR) for the PDATTs ${\tau}_{(11)(00)}$, ${\tau}_{(10)(00)}$, and ${\tau}_{(01)(00)}$. The first panel corresponds to the four experiments in which either only the missing data model (M), only the propensity score (P), only the outcome regression (O), or none of the models are misspecified (None). The second panel corresponds to the four experiments in which two or more models are misspecified.
		\end{tablenotes}
	\end{threeparttable}%
\end{table}%

\section{Empirical applications: Additional details}\label{sec:emp_sc}
\subsection{COVID-19 on voter turnout}
The pre-treatment period in our analysis is August 2018 when there were no COVID cases, therefore $D_0=0$ for all counties. Among counties where confirmed cases data is available, 59\% counties had above-average and 41\% had below-average number of cases. In the final period, 73\% of the counties had above-average number of cases whereas 27\% were below the average. For covariates that predict county-level turnout rates and the number of confirmed cases, we include the percentage of population that is aged 65 years or older, percentage of adults who have completed high school or higher, proportion of population that is black, white, or belong to two or more races, per-capita income, proportion of population speaking languages other than English, and log-transformed population. These are standardized before being used in estimation.

\subsection{Labor market conditions on income and hours worked}
Each month, CPS\footnote{Missing data problems are well-documented in surveys like CPS. First, item non-response can occur where individuals may fail to answer certain questions, resulting in missing values for sensitive variables like income and earnings \citep{bollinger2006match}. Second, unit non-response can result in entire households or individuals to not participate in the survey, resulting in missing rows corresponding to that unit. Third, CPS data are also known to include variables that are top-coded, such as incomes, which can create challenges in analyzing those variables \citep{burkhauser2012recent}. CPS employs imputation techniques (e.g. ``hot-deck'' imputation) to fill in missing data. As \citet{greenlees1982imputation} show, this can introduce potential biases in the resulting estimates.} surveys approximately 60,000 eligible households (or about 110,000 individuals) where households are interviewed for four consecutive months, left out for eight months, and then interviewed again the next four months. 

For each treatment, we examine two outcomes over time: family income (measured in \$1,000) and hours worked. We consider treatment-outcome-year combinations where the treatment is defined as either disability, job certification, or absence. The analysis spans years 2000 to 2024 and includes only those combinations that satisfy the following criteria. First, we restrict the sample to HHs who are not treated in the base period i.e. $D_0=0$. We exclude HHs who are interviewed in only one month or those who are interviewed in two consecutive months in a given year, as we require each household to be observed for at least 3 months. Furthermore, we only retain HHs for whom both the change in the outcome variable between the first and last periods and the treatment status in the first and last periods are observed (i.e., not missing). To ensure appropriate scaling, we use survey weights to expand the dataset, adjusting each individual’s weight to represent approximately 200 people rather than the original 2,000. After applying these criteria, we report percent differences across 4, 5, and 25 outcome-year samples for the disability, job certification, and work absence treatments, respectively. Since some covariates show no variation in some samples, the number of covariates included in the analysis varies across samples. 

Specifically, for disability, we have an average sample of $521,655$ observations across 4 outcome-by-year combinations, with an average of 3 covariates included in the models. In the middle period, an average of 1.04\% HHs are missing disability status. For HHs whose disability status is observed in the middle period, 0.09\% report having some difficulty and 0.15\% report having difficulty in the third period. For job certification, we have an average of $21,544$ observations across 5 outcome-by-year combinations where we control for an average of 5 covariates. In the middle period, around 2.56\% of HHs are missing job certification status. Among those whose status is observed, 3.45\% have job-certification in the middle period and 4.80\% have job-certification in the final period. Finally, for work absence, we consider an average of $418,934$ observations across 25 outcome-by-year combinations where we control for an average of 6 covariates. In the middle period, an average of 3.26\% of the HHs are missing absence status. Among HHs whose absence status is observed in the middle period, 5.75\% reported that they were absent from work and 4.05\% reported being absent from work in the final period.

\end{document}